\documentclass[reprint,amsmath,amssymb,aps,pre,longbibliography]{revtex4-1}

\usepackage[dvipsnames]{color}
\usepackage{graphicx}
\usepackage{dcolumn}
\usepackage{bm}
\usepackage[normalem]{ulem}
\usepackage{stackengine}
\usepackage{braket}

\begin{document}

\preprint{}

\title{Effective temperatures in inhomogeneous passive and active \\
bidimensional Brownian particle systems}

\author{Isabella Petrelli$^1$, Leticia F. Cugliandolo$^{2,3}$, Giuseppe Gonnella$^1$, Antonio Suma$^{1,4}$}
\affiliation{\centerline{$^1$Dipartimento  di  Fisica, Universit\`a  degli  Studi  di  Bari  and  INFN,
Sezione  di  Bari,}\\
\centerline{ via  Amendola  173,  Bari,  I-70126,  Italy} \\
\centerline{$^2$Sorbonne Universit\'e,  Laboratoire  de   Physique  Th\'eorique   et   Hautes   Energies, CNRS UMR 7589,}\\  
\centerline{4  Place  Jussieu,  75252  Paris  Cedex  05,  France}\\
\centerline{ $^3$Institut  Universitaire  de  France,  1  rue  Descartes,  75005  Paris  France}
\\
\centerline{
$^4$Institute for Computational Molecular Science, College of Science and Technology,}\\
\centerline{ Temple University, Philadelphia, PA 19122, USA}
}

\date{\today}

\begin{abstract}
We study the stationary dynamics of an active interacting Brownian particle system.
We measure the violations of the fluctuation dissipation theorem, and the corresponding effective 
temperature, in a locally resolved way. Quite naturally, in the 
homogeneous phases the diffusive properties and effective temperature are also homogeneous.
Instead, in the inhomogeneous phases (close to equilibrium and within the MIPS sector)
the particles can be separated in two groups with different diffusion properties 
and effective temperatures. Notably, at fixed activity strength the effective temperatures in the two phases remain distinct and approximately constant 
within the MIPS region, with values corresponding to the ones of the whole system at the boundaries of this sector of the phase diagram. 
We complement the study of the globally averaged properties with the theoretical and numerical characterization of the fluctuation distributions of the single particle
diffusion, linear response, and effective temperature in the homogeneous and inhomogeneous phases. We also 
distinguish the behavior of the (time-delayed) effective temperature from the 
(instantaneous) kinetic temperature, showing that the former is independent on the friction coefficient. 
\end{abstract}

\maketitle


\section{Introduction}
\label{sec:introduction}

The out of equilibrium dynamics of macroscopic classical systems attract
much theoretical and experimental interest. Basically, there are two ways in which a system can 
evolve out of equilibrium and break ergodicity: its relaxation times 
can exceed the measurable time scales, or external 
and/or internal agents can continuously inject energy and hinder equilibration. 
Glassy systems pertain to the first 
class while active matter is, possibly, the most exciting instance of the latter.

In active matter systems the generalized Stokes-Einstein relation
between injection and dissipation of energy is violated at the microscopic scale.
Energy is thus injected into the samples, they dissipate only part of it, and use the 
rest to perform directed motion. Numerous review articles report different theoretical and experimental aspects 
of active matter systems~\cite{Fletcher09,Menon10,Vicsek12,Ramaswamy13,Marchetti13,elgeti2015physics,Marenduzzo15,gonn15,Bechinger16,Bernheim-Groswasser18,CugliandoloHouches,carenza2019lattice,Bar20,lowen2020inertial}.

Recurrent in the analysis of macroscopic out of equilibrium systems is the search for notions borrowed from
equilibrium statistical physics and thermodynamics, which could guide one towards a better understanding of 
their dynamic behaviour. 
Although essentially out of equilibrium at a microscopic level, the {\it macroscopic}
character of at least some active matter systems is akin to the one of equilibrium systems in some respects.
Moreover, many models of active matter admit a
weak energy injection limit in which the evolution occurs close to equilibrium.
Consequently, small deviations from the equilibrium Gibbs-Boltzmann measure
characterize at least some aspects of the steady state.
For these reasons,  effective measures for the large scale properties of the stationary state, 
that are  close to equilibrium ones, have been proposed;  concerning these features 
a few relevant references are~\cite{Tailleur08,Tailleur09,Takatori15,Marconi15,Fodor16,Cagnetta17,Han17,Solon18,Chiarantoni20}.

A very recent study in this direction is the one of Han {\it et al.} \cite{Vitelli20} who demonstrated that, in a system of active spinners, a 
single effective temperature enters both the Boltzmann distribution and the equation of state. Notably, the same
effective temperature governs the linear response through the Green-Kubo relations for the shear and odd
viscosities. However, this is not a finalised story, and some other works show that extreme care has to be
taken when trying to use equilibrium-like measures to describe the full behaviour 
of active (and for that matter also glassy) systems. See, for instance,~\cite{Solon15,Ginot15,Guioth19}.

In the study of glassy systems, an effective thermal picture of the {\it large scale} dynamics 
can be exactly derived for mean-field models and it has been applied, quite successfully,  to realistic models as well~\cite{CugliandoloKurchanPeliti,CugliandoloKurchan00}.
The notion of an effective temperature defined from the deviations of the 
fluctuation dissipation theorem (FDT) out of equilibrium has been particularly useful in this regard. 
One can naturally wonder whether a similar scenario applies to active matter systems, at least for 
some range of parameters, possibly close to the passive limit. 

More precisely, the idea is to measure the effective temperature~\cite{CugliandoloKurchanPeliti,CugliandoloKurchan00} as 
the parameter that replaces the bath temperature in the fluctuation-dissipation relations 
between the time-delayed correlation and the linear response of the same observables. 
In practice, one exploits the relation (setting $k_B=1$):
\begin{equation}
 2T_{\text{eff}}(\Delta t)\chi(\Delta t)=\Delta^2(\Delta t)
 \; , 
\label{teff}
\end{equation}
which relates spontaneous (mean square displacement of, for example, the position of a tagged particle, $\Delta^2$) 
and induced (linear response to a perturbation applied to the position of the same particle, $\chi$) fluctuations.
In the long time-delay limit, $\Delta t \gg t^*$ with $t^*$ some transient scale, many interesting 
systems presenting collective phenomena, reach a regime in which $T_{\text{eff}}(\Delta t)$ saturates to a 
constant $T_{\text{eff}}$. Under certain conditions, summarised in the review articles~\cite{Cugliandolo11,Puglisi17,Sarracino19}, $T_{\rm eff}$ can be interpreted as an effective temperature, measurable with thermometers, controlling the direction and 
amount of heat flows, partial equilibrations, etc. 
 
Numerical measurements of such effective temperature in different active Brownian interacting systems appeared in~\cite{Loi1,Loi3,LevisBerthier,Preisler16,Szamel,Nandi18,Petrelli19,DalCengio19,flenner20}
(particles),~\cite{Loi2,Loi3} (polymers),~\cite{SumaPRE1,Petrelli18} (dumbbells).
Experimental measurements of the effective temperatures, mainly using tracer particles, 
were presented in, {\it e.g.},~\cite{wu,Mizuno07,Wilhem08,Gallet09,Palacci10,Patteson15,Maggi17,Cao20}.
In living systems, the violation of FDT has been used to characterize the forces generated by, for 
example, intracellular active processes. 
In all these cases, the active systems were in their homogeneous fluid phases. The effective temperatures 
were also considered in simpler effective models in which active homogeneous motion is mimicked by a single particle
generalized Langevin equation with memory~\cite{Szamel14,Fodor16,Wulfert17,Nandi18,Sevilla19b}. 
The relation between fluctuation theorems and FDT 
violations was discussed in~\cite{Zamponi05} and, in the context of active matter, 
some recent studies focused on the relations between the violation of FDT
and the entropy production in Active Ornstein-Uhlenbeck particles~\cite{Fodor16,Caprini19,Chaki18}
and effective coarse-grained stochastic scalar field theories~\cite{Nardini17}. 
Still, the full analysis of the thermodynamic properties linked to this definition, and whether it really behaves as a temperature, 
have not been fully explored in the context of active matter systems.

In this paper we focus on a system of 
active Brownian particles~\cite{Henkes11,Fily12,Fily14,Redner,Siebert17,Digregorio18,Digregorio19} 
with co-existence between dense and dilute phases in two regions of 
its phase diagram: 
close to the passive limit and at high levels of activity where it presents Motility Induced Phase 
Separation (MIPS). We are interested in quantifying the deviations from FDT in a locally resolved fashion
and thus characterizing the effective temperature, as defined in Eq.~(\ref{teff}), 
of the  two phases. To perform this task, we developed a new local (in real space) analysis that allowed us to classify particles according to the phase they belong to during a prescribed time interval, and to consequently measure, for each phase, the effective temperature using the same protocol of ref.~\cite{Petrelli19}. We complemented this analysis with a detailed theoretical and numerical characterization of the per-particle distributions of the displacement, the  linear response, 
and the effective temperature. In particular,  the displacement fluctuations were considered (mostly using tracers as the measuring means) in 
several experiments of active matter systems~\cite{wu,Howse07,Leptos09,Valeriani11,Ortlieb19,Klongvessa19,Lagarde20}, while the other two distributions are new to this field, though they have been studied in the 
context of coarsening and glassy systems~\cite{Castillo02,Castillo03,Roma07,Annibale09,Corberi12}.
Connections and differences with theories for dynamic 
fluctuations in glassy systems will be discussed~\cite{Chamon07}. At the end the paper, we also briefly discuss 
the differences between the effective temperature and the kinetic temperature, providing evidences for the differences between these  two 
objects, which access dynamics at completely different time-scales. More arguments on this topic can be found  in Ref.~\cite{Petrelli19}.

The paper is organized as follows. In Sec.~\ref{sec:methods} we give
the details of the model and the analytical and numerical tools employed. 
In Subsec.~\ref{subsec:model} we briefly describe the Active Brownian Particle 
(ABP) model and its phase diagram. Subsection~\ref{subsec:definitions} contains the
definitions of the mean square displacement, integrated linear response and time-delay dependent effective temperature. Subsection~\ref{subsec:single} presents the 
analytic results for a single ABP that will be used as a reference in the small packing
fraction limit. Subsec.~\ref{subsec:Malliavin}, explains the Malliavin auxiliary variables
numerical method for the evaluation of linear responses. Next, we present our results. Section~\ref{sec:global} is devoted 
to the analysis of the globally averaged diffusive and response properties. In 
particular, in Subsec.~\ref{subsec:homogeneous} we focus on the dynamics in the active homogeneous liquid phase, while Subsec.~\ref{subsec:inhomogeneous} is dedicated to the dynamics in  the inhomogeneous cases, both close to the passive
limit and in MIPS. In the latter case, we study the dynamical properties and the effective temperature of the two phases separately finding strong heterogeneities, hidden by global measurements over 
all the particles.
The fluctuations of individual displacement, response and effective temperature are
studied in great detail in Sec.~\ref{sec:fluctuations}. We present some analytical 
results for the distributions for a single passive and active particle together 
with numerical results both in dense homogeneous and heterogeneous passive and
active systems.
A short Sec.~\ref{sec:kinetic} distinguishes the behavior of the effective temperature from the one of 
the instantaneous kinetic temperature. 
We close the paper with a discussion Section in which we summarize our results.

\section{Methods}
\label{sec:methods}

In this Section we give details on the numerical model we use
and on the protocol we adopt to perform the measurements. We will first recall, in
Subsec.~\ref{subsec:model}, the definition of the ABP model and its phase diagram in
the plane packing fraction {\it vs.} P\'eclet number. We will also provide some details on the numerical
method and the parameters used. In Subsec.~\ref{subsec:definitions} we will define 
the observables chosen to study the dynamics and in 
Subsec.~\ref{subsec:single} we will summarize the known analytical results for
the mean square displacement, the response function and the effective temperature
of a single active Brownian particle. Finally, in Subsec.~\ref{subsec:Malliavin} we will
shortly describe the Malliavin weight sampling technique for the calculation of 
linear responses with simulations of unperturbed systems.

\subsection{The model}
\label{subsec:model}

We consider a system of $N$ identical active Brownian disks (ABPs) 
with mass $m$ and diameter $\sigma_{\rm d}$, confined inside a square box with 
area $A$ and periodic boundary conditions. Their dynamics are described by the Langevin equations:
\begin{eqnarray}
 m\ddot{\bold{r}}_i 
 &=&
 -\gamma\dot{\bold{r}}_i
 +
 \bold{F}^\lambda_i+\bold{F}^{\text{int}}_i
 +
 F_{\text{act}}\bold{n}_i
 +
  \sqrt{2\gamma T}\boldsymbol{\xi}_i \; ,
  \label{eq:Langevin-r}
 \\
 \dot{\theta}_i
 &=&
 \sqrt{2D_\theta}\eta_i \; .
 \label{eq:Langevin-theta}
\end{eqnarray}
($k_B=1$ hereafter). ${\bold{r}_i}$ is the position of the $i$-th particle in the two dimensional space. 
The particle system is supposed to be in contact with an environment that produces the 
friction force in the first term in the right-hand-side of Eq.~(\ref{eq:Langevin-r}) and the 
one in the left-hand-side of Eq.~(\ref{eq:Langevin-theta}), with $\gamma$ the friction 
coefficient. ${\bold{F}^\lambda_i}$ is an applied 
external force depending on a parameter ${\lambda}$. 
${\bold{F}_i^{\text{int}}=-\nabla\sum_{j(\ne i)}U(r_{ij})}$ is the force on the $i$-th particle 
due to the excluded volume repulsive interactions with all other particles.  
${r_{ij}=|\bold{r}_i-\bold{r}_j|}$ is the inter-particle distance and $U$ is a short-ranged 
repulsive potential of the form $U(r)=\left\{4\epsilon[(\sigma/r)^{64}-(\sigma/r)^{32}]+\epsilon\right\}\theta(2^{1/32}\sigma-r)$, 
with ${\sigma}$ and ${\epsilon}$ the parameters that set the length and energy scales 
of the interactions, and $\sigma_{\rm d}=2^{1/32}\sigma$. The self-propulsion is modelled as a constant magnitude force 
$F_{\text{act}}$ along ${\bold{n}_i=(\cos\theta_i,\sin\theta_i)}$. ${\boldsymbol{\xi}_i}$ and ${\eta_i}$ 
are uncorrelated zero-mean and unit variance Gaussian noises. Henceforth we will indicate the 
noise averages with angular brackets $\langle \dots \rangle$. One can 
easily check that the global angular diffusion is characterised by the diffusion coefficient 
${D_\theta=3T/(\gamma\sigma_{\rm d}^2)}$. We will express all relevant physical quantities 
in terms of mass, length and energy units 
given by m, ${\sigma}$ and ${\epsilon}$, with the reference time unit being $\tau=\sqrt{m\sigma^2/\epsilon}$. 

We vary the packing fraction, ${\phi=\pi\sigma_{\rm d}^2 N/(4A)}$, 
and the P\'eclet number, ${\text{Pe}=F_{\text{act}}\sigma_{\rm d}/T}$, that measures the strength of the work performed by the activity with respect to the thermal energy scale, while we keep all other parameters fixed.
Concretely, we fix $\sigma_{\rm d} = 1$ and $T=0.05$. We mainly  use $\gamma=10$,
which ensures the overdamped limit, but we also 
study how the friction coefficient affects the effective and kinetic temperatures. Typically, we collect data from 
systems with $N=256^2$ particles and we use 25-75 independent runs to construct the averages and 
probability distribution functions.

We used a velocity Verlet algorithm for solving Newton’s equations with an additional force term for the Langevin-type thermostat. In order to efficiently parallelize the numerical computation,  
we used the open source software Large-scale Atomic/Molecular Massively Parallel Simulator (LAMMPS), available at github.com/lammps~\cite{plimpton}.

\begin{figure}[h!]
\centerline{
\includegraphics[scale=0.27]{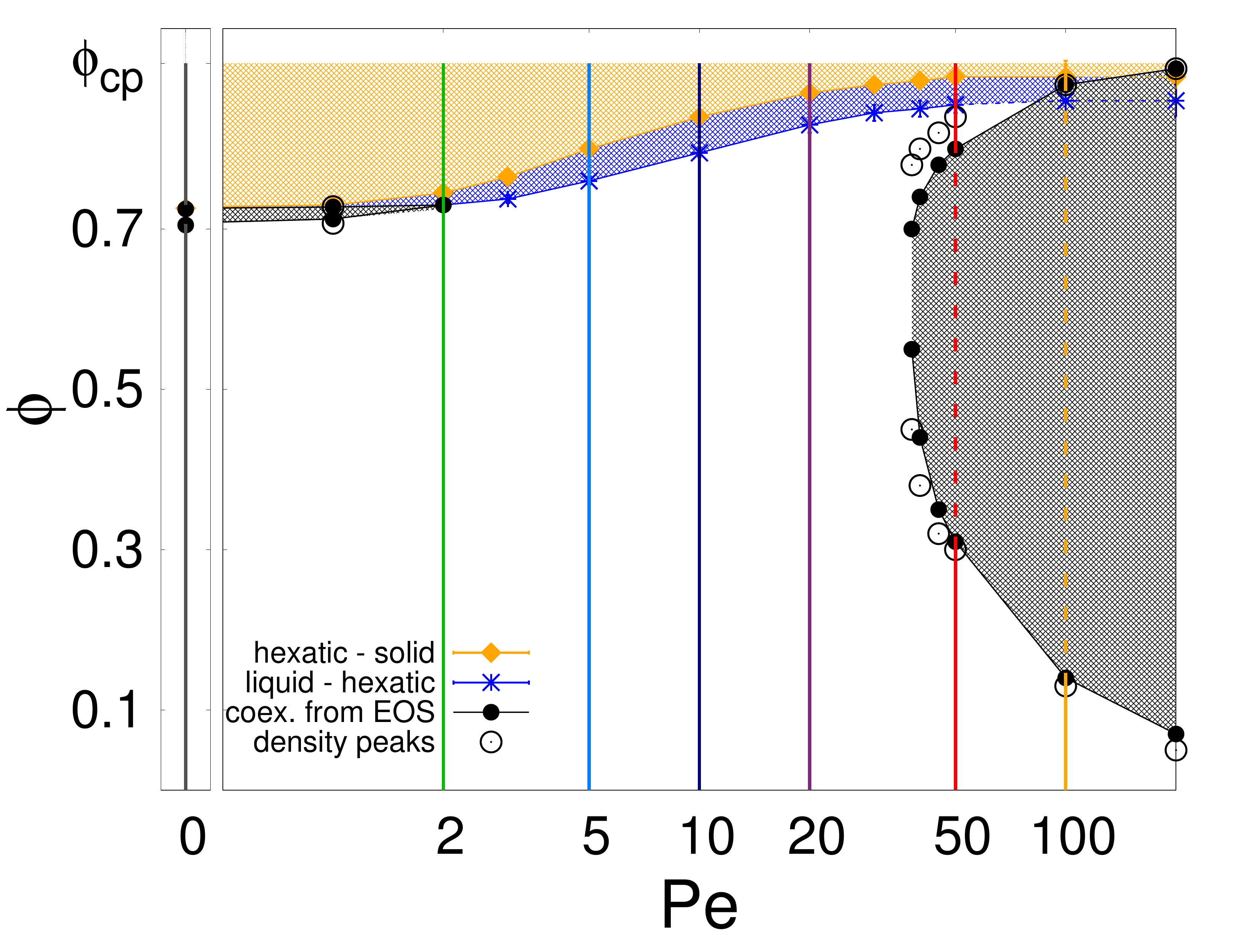}
}
\caption{(Color online.) 
The phase diagram of the ABP model at fixed temperature, $T=0.05$, in lin-log scale, as a function of 
two adimensional parameters: the P\'eclet number Pe  and the packing fraction $\phi$~\cite{Digregorio18}.
The white, blue and orange regions represent the liquid,  hexatic and 
solid phases, respectively. The grey zone at the left end is the co-existence region between 
hexatic and liquid, while the one at the right end is the region with Motility Induced
Phase Separation (MIPS). The lines and points at the border between the different phases 
have been obtained from the study of the hexatic and positional order correlation functions,
the equation of state (EOS) and the bimodal distribution function of the local density, as written
in the key. The vertical lines indicate the directions along which 
we vary $\phi$, at fixed Pe, in order to study the system's effective temperature.
}
\label{fig:phase-diagram}
\end{figure}

We study the dynamic behaviour of the 
system in different regions of the phase diagram. The latter has been
established  in~\cite{Digregorio18,Digregorio19} under  
similar conditions and we reproduce it in Fig.~\ref{fig:phase-diagram}. The phases are represented
with different colors and they correspond to the liquid (white), the hexatic (blue), the 
co-existence region between hexatic and liquid (in grey at the left end of the diagram), the solid (orange)
and the Motility Induced Phase Separation (MIPS) region (in grey at the right end of the diagram). 
The vertical straight lines, 
 added  to the phase diagram, 
 represent the paths along which we vary the parameters in order to study the effective temperature, 
namely, by changing $\phi$ at fixed Pe. The colour code (online) will be the same in all figures.

\subsection{Mean-squared displacement, linear response, and the fluctuation dissipation theorem}
\label{subsec:definitions}

The most usual way of testing the stochastic dynamics of an interacting system 
is to evaluate its global mean-square displacement.
Focusing on a tagged particle, say the $i$th one, the displacement induced by a 
given noise realisation is 
\begin{equation}
\tilde \Delta_i(t,0) = \bold{r}_i(t)-\bold{r}_i(0)
 \; ,
\label{eq:Delta2_i}
\end{equation}
with $\bold{r}_i(t)$ and $\bold{r}_i(0)$ the positions of the selected particle at times $t$ and $0$, 
respectively. 
Henceforth we will assume that the system reached a steady state and that the 
time $0$ is any reference time during this stationary regime.
The mean square displacement of the $i$th particle over the time interval $[0,t]$ 
is then given by the noise average of the above definition,
\begin{equation}
 \Delta^2_i(t,0)=\langle [\bold{r}_i(t)-\bold{r}_i(0)]^2 \rangle
 \; ,
 \label{eq:Delta2}
\end{equation}
and the global mean square displacement by the normalized sum over al particles, 
$\Delta^2 = N^{-1} \sum_{i=1}^N \Delta_i^2$.

Another less usual way of studying the dynamics of a collective system is to measure its response to 
weak perturbations. Such a procedure is put in practice, for example, 
by  applying a constant perturbation 
${\bold{F}^\lambda_i}$, say since ${t=0}$, and then measuring the averaged linear response of the system
to it. More precisely, one defines the instantaneous linear self response
\begin{equation}
 R_{i,\alpha\beta}(t,t')
 =
 \left.
 \frac{\delta\langle{r_{i,\alpha}(t)\rangle_{\bold{F}^\lambda}}}{\delta F_{i,\beta}^\lambda(t')}
 \right|_{\bold{F}^\lambda =0}
\end{equation}
where $\alpha$ and $\beta$ are coordinate indices running from $1$ to $d$, with $d$ the space dimension. 
The time-integrated linear response is then
\begin{equation}
 \chi_{i, \alpha\beta}(t,0)
 =
 \int_0^t dt' \; R_{i,\alpha\beta}(t,t')
 \; , 
 \end{equation}
 the self contributions averaged over all spatial directions is
 \begin{equation}
 \chi_{i}(t,0)
 =
 \int_0^t dt' \; \sum_{\alpha=1}^d R_{i,\alpha\alpha}(t,t')
 \; ,
 \label{eq:self_resp}
 \end{equation}
 and the global function is the average over all particles
 $\chi(t,0) = N^{-1} \sum_{i=1}^N \chi_i(t,0)$. A noise-dependent fluctuating quantity, the average
of which yields the linear response in Eq.~(\ref{eq:self_resp}), will be identified in Sec.~\ref{subsec:Malliavin}.
 
In equilibrium, the mean square displacement and the mean integrated linear response
are related by the model independent equation
\begin{equation}
2T \chi_i(t,0) = \Delta_i^2(t,0)
\; , 
\end{equation}
that states the fluctuation-dissipation theorem (FDT).  

Since the dynamics of this problem can be heterogeneous, with 
different particles behaving differently over the time interval considered, 
$\chi_i$ and $\Delta^2_i$ could have  time 
variations that depend strongly on the particle considered. 
Still, under equilibrium conditions the relation above 
should  remain particle independent.

 \subsection{A single active Brownian particle}
 \label{subsec:single}

In the numerical  analysis of the linear response function we 
apply a force along the $x$ direction of space and 
we focus on the linear response of the system along this direction.
Concerning the mean-square displacement, we measure it 
according to the definition in Eq.~(\ref{eq:Delta2}), that is to say, 
by summing over all directions of space, which yields, on average,
$\Delta_i^2(t,0)=\langle [\bold{r}_i(t)-\bold{r}_i(0)]^2 \rangle=d\langle [x_i(t)-x_i(0)]^2 \rangle$. For this reason, there will be 
some unusual $d$ factors in the relations appearing in the rest of this paper. They 
take care of the different ways of calculation linear responses and displacements.
In this Section we use the same protocol.

For a single active Brownian particle a straightforward calculation leads to
the integrated linear response along one spatial direction
\begin{equation}
 \chi(t,0)
 =\frac{t}{\gamma}\equiv\mu_{\rm sp}t,
 \; ,
\end{equation}
with $\mu_{\rm sp}$ the single particle mobility.
The mean square displacement reads
\begin{equation}
 \Delta^2(t,0)
 =
 \frac{2}{D_\theta^2}\frac{F_{\text{act}}^2}{\gamma^2}
 \; 
 \left( D_\theta t+e^{-D_\theta t}-1 \right) 
 +\frac{4T}{\gamma} \, t
 \; , 
 \label{delta_sp}
\end{equation}
where we have taken into account the $d=2$ factor, 
and in the long time limit it approaches the normal diffusion form
 \begin{equation}
 \Delta^2(t,0)
 \to 2d D_{\rm sp} t
 \; , 
 \end{equation}
 with the single particle diffusion coefficient 
 \begin{equation}
D_{\rm sp}=\frac{T}{\gamma}\left(1+\frac{\text{Pe}^2}{6}\right)
\; .
\end{equation}

The effective temperature of a single active Brownian particle 
is then immediately derived by taking the asymptotic value of the ratio 
between displacement and response
\begin{eqnarray}
T_{\rm eff} &=& \frac{\lim_{t\gg1}\Delta^2(t,0)}{2d\chi(t,0)} = \frac{D_{\rm sp}}{\mu_{\rm sp}}
\nonumber\\
&=& 
T\left(1+\frac{\text{Pe}^2}{6}\right)
\; . 
 \label{eq:Teff-single}
\end{eqnarray}

\subsection{Malliavin weights}
\label{subsec:Malliavin}

In simulations, it can be very hard to control the vanishing perturbation limit needed to 
calculate the linear response function. 
Fluctuations and numerical errors become increasingly large as the perturbation approaches zero.  
For this reason, techniques that allow one to calculate linear 
responses with simulations of the unperturbed system only have been developed for some 
kinds of systems. The trick is to find an exact relation between the linear response and a correlation between some variables evaluated in the unperturbed system that could be, though, complicated (even non-local in time) but still manageable. 
The advantage of these methods is that the zero perturbation limit is taken analytically and it does not introduce strong numerical uncertainties.
Such methods were introduced by Chatelain \cite{Chatelain} and Ricci-Tersenghi 
\cite{Ricci-Tersenghi} for Ising spin systems evolving with Monte Carlo dynamics. These ideas  were then generalised by Corberi \emph{et al.} \cite{Corberi} to treat 
similar systems evolving in discrete time via a stochastic 
non-equilibrium Markov process. Other relations of this kind, 
though not necessarily presented with the aim of simplifying the numerical computation of the 
linear response, can be found in~\cite{CuKuPa,Gradenigo12,Speck,Maes,DalCengio19} to cite a few papers 
where these ideas were explored.

Lately, Warren and Allen \cite{WarrenAllen1,WarrenAllen2} presented a similar approach for
interacting Brownian particles of the type we are dealing with here.
The method involves  tracking, in an unperturbed system,  auxiliary stochastic variables, termed \textit{Malliavin weights}. More recently, Szamel \cite{Szamel} 
applied it to systems of active particles propelled by a persistent (colored) noise and 
interacting {\it via} a screened Coulomb potential. We have already used this method in a preliminar
study of active Brownian disks~\cite{Petrelli19}.

For the model and perturbation considered in this paper, 
the Malliavin weights sampling (MWS) technique is particularly simple. It
 consists in evaluating the correlator between the position of the particle 
 and the thermal white noise acting on it. Indeed, 
denoting by  ${P(\{\bold{r}_i\};t)}$  the joint probability distribution function (pdf) of all particles' positions, one 
defines
\begin{equation}
Q_\lambda(\{\bold{r}_i\};t)\equiv
\frac{\partial \ln P(\{\bold{r}_i\};t)}{\partial \lambda}
\; ,
\end{equation}
with ${\lambda}$ the strength of the perturbation.
Next, one  introduces the  ``Malliavin weight'',
that is to say an auxiliary stochastic variable, 
${q_\lambda}$ and lets it evolve, starting from 
${q_\lambda=0}$, according to the rule
\begin{equation}
 q_\lambda(t')=q_\lambda(t)+\frac{\partial \ln W(\{\bold{r}'_i\};t'|\{\bold{r}_i\};t)}{\partial \lambda}
 \; ,
 \label{updateq}
\end{equation}
where ${W(\{\bold{r}'_i\};t'|\{\bold{r}_i\};t)}$ is the propagator, \emph{i.e.} the probability of finding the particles at the positions ${\{\bold{r}'_i\}}$ at ${t'=t+\Delta t}$ given the positions ${\{\bold{r}_i\}}$ at $t$. 
Given the previous updating rule, it is then readily shown that
$  \langle q_\lambda\rangle_{\{\bold{r}_i\}}  = Q_\lambda(\{\bold{r}_i\};t)$ \cite{WarrenAllen1}.
Let ${A(\{\bold{r}_i\})}$ be a generic function of the coordinates. Its response to a change of  $\lambda$ is given by its average, in the unperturbed system, 
weighted by the appropriate ${q_\lambda}$:
\begin{equation}
\frac{\partial\langle A\rangle_\lambda}{\partial\lambda}=\langle Aq_\lambda \rangle
\; .
\end{equation}

In general, one cannot calculate the exact propagator but this method can still be applied  
numerically, by enforcing the relation \eqref{updateq} at each time-step with an explicit integration scheme.
If we integrate the equations of motion  using a standard Euler-Maruyama scheme, the propagator reads 
\begin{widetext}
\begin{eqnarray}
&&
W(\{\bold{r}'_i,\theta'_i\};t'|\{\bold{r}_i,\theta_i\};t)
=
\Big[(2\pi)^3\left(\frac{2T}{\gamma}\right)^22D_\theta(\Delta t)^3\Big]^{-N/2}
\;
\nonumber\\
&&
\qquad
\times \; \prod_{i=1}^N
\exp\Bigg\{
-\frac{(x'_i-x_i-\frac{F_\text{act}}{\gamma}\cos\theta_i\Delta t-\frac{F_{i,x}^{\text{int}}}{\gamma}\Delta t-\frac{F_{i,x}^\lambda}{\gamma}\Delta t)^2}{2\frac{2T}{\gamma}\Delta t}
-\frac{(y'_i-y_i-\frac{F_\text{act}}{\gamma}\sin\theta_i\Delta t-\frac{F_{i,y}^\text{int}}{\gamma}\Delta t-\frac{F_{i,y}^\lambda}{\gamma}\Delta t)^2}{2\frac{2T}{\gamma}\Delta t}
\nonumber\\
&& 
\qquad\qquad\qquad\quad\;\;
-\frac{(\theta'_i-\theta_i)^2}{2D_\theta\Delta t}\Bigg\}
\label{propagator}
\end{eqnarray}
\end{widetext}
and  we just need to let the 
Malliavin weight evolve according to Eq.~\eqref{updateq}
using the expression in Eq.~(\ref{propagator}) for the propagator. For  ${\bold{F}_i^\lambda=\lambda\bold{e}_x}$
one deduces
\begin{widetext}
\begin{eqnarray}
q'_\lambda
=
q_\lambda+\frac{x'_i-x_i-\frac{F_\text{act}}{\gamma}\cos\theta_i\Delta t-\frac{F_{i,x}^\lambda}{\gamma}\Delta t-\frac{F_{i,x}^\text{int}}{\gamma}\Delta t}{2T}\frac{\partial F^\lambda_{i,x}}{\partial\lambda}
=
q_\lambda+\sqrt{\frac{\Delta t}{2\gamma T}}
\, \omega_i^x,
\label{rule}
\end{eqnarray}
\end{widetext}
where ${\omega_i^x}$ are the normal random variables with expected 
value zero and unit variance used in the algorithm to simulate the noise.
Notice that all the derivatives with respect to $\lambda$ are evaluated at $\lambda=0$, corresponding to the unperturbed evolution.
Finally, 
\begin{eqnarray}
 \chi_i(t,0)
 &=&
 \frac{\partial\langle x_i(t)\rangle_\lambda}{\partial\lambda}
 = \langle x_i(t)q_\lambda(t)\rangle
 \nonumber\\
 &=&
 \frac{1}{{\sqrt{2\gamma T}}} \int_0^t dt' \, \langle x_i(t)\xi_{i,x}(t')\rangle
 \; , 
\label{chi}
\end{eqnarray}
in terms of a correlation with  the noise, that we wrote  in continuous time notation.
Therefore, for the model and perturbing forces considered here, the Malliavin weights sampling technique 
reduces to the evaluation of the correlation between the position of the tagged particle and the 
thermal noise acting on it. In practice, we imagine that the perturbation is acting on each particle and we evaluate 
${\chi}$ considering an average over the selected particles we will be interested in. In heterogeneous cases we 
will implement a method to distinguish those in dilute and dense phases and we will study them separately.

Besides, we will be interested in studying the spatial fluctuations of the particle displacement 
and the quantity 
\begin{equation}
\tilde \chi_i(t,0) = \frac{1}{\sqrt{2\gamma T}} \int_0^{t} dt' \, x_i(t) \xi_{i,x}(t')
\label{eq:susc_i}
\end{equation}
 that, once averaged and time integrated, gives rise to the 
linear susceptibility.

\section{Global properties}
\label{sec:global}

In this Section we study the globally averaged diffusive and
response properties of the ABP model defined in Subsec.~\ref{subsec:model}, and we compare them to the 
equilibrium limit. We start our analysis in the homogeneous region of the 
phase diagram, Subsec.~\ref{subsec:homogeneous}. In 
Subsec.~\ref{subsec:inhomogeneous} we move to  the coexisting 
regions, either in the passive limit or within the MIPS sector. In the latter cases
we separate the particles which belong to the dense phases from  the ones 
belonging to the dilute phases and we find evident heterogeneities which are
hidden when flat global averages are performed.

\subsection{Homogeneous phases}
\label{subsec:homogeneous}

\begin{figure}[b!]
\hspace{-3.5cm} (a) \hspace{4.25cm} (b) \\
\centerline{
\includegraphics[width=4.25cm]{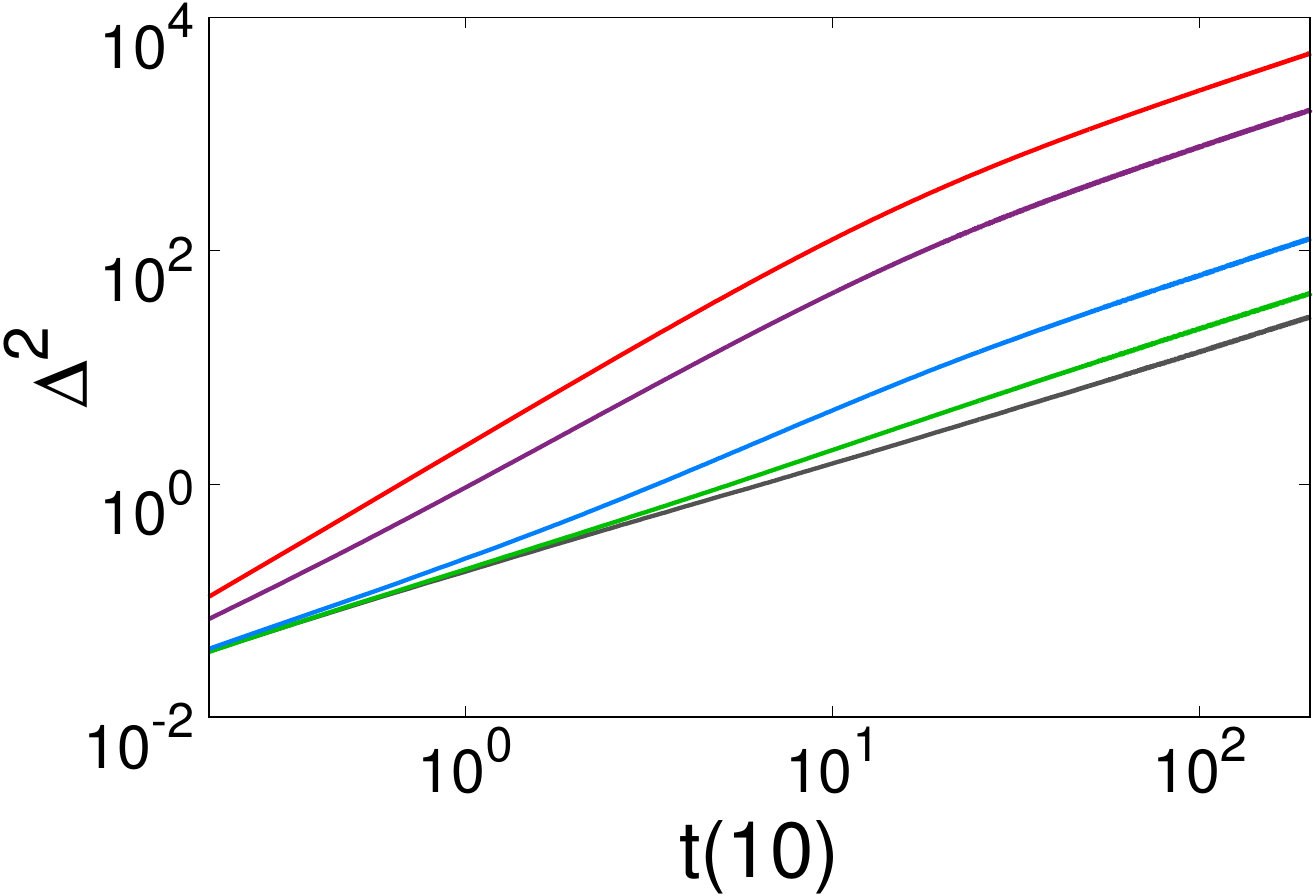}
\includegraphics[width=4.25cm]{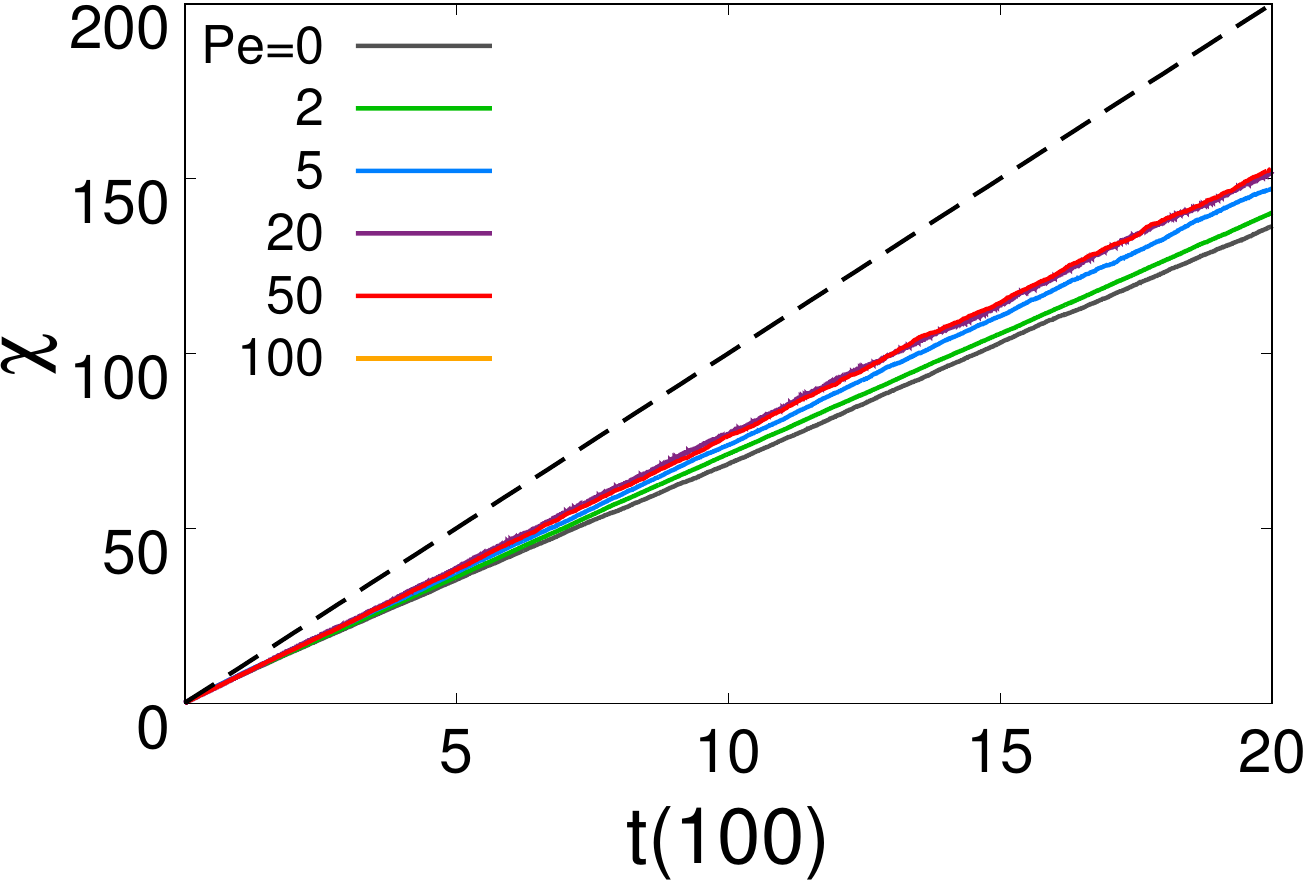}
}
\hspace{-3.5cm} (c) \hspace{4.25cm} (d) \\
\centerline{
\includegraphics[width=4.25cm]{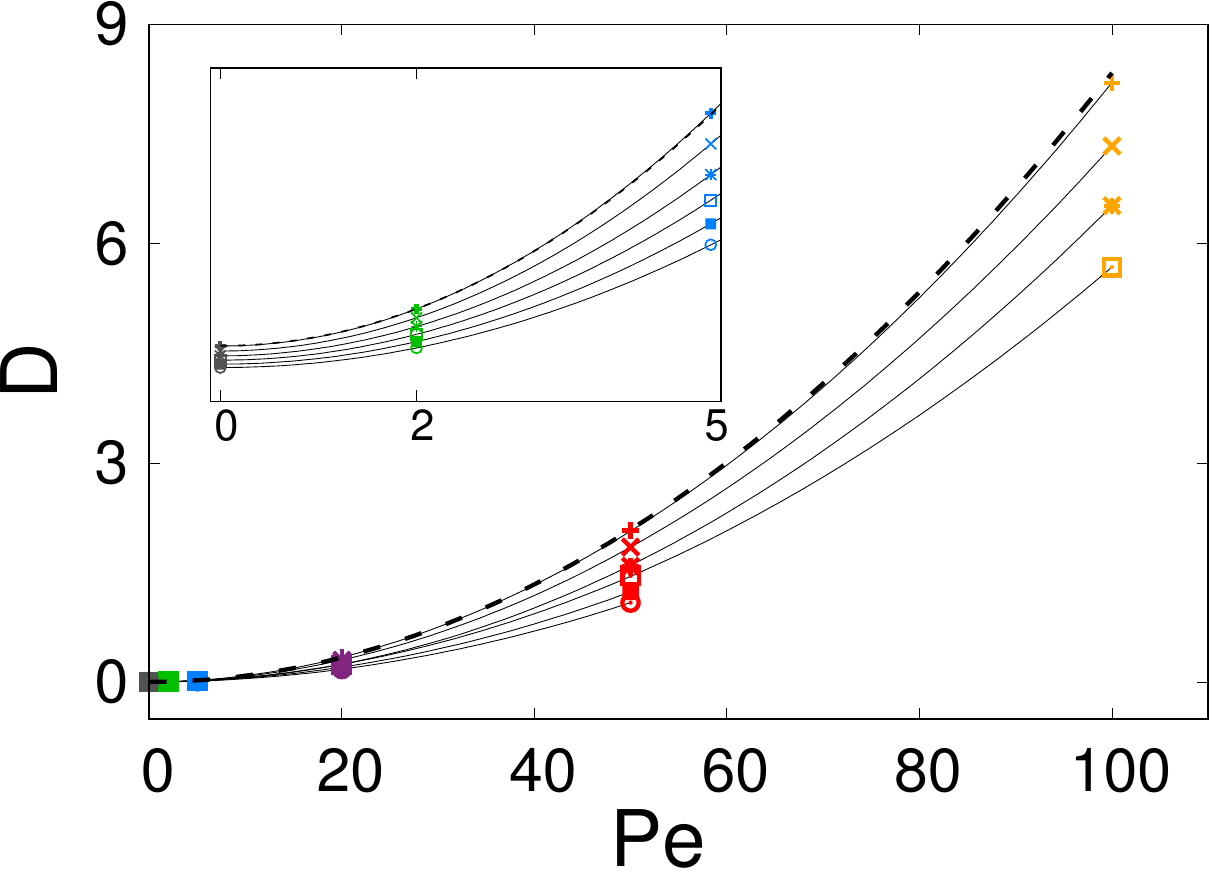}
\includegraphics[width=4.25cm]{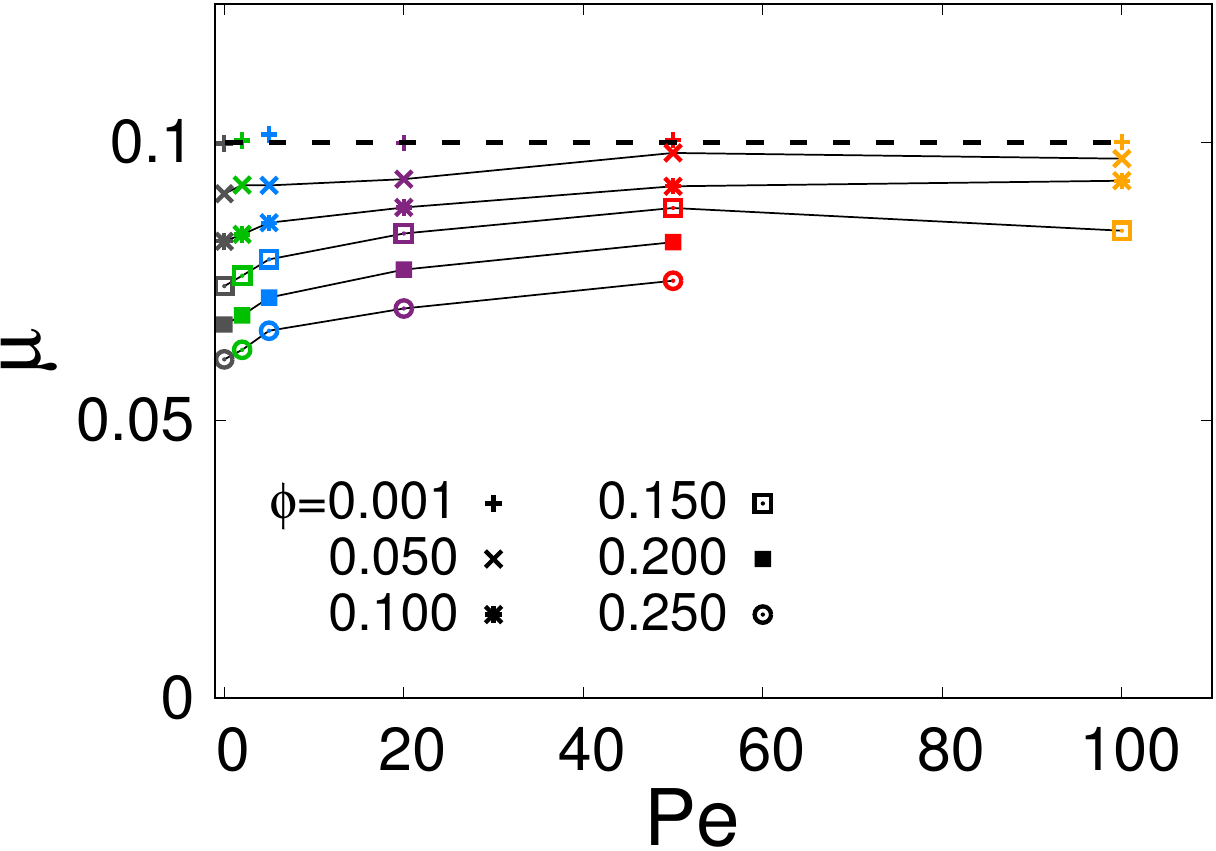}
}
\caption{(Color online.) 
Homogeneous liquid phase.
Global 
(a) mean squared displacement, in log-log scale, and (b) integrated linear 
response function as a function of delay time 
for packing fraction $\phi=0.200$ and P\'eclet values
Pe = 0, 2, 5, 20, 50, plotted with the line code given in the key,
see also Fig.~\ref{fig:phase-diagram} where this Pe values are indicated as
vertical lines with the same color. 
The diffusion constant (c) and mobility (d) as a function of the 
packing fraction $\phi$ for six different values of Pe, see the key in panel (b). 
The dashed lines are the single particle behavior
and the key to the datapoints is in panel (d).
The data shown in panel (c) are fitted using a quadratic
function, while the solid line in panel (d) are guides to the eye.
(Some points are missing because for $\phi=0.200, \ 0.250$ and 
Pe = $100$ the system is already in the MIPS region.)
}
\label{fig:DeltaChiDmu-homo}
\end{figure}

\begin{figure}[h!]
(a) $\;$ \hspace{5cm} $\;$ \\
\vspace{0.3cm}
\centerline{
\includegraphics[width=7.4cm]{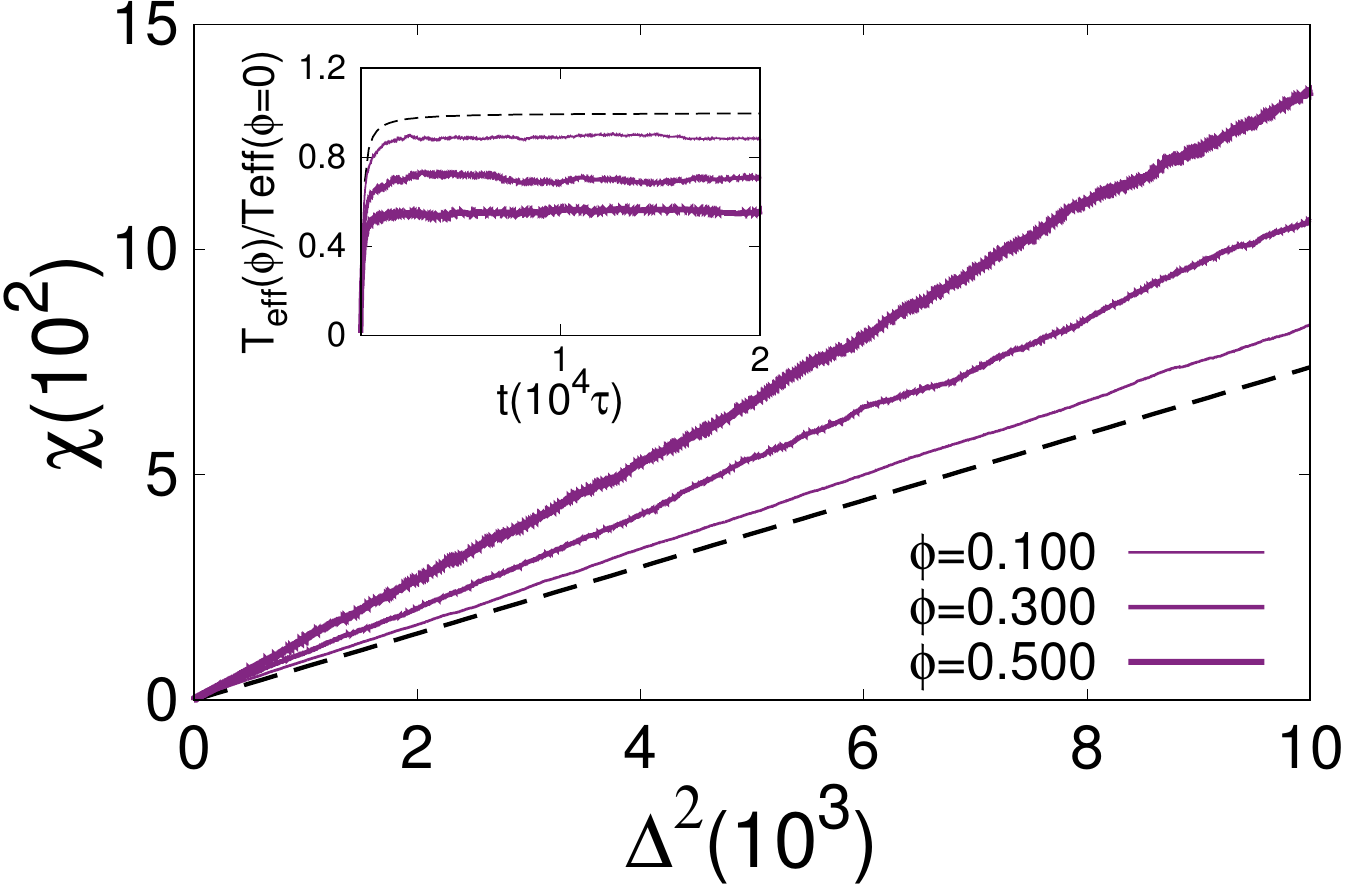}
}
(b) $\;$ \hspace{5cm} $\;$ \\
\vspace{0.3cm}
\centerline{
\includegraphics[width=7cm]{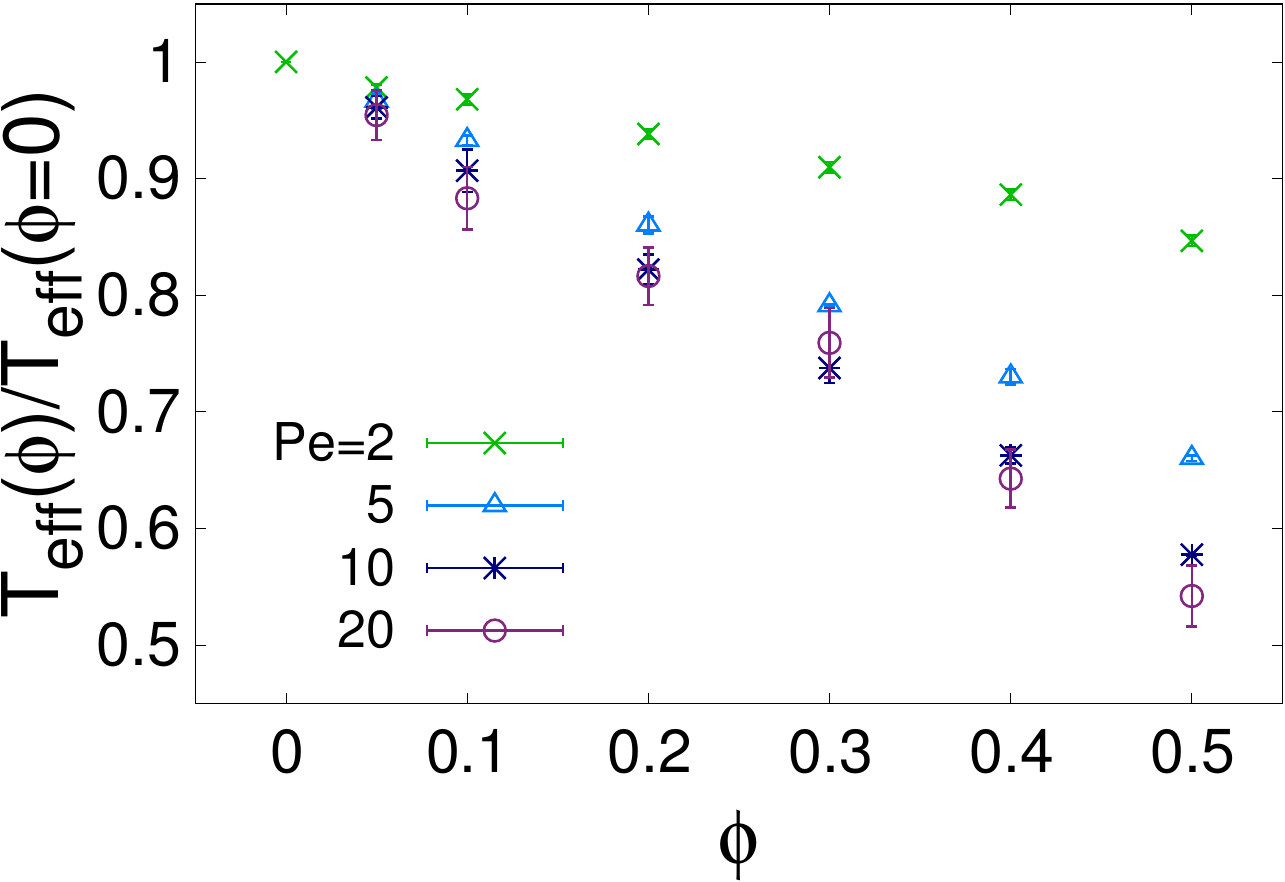}
}
\caption{(Color online.) 
Homogeneous liquid phase.
(a) The global integrated linear response function \emph{vs.} 
the mean squared displacement (main plot) and the effective temperature normalized 
by the one of a single active particle (inset) for 
Pe = $20$ and the packing fractions values given in the key. 
The dashed line is the single particle behavior.
(b) The long-time effective temperature \emph{vs.} 
the packing fraction for the activities given in the key. 
}
\label{fig:Teff-homo}
\end{figure}

As long as we stay in the homogeneous liquid phase, both the 
global mean square displacement, shown in Fig.~\ref{fig:DeltaChiDmu-homo}(a) for $\phi=0.2$,  
and the global linear response, shown in Fig.~\ref{fig:DeltaChiDmu-homo}(b) for the same $\phi$, 
increase with increasing P\'eclet number, and slow down 
when the density is increased. 
At the low densities of this plot, the dynamics reach a normal 
diffusive regime and
in Fig.~\ref{fig:DeltaChiDmu-homo}(c)-(d)
we see that both the diffusion coefficient (defined through the long-time limit of the mean squared displacement $D=\lim_{t\gg 1}\Delta^2(t,0)/(2dt)$) and the mobility (defined as $\mu=\lim_{t\gg1}\chi(t,0)/t$) decrease when the packing fraction is 
increased for all values of the P\'eclet number such that the system remains in the
homogeneous liquid phase. The combination of these behaviors also results in a 
decrease of the effective temperature  with increasing packing 
fraction. These claims were already shown in~\cite{Petrelli19} and 
are supported by the numerical data displayed in Figs.~\ref{fig:DeltaChiDmu-homo} 
and \ref{fig:Teff-homo}. In the latter, the parametric construction $\chi(\Delta^2)$
is presented in the main plot of panel (a) for Pe = 20 and several packing fractions. 
Extracting the effective temperature from these measurements, 
one sees that after a transient with a non-trivial dependence on the time delay, 
there is a time scale beyond which $T_{\rm eff}$ saturates to a constant, 
see the inset of the same panel, where the effective temperatures measured from the long
time limit, and normalized by the effective temperature at Pe = 0, are shown.
The dependence of this long-time value on the packing fraction of the system
is displayed in Fig.~\ref{fig:Teff-homo}(b).
The effective temperature is a monotonic decreasing function 
of the global density (a similar trend was found in~\cite{LevisBerthier}). 

Nandi and Gov proposed a simple picture for the long-term dynamics 
of active systems, in which the behaviour of the global system is reduced to the one of 
a single active particle inside a visco-elastic fluid,  that leads to a dependence of the effective temperature 
of the form $A/(1+B\phi^2)$~\cite{Nandi18}. Although our numerical data are compatible with this form, 
they are also rather well described by pure linear decays, and it is hard to establish beyond doubt which 
of the two functions describes more accurately the intermediate packing fraction behaviour, $\phi\stackrel{<}{\sim} 0.5$,
that we study here.

The curves in Fig.~\ref{fig:Teff-homo}(b) tend to 1 at vanishing packing fraction 
indicating that for dilute enough systems $T_{\rm eff}$ grows as Pe$^2$ similarly to what was 
found in~\cite{Loi1,Loi2,Loi3} for other interacting active particle and molecular models,
and in~\cite{SumaPRE1} for an active dumbbell system. This dependence is conserved at
finite relatively low
density as can be guessed from the Pe$^2$ dependence of the diffusion coefficient and 
almost constant behavior of the mobility shown in Fig.~\ref{fig:DeltaChiDmu-homo}.

\subsection{Inhomogeneous phases}
\label{subsec:inhomogeneous}

When entering the coexisting regions, either between the fluid and the hexatically 
ordered phase at low Pe, or within the MIPS sector of the phase diagram at high Pe, 
the scenario becomes richer and more complex. 
On the one hand, we can perform global averages over all particles, 
be them in dense or dilute regions of the sample. On the other hand, we can
go beyond these naive measurements and differentiate the dynamics of the two phases separately. 

More precisely, we will demonstrate  that one can attribute a dilute/dense character to the particles, 
over a chosen time interval, and that these behave as if they were characterized by 
two different effective temperatures. The latter fact is hidden when flat global averages are performed.
We will then compare this kind of heterogeneities to the ones already studied in different glassy 
materials.

\subsubsection{Method for particle distinction}

In heterogeneous systems, we proceed as follows to separate the particles in two groups. 
First of all, we evaluate the hexatic local order parameter, $\psi_{6i} = (1/N^i_{nn}) \sum_{j(i)} e^{6{\rm i} \theta_{ij}}$
where $\theta_{ij}$ is the angle formed by the bond between the selected particle $i$ and 
its first neighbor $j$, and a reference axis, say the horizontal one. It is clear that, in order to 
define these bonds, a notion of neighborhood needs to be introduced. This is done 
by performing a Voronoi tessellation of space. The sum $\sum_{j(i)}$  runs over nearest neighbors ($N^i_{nn}$ in total), in the 
Voronoi sense,  of the selected
particle. In this way, each particle acquires a vector $\psi_{6i} $ that is attached to it.
Next, we perform a time-average of the absolute value of the  local hexatic order parameter over a time window of 
duration $\Delta t$, 
$(1/\Delta t) \, \int_t^{t+\Delta t} dt' \, |\psi_{6i}(t')|$.
The statistics of the thus constructed mean local hexatic order parameter shows a bimodal structure. 
We use the central minimum of the probability distribution function, located between the two local maxima, to separate the particles in two subgroups: 
the ones which spent most of their lifetime $\Delta t$ in the dilute phase and those which 
spent most of their lifetime $\Delta t$ in the dense phase. 

\begin{figure}[h!]
\hspace{-2.25cm} (a) \hspace{4cm} (b)  \hspace{3cm} 
\centerline{
\includegraphics[width=4.5cm]{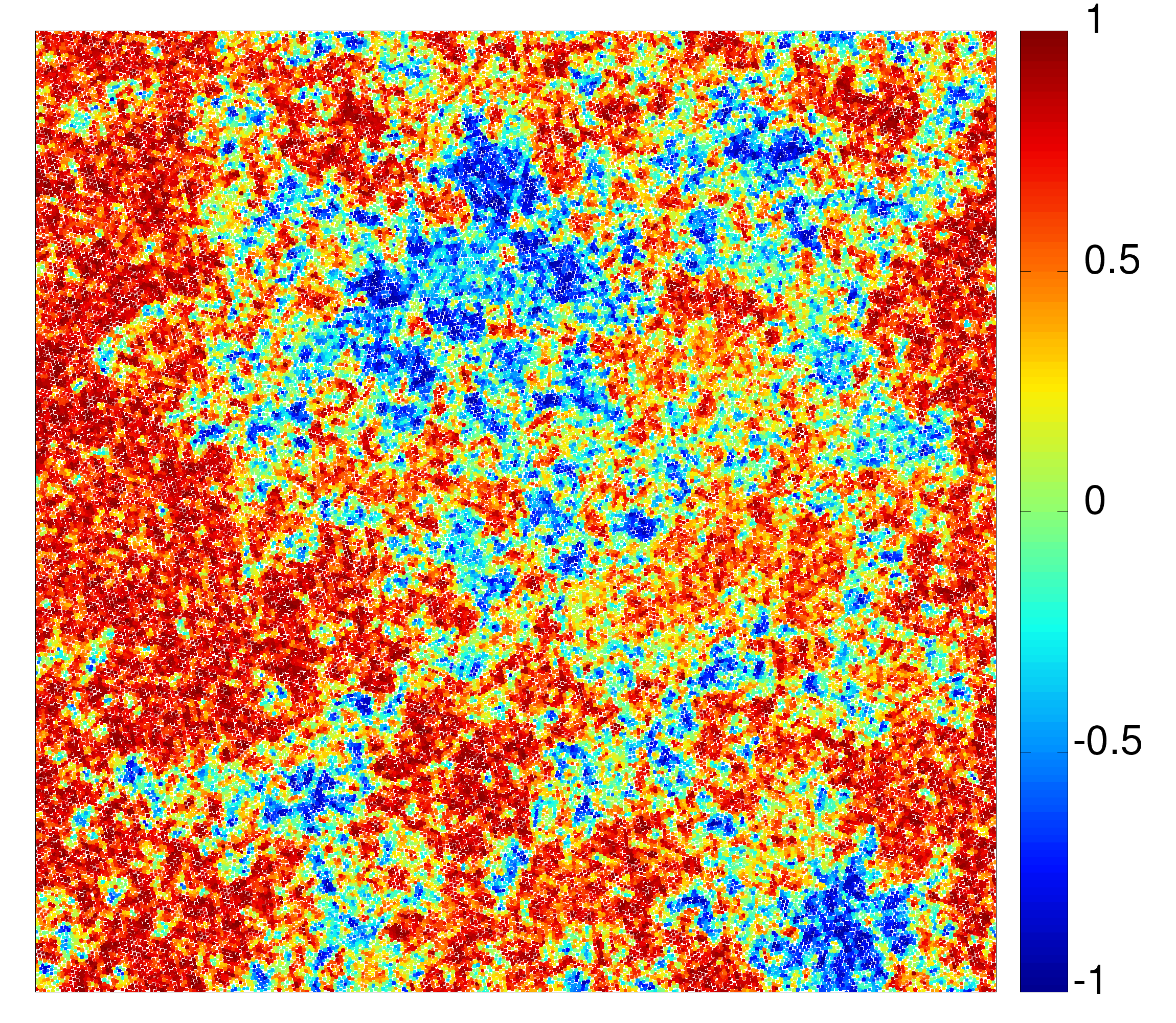}
\hspace{-0.2cm}
\includegraphics[width=3.95cm]{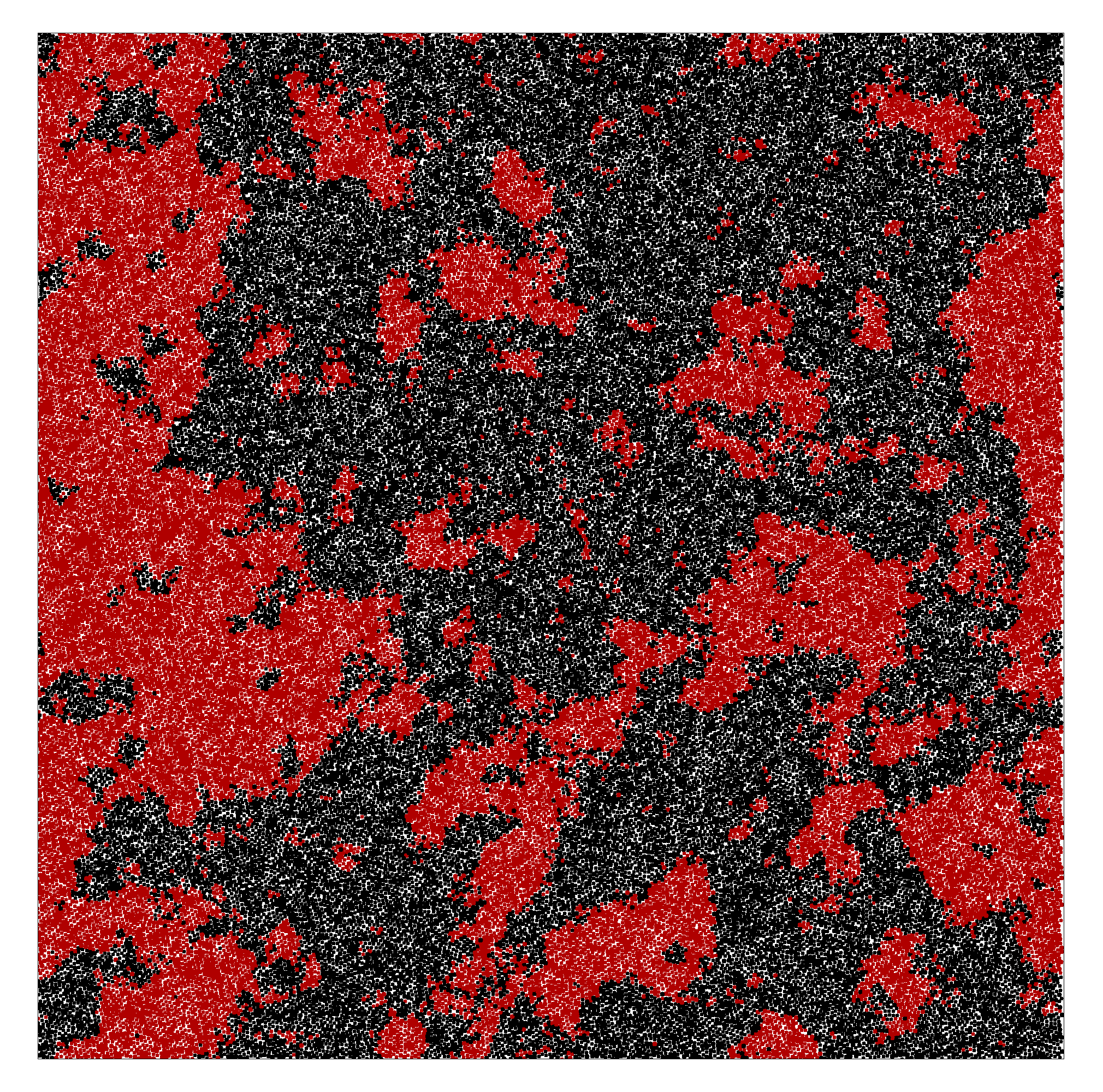}
\vspace{0.25cm}
}
 (c) \hspace{6cm}  $\;$
\centerline{
\includegraphics[width=\linewidth]{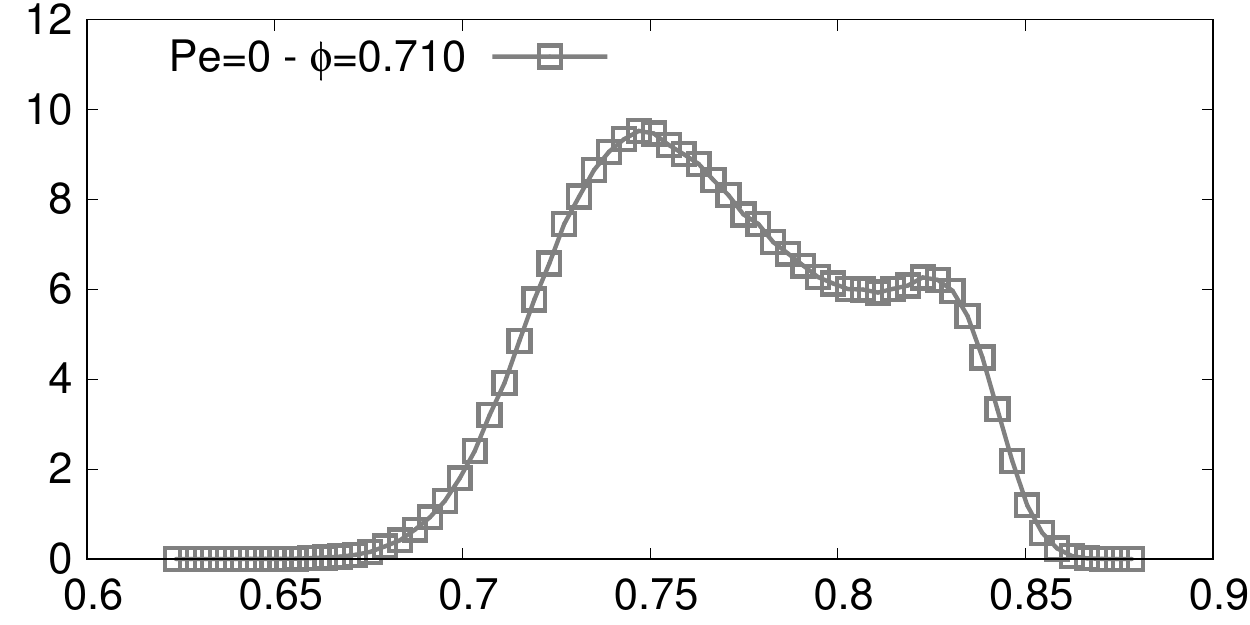}
}
\caption{(Color online.) 
Example of particle classification in a passive system (Pe = 0) in the coexistence region, $\phi=0.71$.  (a)
Map of local hexatic order parameter at a given time $t$ within the measuring interval $\Delta t$,  
projected on the direction of its global average. Particles 
painted red have unit projection on the average value and the rest of the colour code is given by the right vertical 
bar where the numerical values are the outcome of this projection.  (b) 
Particles in the dense  phase are painted red and those in the dilute  phase are colored black, 
as identified using the criterium explained in the main text. In this case $\Delta t =10^3$.
(c) Bimodal probability distribution function of the hexatic order 
parameter averaged over $\Delta t =10^3 $.
}
\label{fig:classification}
\end{figure}

We show an example of this construction in Fig.~\ref{fig:classification}. In panel (a) we present the 
map of the instantaneous local hexatic order parameter with the convention that red corresponds to the average hexatic direction (in complex space) and blue to the one opposed to it (the hexagonal lattice is rotated by $\pi/6$ with respect to the one of the dark red regions), with the scale given next to the plot for the 
intermediate directions. In general, regions of the same color identify clusters having the same hexatic value.
This is the convention used in~\cite{Digregorio18,Digregorio19}.
In  (b) we show which particles belong to the dense cluster (red) or the dilute phase (black) 
after an averaging time of $\Delta t=10^3 $, using the criterion described above. 
We see a correlation with the instantaneous map in panel (a) although it is 
not perfect, because of the time average. 
In panel (c) we show the probability distribution function of the
hexatic order parameter averaged over the same time interval. 

\subsubsection{The passive case}

We have explained the method that we use to identify the particles that, on the one hand, 
were and stayed in the dense/hexati\-cally ordered clusters, 
and the ones that,  on the other hand,  were and stayed in the dilute/disordered phase,
between two selected times.

\begin{figure}[h!]
\hspace{-6.5cm} (a) \\
\centerline{
\includegraphics[width=7.5cm]{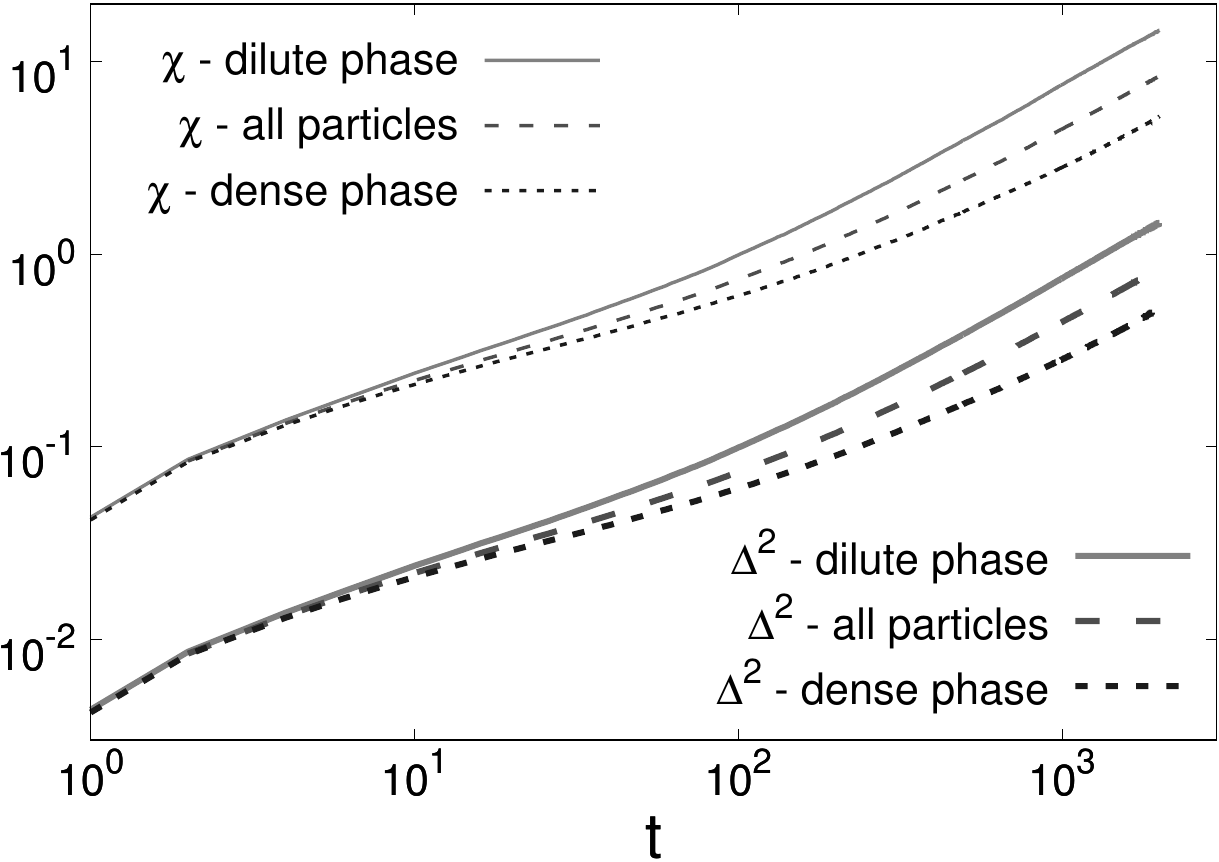}
}
\hspace{-6.5cm} (b) \\
\centerline{
\includegraphics[width=8cm]{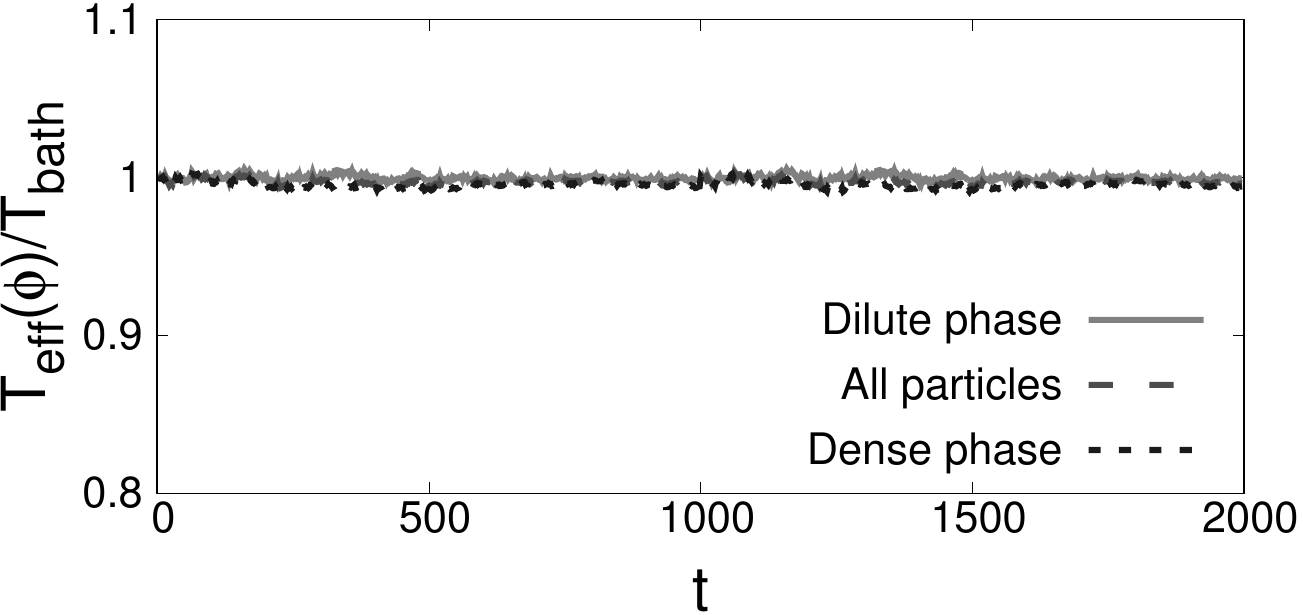}
}
\caption{(Color online.)
Passive inhomogeneous system. ${\text{Pe}=0}$ and ${\phi=0.710}$.
(a) Time-delay dependence in log-log scale of the mean square displacement (thick lines) 
and integrated linear response (thin lines) calculated by averaging over the particles which spent enough time in the gaseous phase (solid), in the cluster phase (dotted) or over all particles 
in the system (dashed). (b) Effective temperature normalized by the temperature 
of the bath calculated considering all the particles (dashed), 
only the ones belonging to the dense phase (dotted) or to the dilute 
phase (solid).
}
\label{pe0_teff}
\end{figure}

As already explained, in Fig.~\ref{fig:classification} we show  an example of the outcome of this 
classification in a passive system with coexistence. 
Concretely, we show data for ${\text{Pe}=0}$ and ${\phi=0.710}$ in order to be approximately in the middle 
of the coexistence region. 
From Fig.~\ref{pe0_teff}, we infer that even though the mean square displacement 
and the integrated linear response of the particles in the dense and dilute phases are very different, 
the effective temperature of both is the same (within numerical errors). Consistently with the expectations, 
since the system is in thermal equilibrium, the effective temperature is homogeneous and it equals 
the temperature of the bath 
at all time delays, even when the displacement and linear response 
show non-trivial time-dependencies.

\subsubsection{The active case}

Close to the passive limit Pe = 0, in a very narrow region of the phase diagram, the co-existence between 
hexatic and liquid phases survives. It is, however, quite difficult to see the effects of activity here, since the 
Pe values are very small and their effect is very weak. Data are not considerably different from the ones 
in the passive case.  We do not show them here.

Once strong activity is applied, the system has the possibility of undergoing MIPS 
in between two limiting packing fractions, say $\phi_<({\rm Pe})$ and $\phi_>({\rm Pe})$.
In MIPS, two sets of particles can be identified, those in the gaseous phase and those in 
dense cluster, the former with packing fraction $\phi_<$ and the latter with packing fraction
$\phi_>$.

\begin{figure}[h!]
\hspace{0.25cm}
\hspace{-4cm} (a) \hspace{4cm} (b)  \hspace{3cm} 
\centerline{
\includegraphics[width=4.5cm]{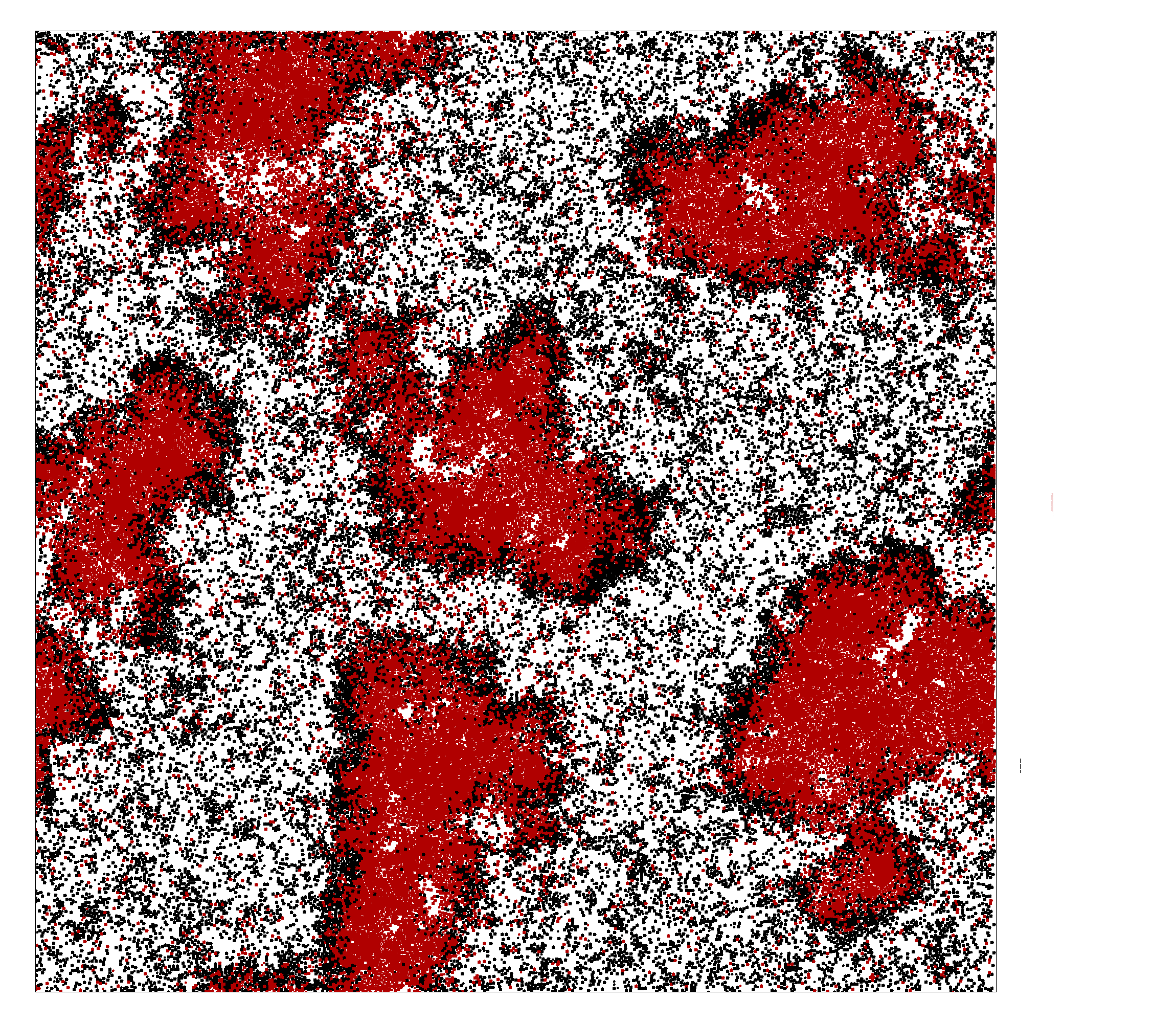}
\hspace{-0.5cm}
\includegraphics[width=4.5cm]{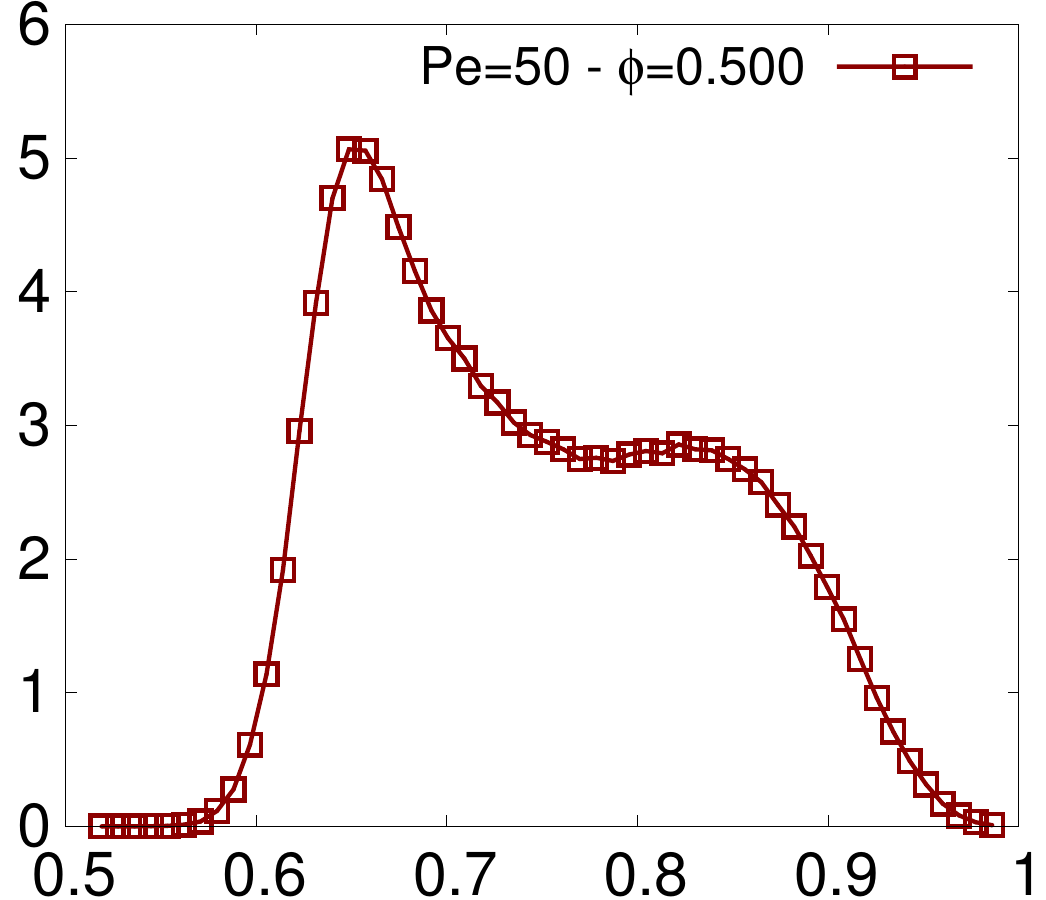}
}
\caption{(Color online.) 
Example of particle classification in the MIPS phase, $\phi=0.5$ and Pe = 50. (a)
Particles coloured red (black) belong to the dense (dilute) phase during the time interval $\Delta t = 10^3 $. The configuration is taken at the end of the 
measuring time: most of particles at the interfaces are coloured in black because
either the instantaneous value of the hexatic parameter is lower than the 
chosen threshold to separate the cluster or they have not spent enough time in the
 dense region to be considered belonging to it.
(b) Probability distribution function of the averaged hexatic 
order parameter.
}
\label{fig:separation-MIPS}
\end{figure}

\begin{figure}[t!]
\centerline{
\includegraphics[width=8cm]{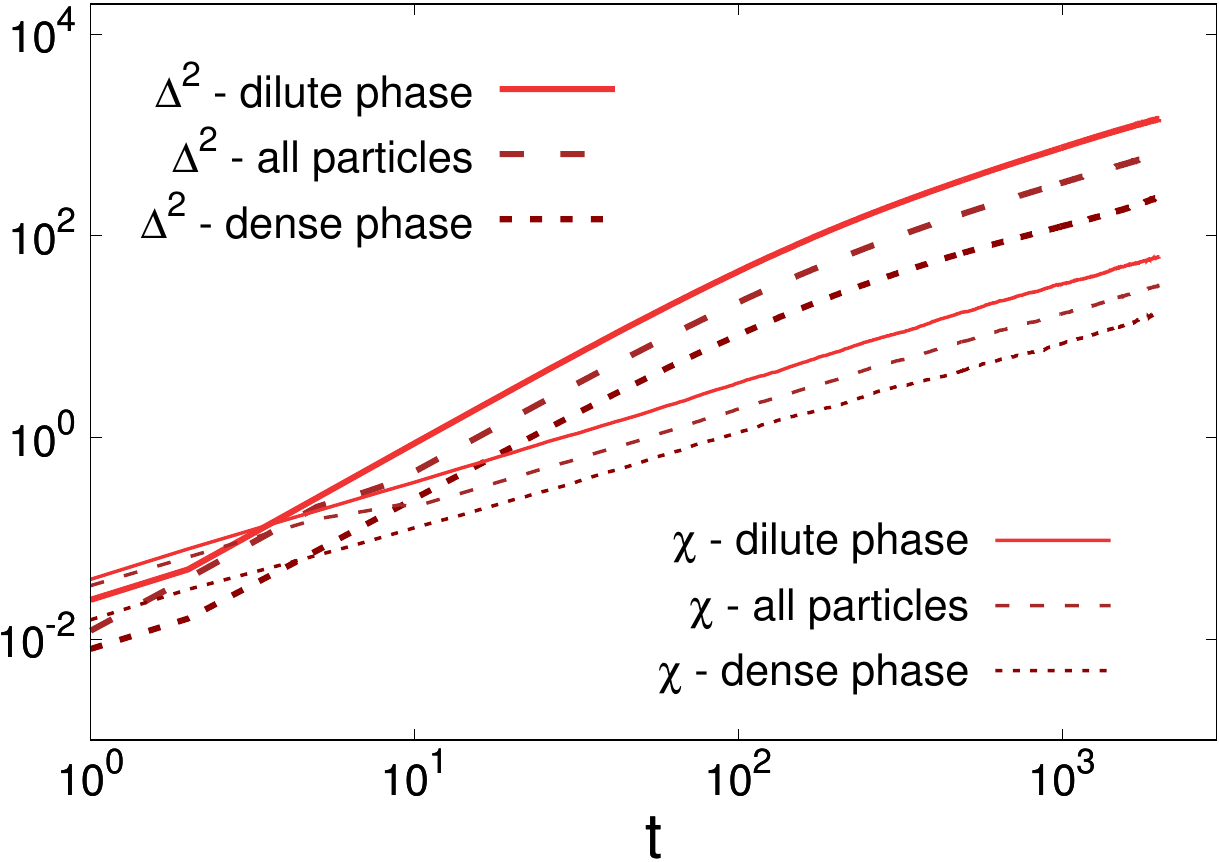}
}
\caption{(Color online.) 
Active system within MIPS. Pe = 50 and $\phi=0.5$.
Time-delay dependence in log-log scale of the mean square displacement (thick lines) 
and integrated linear response (thin lines) calculated averaging only over the particles which spent enough time in the gaseous phase (solid), in the cluster phase (dotted) or over all particles 
in the system (dashed). 
}
\label{fig:MSDPe50_1}
\end{figure}

The separation of the particles according to whether they mostly belong to the dilute or 
dense phase during a pre-defined time-interval, looks like what is shown in Fig.~\ref{fig:separation-MIPS}.
In panel (a) we show a typical configuration of the system:
the particles which spent most of their lifetime in the dense phase are depicted in
red, while the particles of the dilute phase are depicted in black.
Panel (b) illustrates the criterium used to separate the particles in the two groups. 
The pdf of the hexatic order parameter 
averaged over a time interval $\Delta t = 10^3 $ is bimodal
and we chose its minimum as the threshold to separate the
particles in the two phases. 
In Fig.~\ref{fig:MSDPe50_1} we show  the mean square displacements and linear responses
that lead to the effective temperatures in the two phases. First of all, we note that in the long 
time delay limit the integrated linear responses lie below the corresponding 
mean square displacements,  contrary to what happens in the passive case, 
cfr. Fig.~\ref{pe0_teff}. This implies $2T_{\rm eff}>1$ (notice that $k_B=1$) for dilute and dense components
and, moreover, $T_{\rm eff}>T$ in both cases. 

\begin{figure}[t!]
\centerline{
\includegraphics[width=6cm]{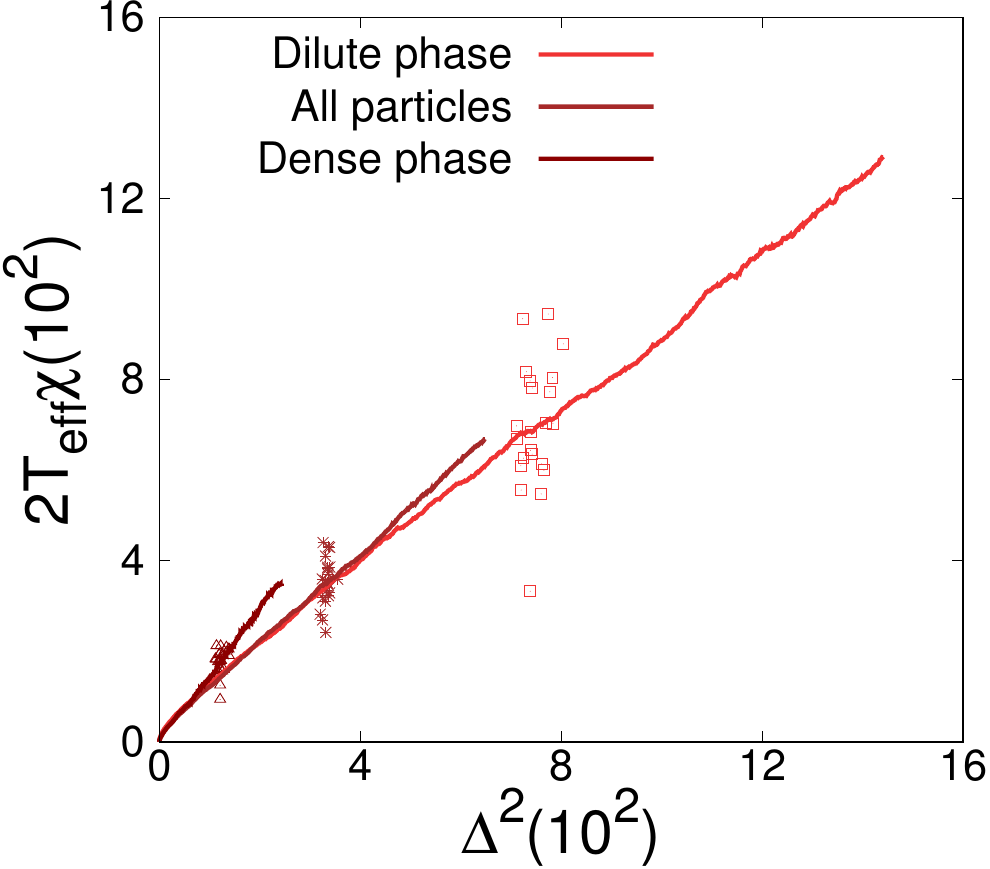}
}
\caption{(Color online.) 
Active system within MIPS. Pe = 50 and $\phi=0.6$. 
The parametric construction for particles in the dilute
and dense phases. The global data are also shown. In all cases the integrated linear 
susceptibility is multiplied by the global $2T_{\rm eff}$ so as the two 
quantities vary over the same interval. The data-points are the results for 
each of the 22 runs performed.  
}
\label{fig:MSDPe50}
\end{figure}

The parametric construction leading to the global effective temperature and the ones 
of the two MIPS phases is presented in Fig.~\ref{fig:MSDPe50} ($\phi=0.6$ and Pe = 50).
The figure also gives us an idea of the extent of the run-to-run fluctuations in the global data as well as 
in the two co-existing phases. The solid lines are the parametric constructions and the data points are the 
data for the single runs at a chosen time-delay $t$. The data are presented in the form $(\Delta^2, 2T_{\rm eff} \chi)$ 
in such a way that the scale is the same in both axes. It is clear that the fluctuations in the vertical direction are wider 
than the ones in the horizontal direction. A similar analysis of the noise-induced fluctuations in the 3$d$ 
Edwards-Anderson spin-glasses can be found in~\cite{Castillo03}.

\begin{figure}[t!]
(a) \hspace{4cm} (b) \hspace{2.5cm} $\;$
\vspace{0.2cm}
\\
\centerline{
\includegraphics[height=2.5cm]{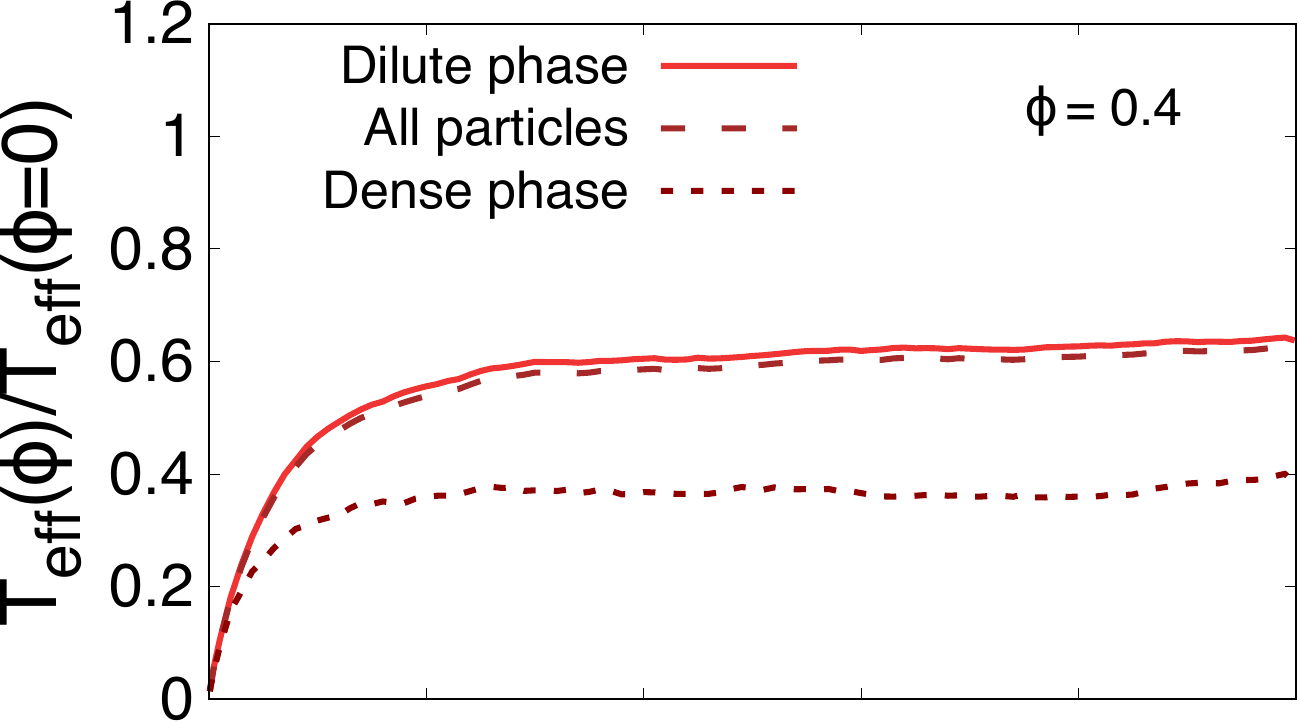}
\raisebox{0.05cm}{
\includegraphics[height=2.38cm]{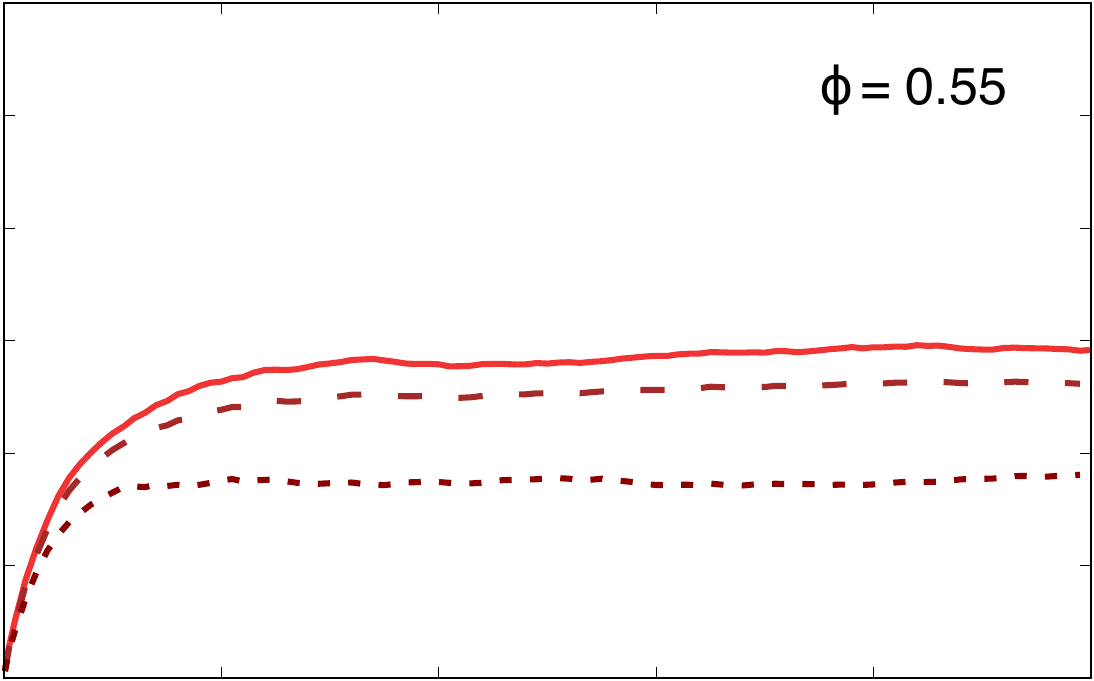}}
}
\vspace{0.2cm}
\centerline{
\hspace{0.01cm}
\includegraphics[height=3.05cm]{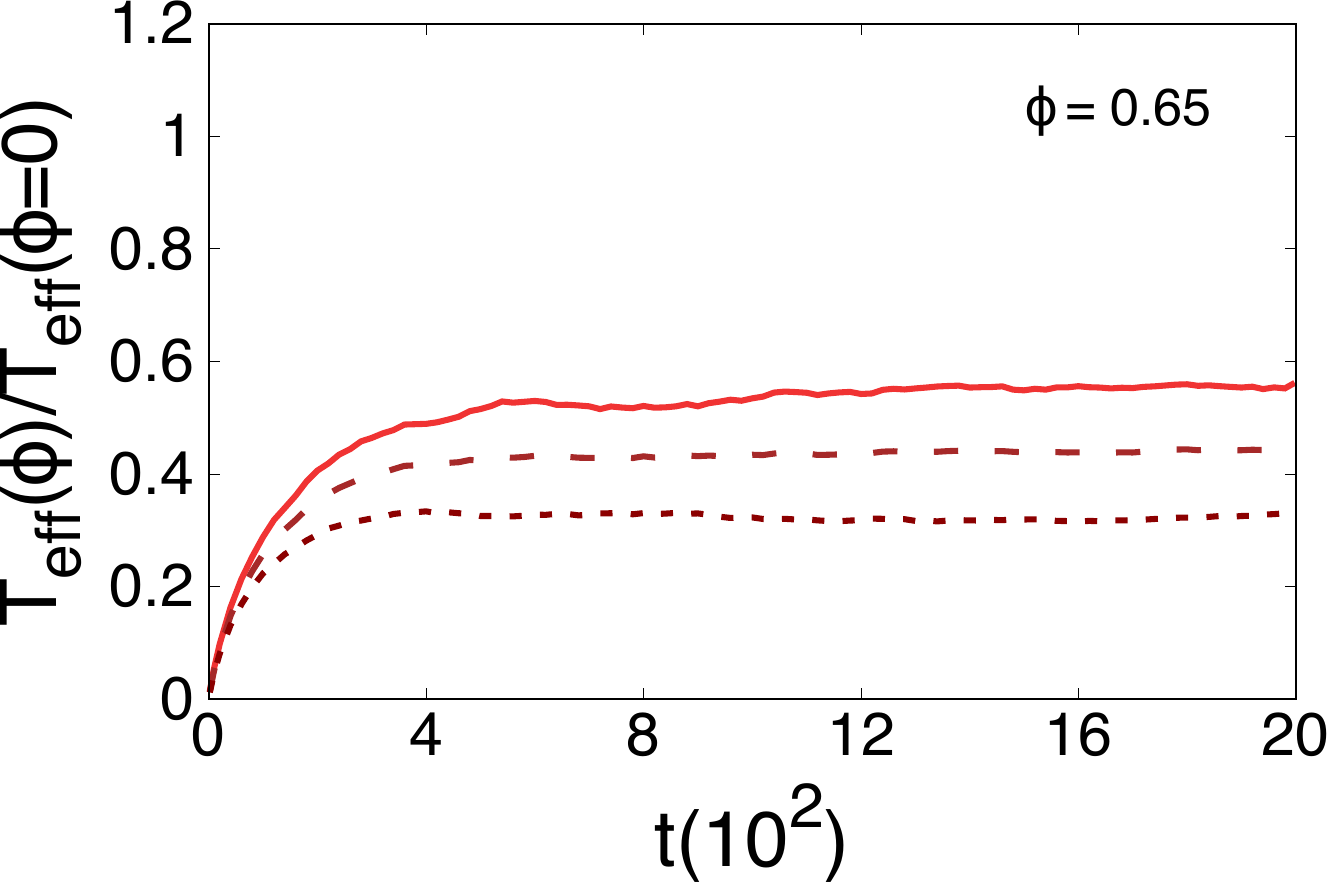}
\hspace{-0.1cm}
\includegraphics[height=3.0cm]{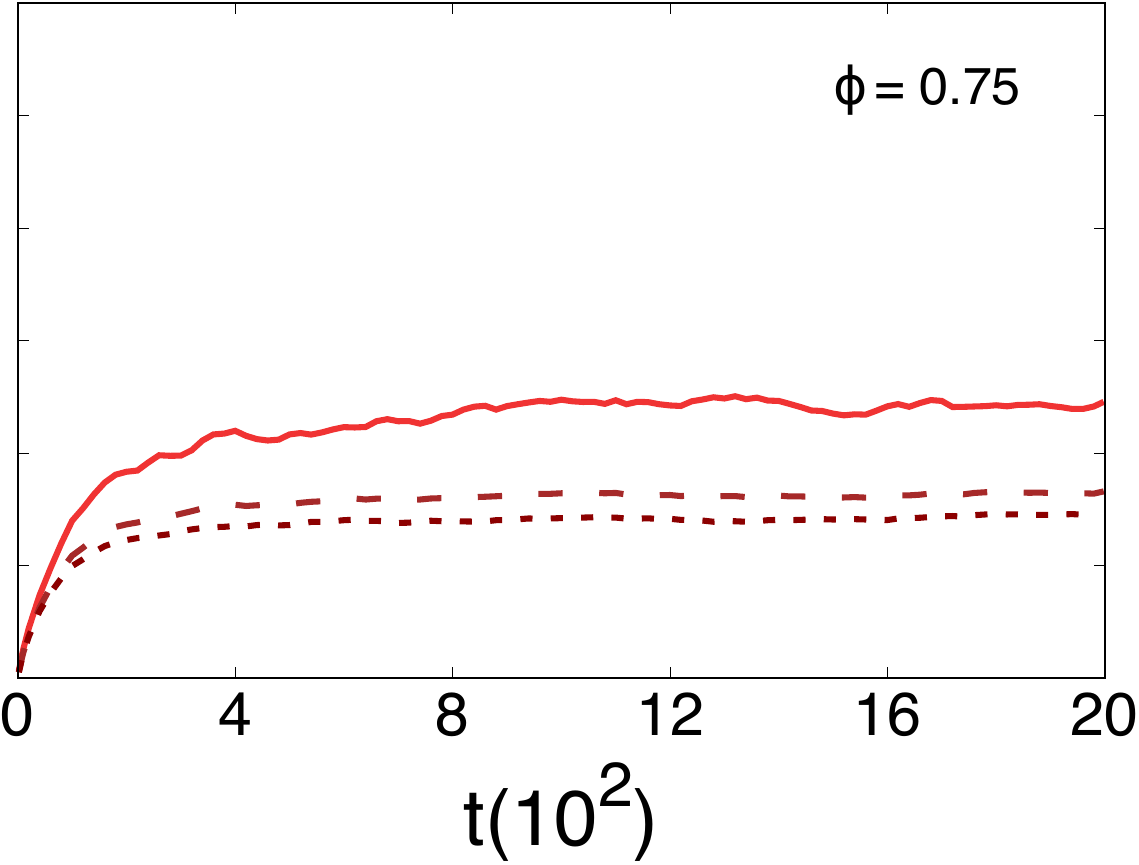}
}
(c) \hspace{4cm} (d) \hspace{2.5cm} $\;$
\caption{
(Color online.)
Active system across MIPS. 
The effective temperatures of the dilute and dense subsystems 
(continuous and dotted lines, respectively), together with the one of the
 full system (dashed lines) at Pe = 50, all normalized by the single particle value  at this Pe. The densities considered are $\phi= 0.4$ (a),  $\phi= 0.55$ (b), $\phi= 0.65$ (c) and  $\phi= 0.75$ (d).
Data are averaged over 60 independent runs.
}
\label{fig:4Teff-phi-inhomog}
\end{figure}

The question is, now, how does the effective temperature depend on the packing fraction, when this one varies from 
one end to the other end of MIPS at fixed Pe. Plots of $T_{\rm eff}(\phi)/T_{\rm eff}(0)$ as a function of time delay 
for four representative values of the packing fraction going from the lower to the higher are displayed in 
Fig.~\ref{fig:4Teff-phi-inhomog}. For all $\phi$,  
after a transient of roughly $400$,  
the asymptotic plateau is at different heights for the dilute and 
dense phases and, consequently, the global value is in between these two.
The data also demonstrate that the effective temperature of the whole system 
progressively changes from being equal to the one of the dilute phase, at low $\phi$, 
to reaching the one of the dense phase, at high $\phi$. We also see that the
value of the effective temperature of each of the two phases does not change much with $\phi$.
Clearly, as the fraction of particles belonging to the dilute phase diminishes, the 
corresponding $T_{\rm eff}$ data becomes noisier.

\begin{figure}[h]
\vspace{0.25cm}
\centerline{
\includegraphics[width=7.5cm]{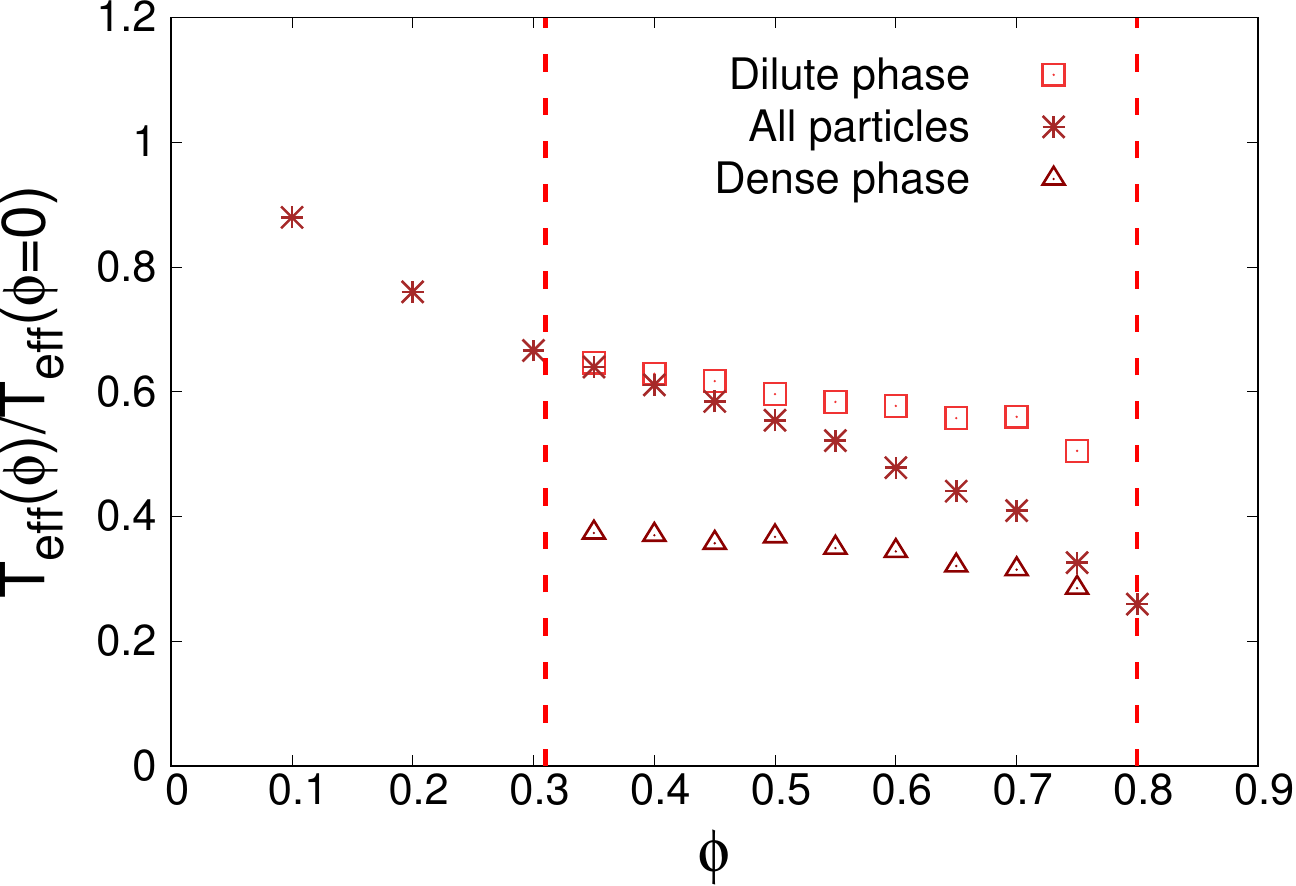}
}
\caption{
(Color online.)
Active system across MIPS.
The global packing fraction dependence of the
effective temperature of the dense (triangles) and dilute (squares) regions, compared to the one of the 
whole system (stars), at Pe = 50. The dotted vertical lines indicate the limits of MIPS
for this Pe value, $\phi_<$ and $\phi_>$.  
}
\label{fig:Teff-phi-inhomog}
\end{figure}

Figure~\ref{fig:Teff-phi-inhomog} summarizes the picture that emerges. 
The figure presents the global packing fraction dependence
of the effective temperatures of particles belonging to the dense and dilute phases 
in between the two measuring times. The data are presented normalized by the 
effective temperature of the single particle with the same value of Pe.
At the limits of co-existence the effective temperature of the majority 
phase joins the one of the whole system. The limiting values $\phi_<$ and $\phi_>$ thus 
measured (showed with vertical dashed lines in the figure) 
coincide with the ones measured more conventionally to delimit the MIPS region of the 
phase diagram (within numerical accuracy)~\cite{Digregorio18}. We notice that the effective temperature of the two phases are approximately 
constant within MIPS, with deviations appearing at the border which are due, presumably, to 
the fact that the fraction of system occupied by one of the phases approaches zero. 
The value of the effective temperature of the whole system, instead, changes and this is 
because the portion of particles belonging to each phase vary going from a purely 
dilute to a purely dense limit as $\phi$ increases.

\section{Fluctuations}
\label{sec:fluctuations}

In this Section we aim to complement the study of globally averaged diffusive and
response properties  with the one of their fluctuations.
Having access to the individual square displacement and the product of the
position and time-integrated noise, which once averaged over the latter yields the
susceptibility, allows us to study their statistical properties. Also, it permits us to 
correlate them with the local structure of the system.

\begin{figure}[b!]
\centerline{
\includegraphics[width=7.5cm]{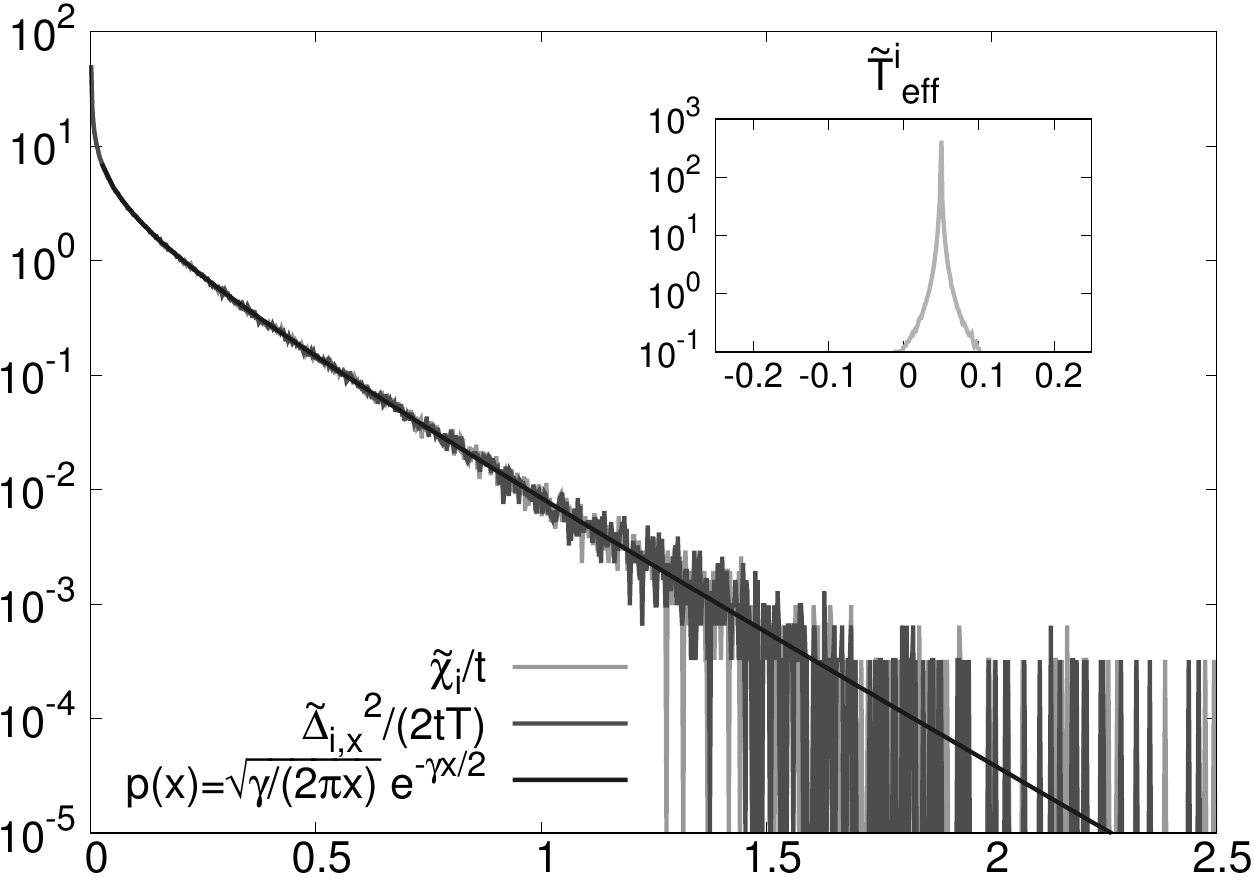}
}
\caption{
(Color online.)
Fluctuations in the passive and very dilute system, 
Pe = 0 and $\phi=0.001$, with $t_w=0$ and $x(0)=0$. Main plot:
probability distribution functions of the individual 
square displacement, $\tilde \Delta_{i,x}^2$, see Eq.~(\ref{eq:fluct-Deltai}), 
normalized by $2tT$  with $T$ the temperature of the bath,
and time integral of the product of position and noise, 
$\tilde \chi_i$ defined in Eq.~(\ref{eq:susc_i2}), 
normalized by $t$. The measurements were done using a time-delay 
$t \simeq 5\times 10^4$. 
The solid black line is the analytical prediction for the single passive
particle distribution, Eq.~(\ref{eqn:pdf_mu_tw0}). The average values of $\tilde \chi_i/t$ and $\tilde \Delta_{i,x}^2/(2tT)$ 
equal $1/\gamma=0.1$ within numerical accuracy, as consistently predicted by 
Eq.~(\ref{eqn:pdf_mu_tw0}).
Inset: probability distribution of the effective temperature defined
in  Eq.~(\ref{eq:single-temperature0}). With the choice $t_w=0$ and for $\phi=0$ the distribution
should collapse on a delta function centered on the bath temperature
 ($T=0.05$ in simulation units), since the response and the squared
  displacement divided by $2T$ become just identical in this particular case.
}
\label{fig:pdfs_Pe0_fi001}
\end{figure}

As noted in~\cite{Corberi12}, for a system with dynamics 
ruled by a Langevin equation, relations of the type in 
Eq.~(\ref{chi}) give two fluctuating fields, $\partial x_i^\lambda(t)/\partial\lambda|_{\lambda=0}$ (the superscript $\lambda$ indicates that the dynamics is perturbed) and 
${\int_0^t dt' x_i(t)\xi_i(t')/\sqrt{2\gamma T}}$ (whose noise averages yield the linear response function) as the 
candidates to define the fluctuating part. However,  
even though these objects have the same thermal average, they may have 
different fluctuation spectra and higher order correlations. 
The second field exhibits interesting properties in the case of 
aging spin-glasses and ferromagnetic systems~\cite{Castillo02,Castillo03,Chamon07}. 
The fact that Eq.~(\ref{chi}) holds also in a systems of active Brownian 
particles suggests to associate the fluctuations of the response function to  this same field.
The Malliavin method allows us to use the same noise realisation and the same 
stochastic trajectory to compute both this fluctuating quantity and the displacement fluctuations.

The strategy we will follow in the rest of this Section is the  following.
First of all, we select a time-delay $t-t_w$ such that the global (and noise-averaged) 
mean square displacement is (approximately) 
diffusive and the (also global and noise averaged) 
integrated linear response is linear in time. We then analyze the statistics of 
the single particle displacement, 
the van Hove function,  
 the ones 
of the single particle square displacement
\begin{equation}
\tilde \Delta_{i,x}^2(t,t_w)\equiv [x_i(t) - x_i(t_w)]^2
\; , 
\label{eq:fluct-Deltai}
\end{equation}
 and the ones of integrated linear response
 \begin{equation}
 \tilde \chi_i(t,t_w) = \frac{1}{\sqrt{2\gamma T}} \int_{t_w}^t dt' \, x_i(t) \xi_i(t')
 \; . 
 \label{eq:susc_i2}
 \end{equation}
  To avoid using wide horizontal intervals in the plots, and to compare the two fluctuating 
  quantities in the manner imposed by the FDT,  we 
divide $\tilde\Delta_{i,x}^2(t,t_w)$  by $2(t-t_w)T$ and $\tilde \chi_i(t,t_w)$ by $t-t_w$.
The reason why in this section we let $t_w$ be greater than zero will become clear soon.
We also define a local effective temperature as
\begin{equation}
\tilde T_{\rm eff}^i(t,t_w) \equiv 
\frac{
\tilde\Delta_{i,x}^2(t,t_w)}{
2\left[\int_{t_w}^t x_i(t) \xi_i(t')dt'/
\sqrt{2\gamma T}\right]
}
\; . 
\label{eq:single-temperature0}
\end{equation}
Note that the average over $i$ of the latter is not 
necessarily equal to the global effective temperature.
 We also use a normal representation in 
which we subtract the global averages and divide by the standard deviation.

\subsection{A single passive particle}

For a single passive Brownian particle in the over-damped regime, 
the explicit solution of the Langevin equation implies
\begin{equation}
\tilde \Delta(t,t_w) =
x(t)-x(t_w)=\frac{\sqrt{2\gamma T}}{\gamma}\int_{t_w}^t \xi(s)\;ds
\; .
\end{equation}
Consequently,  the fluctuating time-integrated linear response function can also be written as 
\begin{equation}
\tilde \chi(t,t_w)=\frac{1}{2T} \, x(t) \, [x(t)-x(t_w)]
\; ,
\label{eqn:response_simplified}
\end{equation}
that is {to say}, the product of two Gaussian random variables. For the particular choice $t_w=0$
and $x(t_w=0)=0$,  the time-integrated response function 
is just proportional to the square of $x(t)$, and $\tilde \chi(t,0)$ becomes identical to $\tilde\Delta_{i,x}^2(t,0)/2T$. 
The fluctuating linear response (divided by $t$) and the square displacement 
(divided by $2tT$) are equally distributed according to
\begin{equation}
p(x) = \sqrt{\frac{\gamma}{2\pi x}} \ e^{-\gamma x/2}
\label{eqn:pdf_mu_tw0}
\end{equation}
for $x\geq 0$.
This is in agreement with one of the results shown in~\cite{Corberi12} and it could
also be derived exploiting the symmetries exhibited by the joint probability 
distribution functions derived in that paper.
Notice that setting $t_w=0$, the effective temperature becomes identically equal to
the temperature of the bath and therefore its probability distribution is just {a 
delta function centered at $T$}.
Figure~\ref{fig:pdfs_Pe0_fi001} shows that in the dilute passive case the distribution
of the effective temperature resembles a delta function (inset; note that the small distribution spread is solely due to inter particle interactions, and disappears when considering a single particle), while the pdfs of the
response and the squared displacement are numerically equal and follow very
 closely Eq.~(\ref{eqn:pdf_mu_tw0}) for the single over-damped
 Brownian passive particle.
  
 The mean squared 
 displacement is  independent on the particular value of $t_w$ since it is a function of 
 the time delay alone. It is clear from the same definition of the squared displacement, 
  Eq.~(\ref{eq:fluct-Deltai}), that also its distribution function depends only on 
 $\Delta t=t-t_w$ and not on its average. 
 
 These considerations are no longer valid when we consider the distribution of the
 response function. When setting $t_w\ne 0$, the fluctuating part of the response 
 becomes the product of two correlated Gaussian variables with correlation 
 coefficient strictly less than one, see Eq.~(\ref{eqn:response_simplified}). 
 Since the joint probability distribution function of each of these two variables is 
 also Gaussian, it is possible to evaluate exactly the pdf of their product; it reads
 \begin{eqnarray}
 &&
 p_{\left.\frac{\tilde\chi(t,t_w)}{t-t_w}\right|_{t_w\ne0}}(x)
 \nonumber\\
 && \qquad =
 \frac{\gamma(t-t_w)\exp\left[\frac{\gamma (t-t_w) x}{t_w}\right]}{\pi\sqrt{t(t-t_w)-(t-t_w)^2}}
 \nonumber\\
 && \qquad\qquad \times \; K_0\left(\sqrt{\frac{\gamma^2(t-t_w) t}{t_w^2}}\, |x|\right)
 \; ,
 \label{eqn:pdf_mu}
 \end{eqnarray}
where $K_0$ is the modified Bessel function of the second kind of order zero.
In the limit $t\gg t_w$  and $x>0$ one recovers the form in Eq.~(\ref{eqn:pdf_mu_tw0}).
 The distribution of the response does depend explicitly on the waiting time. Only
 if $t_w=0$ the distribution exhibits a single branch for positive values, since in this
 particular case we are evaluating the pdf of the product of two perfectly correlated
 Gaussian random variables, see Fig.~\ref{fig:pdfs_Pe0_fi001}. As soon as $t_w$ becomes greater than zero, the pdf  develops a negative branch. 
 Notice that this particular feature does not affect the average of the distribution
 which equals $1/\gamma$ for every value of the waiting time.
 
\begin{figure}[h!]
\centerline{
\hspace{-0.25cm}
\includegraphics[width=0.5\linewidth]{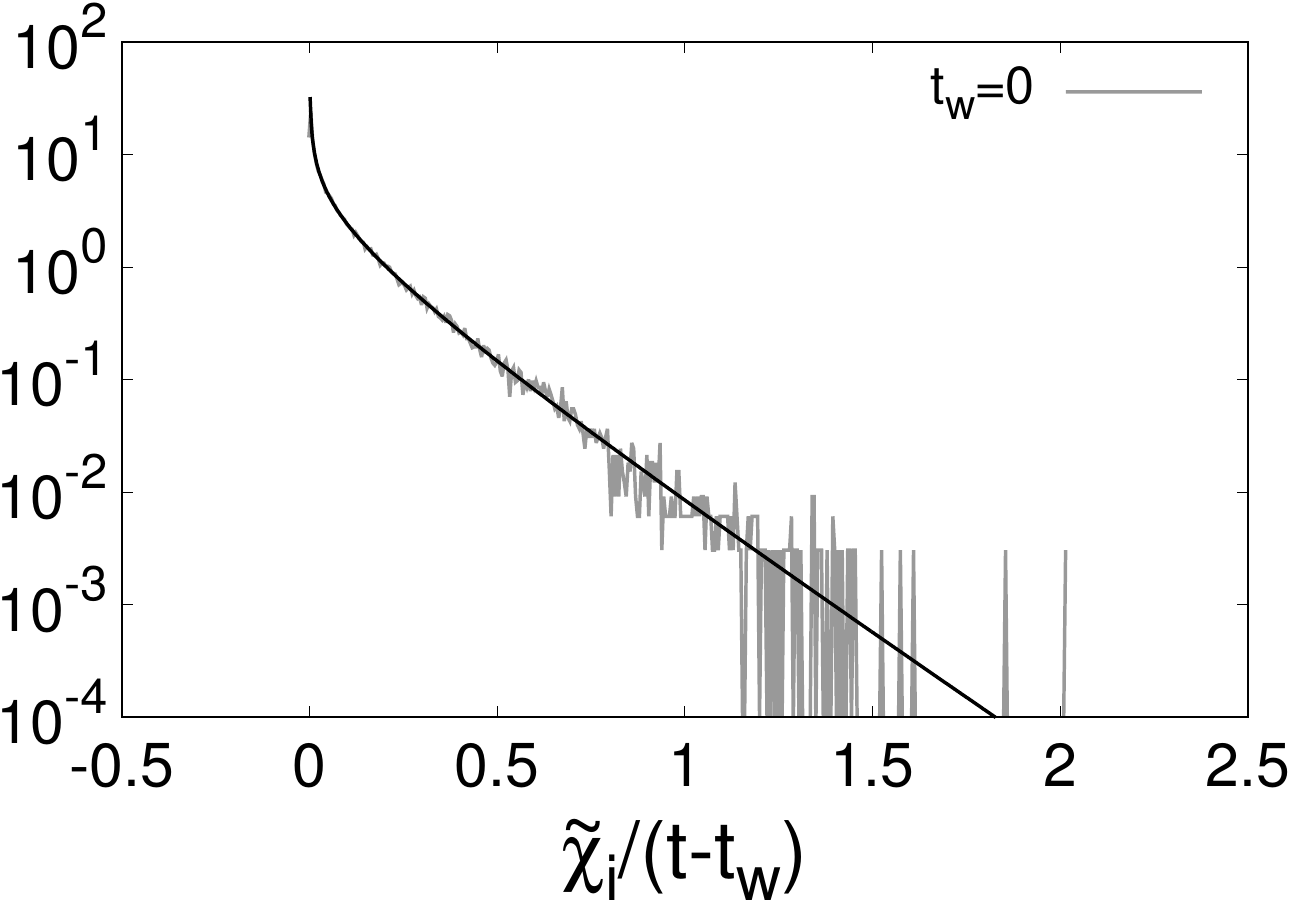}
\hspace{-0.13cm}
\includegraphics[width=0.5\linewidth]{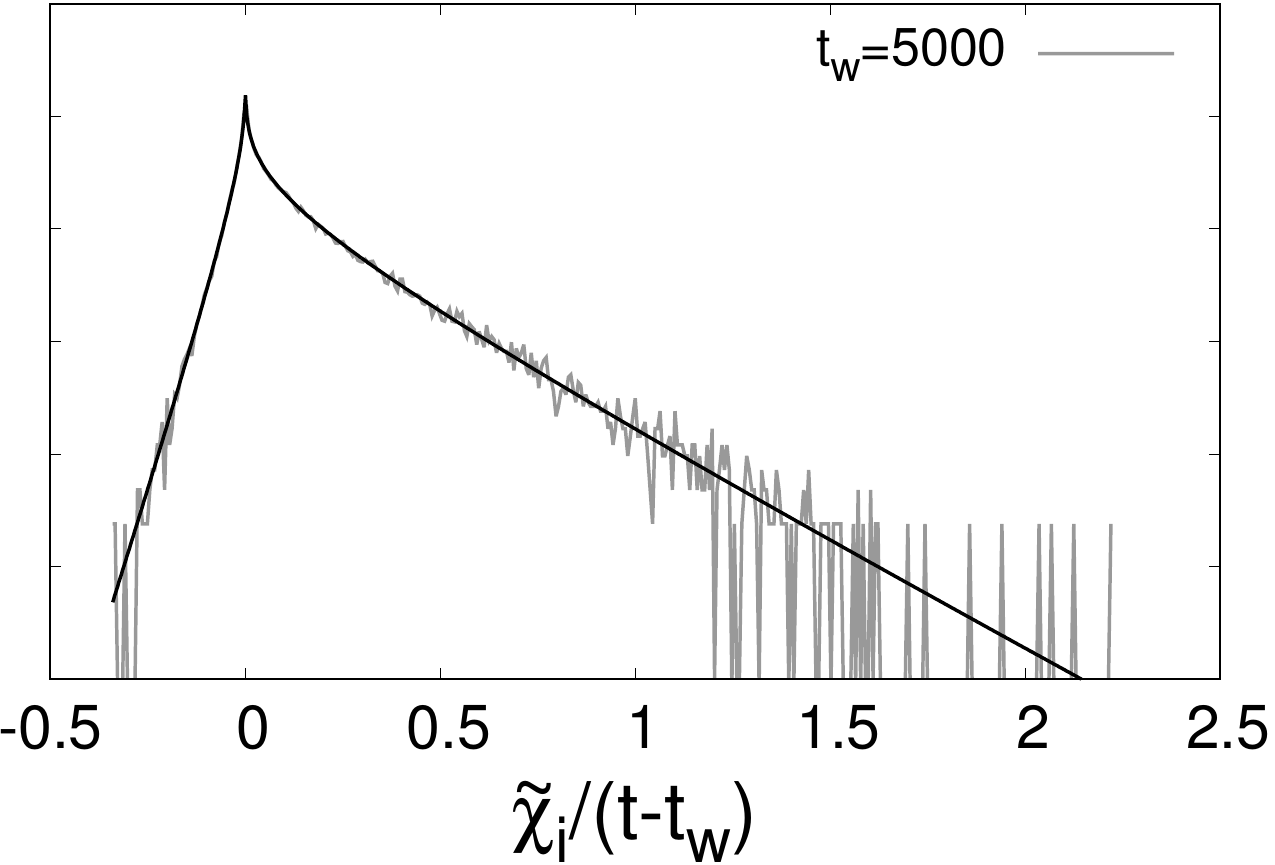}
}
\caption{Probability distribution function of $\tilde\chi(t,t_w)/(t-t_w)$ at Pe = 0 and $\phi=0.001$ evaluated at fixed $\Delta 
t=5\times 10^3 $ and waiting times given in the caption. Black lines: analytical predictions in Eq.~(\ref{eqn:pdf_mu_tw0})
for $t_w=0$ and Eq.~(\ref{eqn:pdf_mu}) for $t_w\neq 0$.
}
\label{fig:pdfs_mu_multiple_tw}
\end{figure}

 We checked this calculation numerically by comparing the distributions of the 
 time-integrated response in a very dilute ($\phi=0.001$) system of passive Brownian particles and 
 different waiting times with the analytical predictions, see Fig.~\ref{fig:pdfs_mu_multiple_tw}.
 
\begin{figure}[h]
\centerline{
\hspace{-0.25cm}
\includegraphics[width=0.5\linewidth]{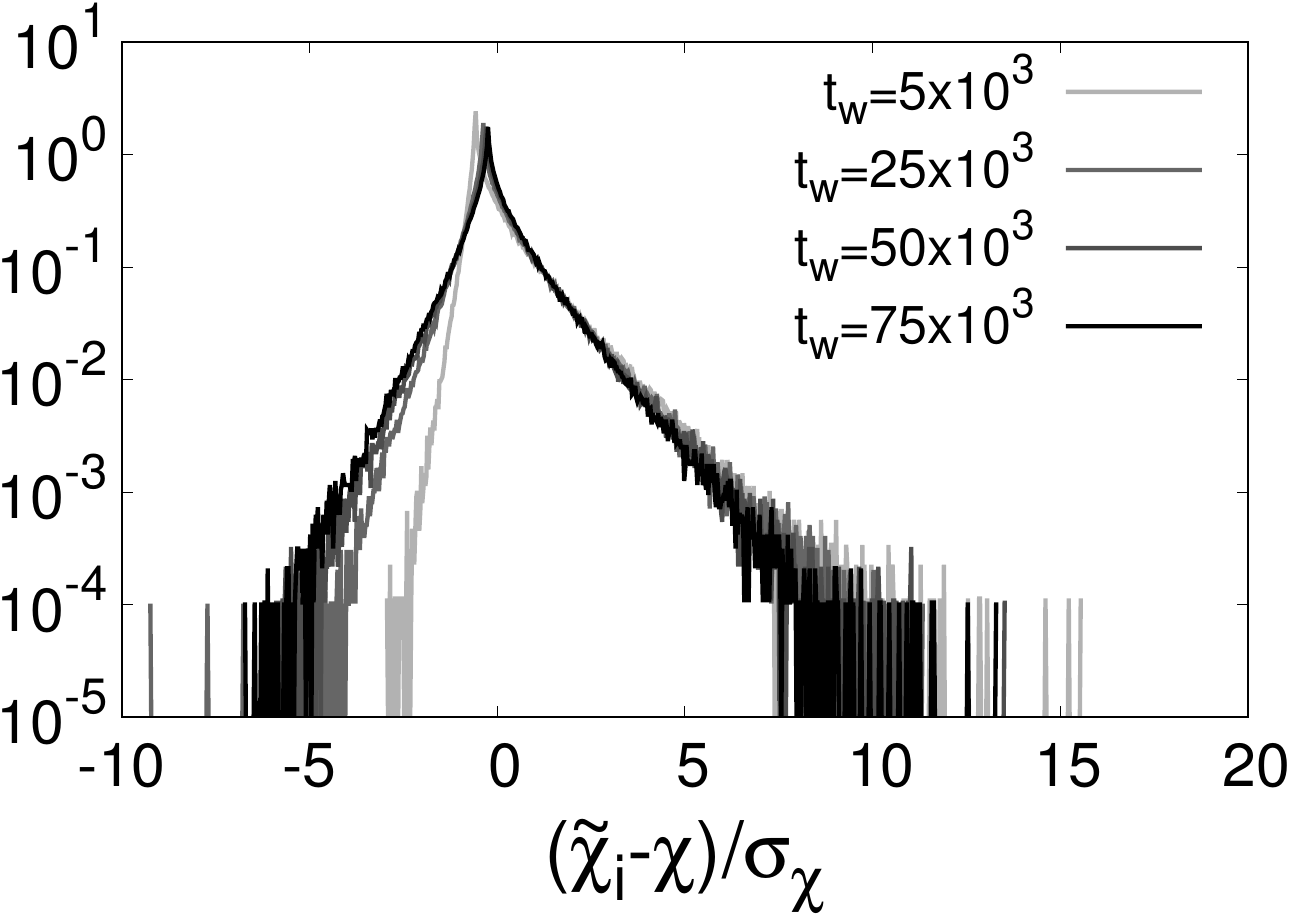}
\hspace{-0.13cm}
\includegraphics[width=0.5\linewidth]{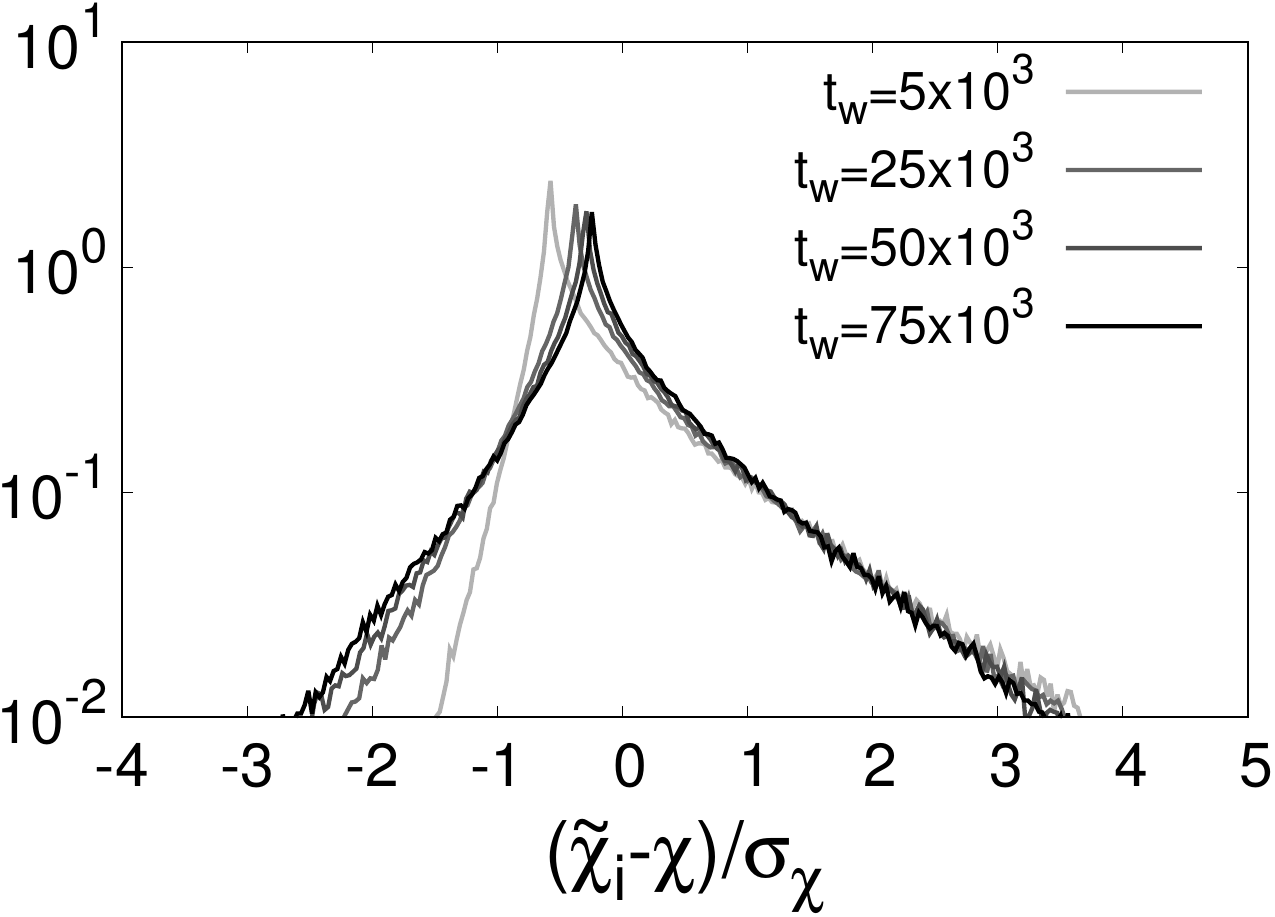}
}
\hspace{3cm}(a) \hspace{4cm} (b)
\caption{(a) Probability distribution function of the time-integrated linear response in
normal form at Pe = 0 and $\phi=0.001$ for $\Delta 
t=5\times 10^3$ and different values of the waiting time
given in the key. (b) Enlargement of the distribution peaks.
For increasing $t_w$ the distributions tend to collapse on the same curve.
}
\label{fig:pdf_mu_tw_norm}
\end{figure}

We want to underline that the distributions of $\tilde \chi_i$ for different
values of the waiting time do not collapse on top of each other even when plotted in normal form. In spite of this, 
as the waiting time grows, the dependence on $t_w$ becomes less
evident, see Fig.~\ref{fig:pdf_mu_tw_norm}. We will exploit it when studying the 
 active particle and interacting finite density problems.
 
The single particle effective temperature, 
Eq.~(\ref{eq:single-temperature0}), can be simplified in the same way by using
 the explicit solution of the Langevin equation
 \begin{equation}
 \tilde T_{\rm eff}(t,t_w) = T\; \frac{x(t)-x(t_w)}{x(t)}
\; . 
 \end{equation}
 Therefore, for a single passive particle, the effective temperature is 
 given by the ratio of two correlated Gaussian variables with known joint probability distribution. A quite 
 long but straightforward calculation leads to
 \begin{eqnarray}
 &&
 p_{ \tilde T_{\rm eff}(t,t_w)}(x)=
 \frac{1}{\pi} \frac{A}{A^2 + (x-x_0)^2}
 \label{eqn:pdf_teff}
 \end{eqnarray}
 with
 \begin{eqnarray} 
 \begin{array}{rcl}
 A &=& T \sqrt{(t-t_w)t_w/t^2}
 \;  ,
 \vspace{0.2cm}
 \\
  x_0 &=& T(t-t_w)/t
 \; ,
 \end{array}
 \label{eq:pdf_teff1}
 \end{eqnarray}
 that is to say, a Cauchy distribution centered at the maximum $x=x_0$ with height $1/(\pi A)$.
 For $t\gg t_w$, $A\to 0$, $x_0 \to T$, and the form approaches $\delta(x-T)$. This last 
 limit also applies in the case $t_w=0$ and $t\neq 0$.
 
  Notice that, since neither
 the mean value of the squared displacement nor the average of the response function
 depend on the choice of $t_w$,
the value of the effective temperature, $T_{\rm eff}$, is not affected by its particular choice. Instead, 
the distribution of $ \tilde T_{\rm eff}^i(t,t_w) $ does depend on 
the waiting time when $t$ is not much larger than $t_w$.

\begin{figure}[h!]
\centerline{
\includegraphics[scale=0.35]{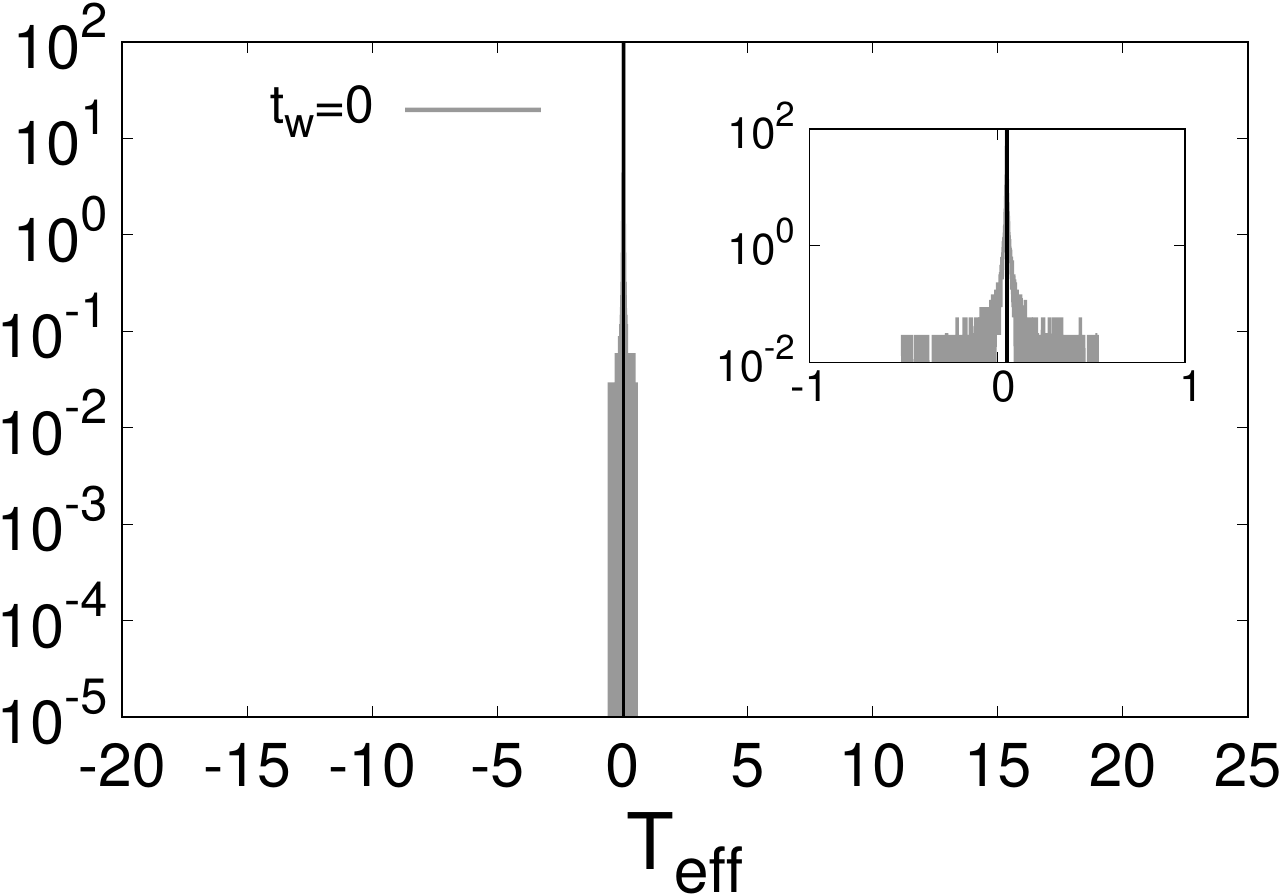}
\hspace{0.0cm}
\includegraphics[scale=0.35]{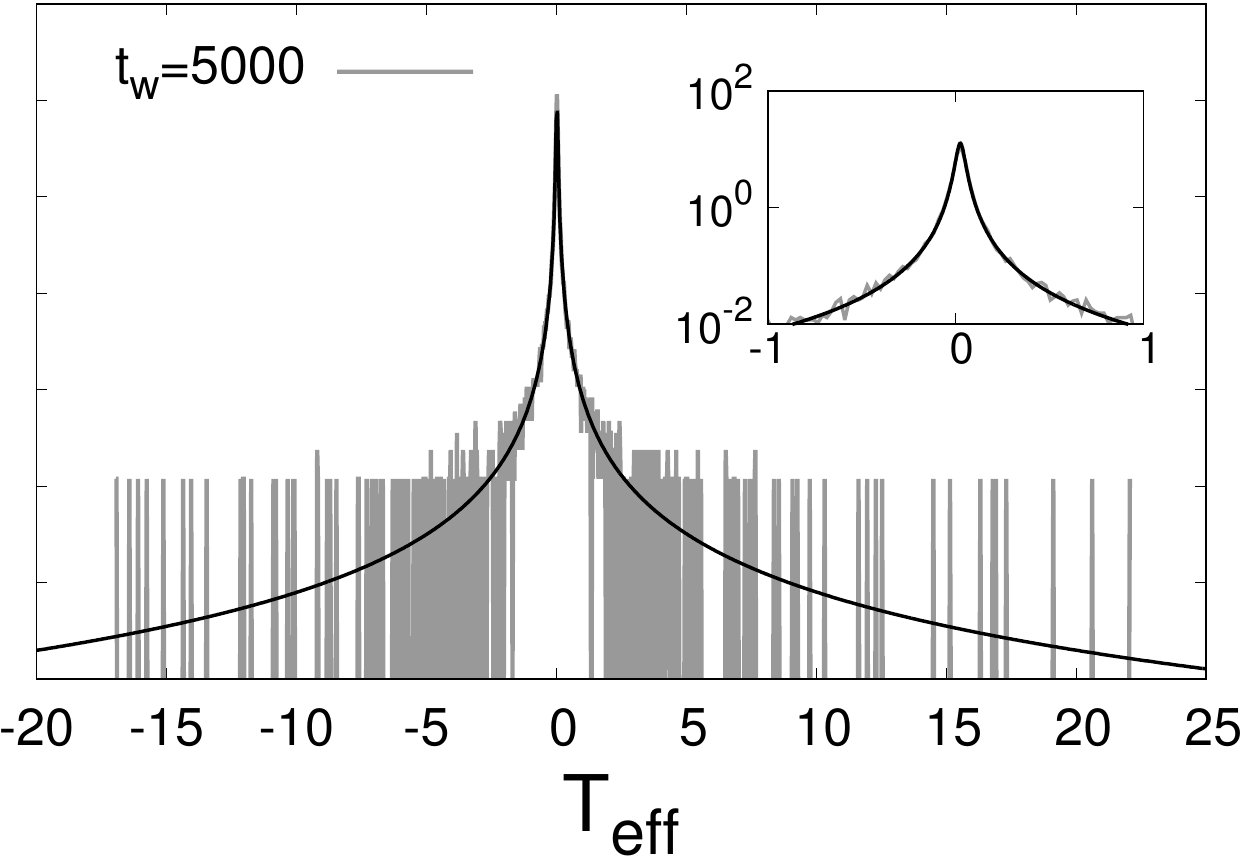}
}
\caption{Probability distribution function of 
$\tilde T_{\rm eff}^i(t,t_w)$ at Pe = 0 and $\phi=0.001$  evaluated at fixed 
$\Delta t=5\times 10^3 $ and waiting times given in the keys. 
Black lines: analytical predictions Eq.~(\ref{eqn:pdf_teff}) for $t_w\neq 0$
and $\delta(x-T)$ for $t_w=0$. In the inset, zooms over the peaks centered at 0.
}
\label{fig:pdfs_teff_multiple_tw}
\end{figure}

We evaluated numerically the distributions of the single particle effective temperature
with different waiting times, again in a very dilute system of passive particles, and we
compared the results with the analytical prediction in Eqs.~(\ref{eqn:pdf_teff})-(\ref{eq:pdf_teff1}), see 
Fig.~\ref{fig:pdfs_teff_multiple_tw}, finding very good agreement between numerical and analytic 
results.

\subsection{A single active particle}

For a single active particle in the overdamped limit, the distribution
of the displacement in a fixed direction is still Gaussian both at short and long
times with a variance determined by the temperature of the bath in the first case
and by the effective temperature in the latter one~\cite{Sevilla15}.
 Accordingly, the distribution of the squared displacement (divided by 
 $2T_{\rm eff}$) is given by Eq.~(\ref{eqn:pdf_mu_tw0}).
For the same reason, at long time,  the response function is the product of two
Gaussian variables: the variance of the first one (the position of the particle) depends
on the effective temperature and the whole time interval $t$, while the variance of 
the second one (the integral of the thermal noise) depends on the temperature of 
the heat bath and the measuring time $t-t_w$. Therefore, the probability distribution function of the response
in the active case is the analogue of Eq.~(\ref{eqn:pdf_mu}) after a proper
 mapping of the coefficients:
 \begin{eqnarray}
 &&
 p_{\frac{\tilde\chi(t,t_w)}{t-t_w}}(x)=
\frac{\sqrt{T}}{\pi}\frac{\gamma(t-t_w)}{\sqrt{T_{\text{eff}} \, t(t-t_w)- T \, (t-t_w)^2}}
 \nonumber\\
 && \qquad \times \; \exp\left[ \frac{T \, \gamma \, (t-t_w)x}{T_{\text{eff}} \, t- T \, (t-t_w)}\right] 
 \nonumber\\
 && \qquad \times \; 
 K_0\left(\sqrt{\frac{T_{\text{eff}} \,T\, \gamma^2 \, t(t-t_w)}{\left[T_{\text{eff}} \, t- T \, (t-t_w)\right]^2}}\,|x|\right)
 \; .
 \label{eqn:pdf_mu_active}
 \end{eqnarray}
 Notice that, since $T_{\text{eff}}>T$, 
 the determinant of the correlation matrix of the two Gaussians does not vanish even if  $t_w=0$. Thus, in the active case a negative branch is always present. Indeed, since the active particle is not moving solely under the influence of the thermal noise, even when the measuring time coincides with the whole time interval there is the possibility that the
displacement of the particle is opposite to the integrated noise. The weight of this
negative branch is enhanced for higher activity and we do expect that the presence of
interparticles forces would produce similar effects. 
 
The evaluation of the probability distribution function of the effective temperature in the active
 case is not straightforward since the simplification we performed in the passive case
 exploits the explicit solution of the equation of motion and it is no longer valid. 
In spite of that, if we restrict the analysis to the case $t_w=0$ with $x(0)=0$ (since 
only the
 shape of the distribution is affected by the particular choice of the waiting time, while the average is not),
 it is possible to simplify the expression for the effective temperature,
 \begin{equation}
 \tilde T_{\text{eff}}(t,0)=\sqrt{\frac{T\gamma}{2}} \frac{x(t)}{\int_0^t dt'  \, \xi(t')}
 \; . 
 \end{equation}
 and its statistics are given by another Cauchy distribution with coefficients
 \begin{eqnarray} 
 \begin{array}{rcl}
 A &=& \dfrac{T}{T_{\rm eff}} \sqrt{T (T_{\rm eff}-T)}
 \;  ,
 \vspace{0.2cm}
 \\
  x_0 &=& T^2/T_{\rm eff}
 \; ,
 \end{array}
 \label{eq:pdf_teff1_act}
 \end{eqnarray}
 that reduce to the values in Eq.~(\ref{eq:pdf_teff1}) in the passive limit $T_{\rm eff}=T$.

 \begin{figure}[h]
\centerline{
\hspace{-0.25cm}
\includegraphics[width=0.49\linewidth]{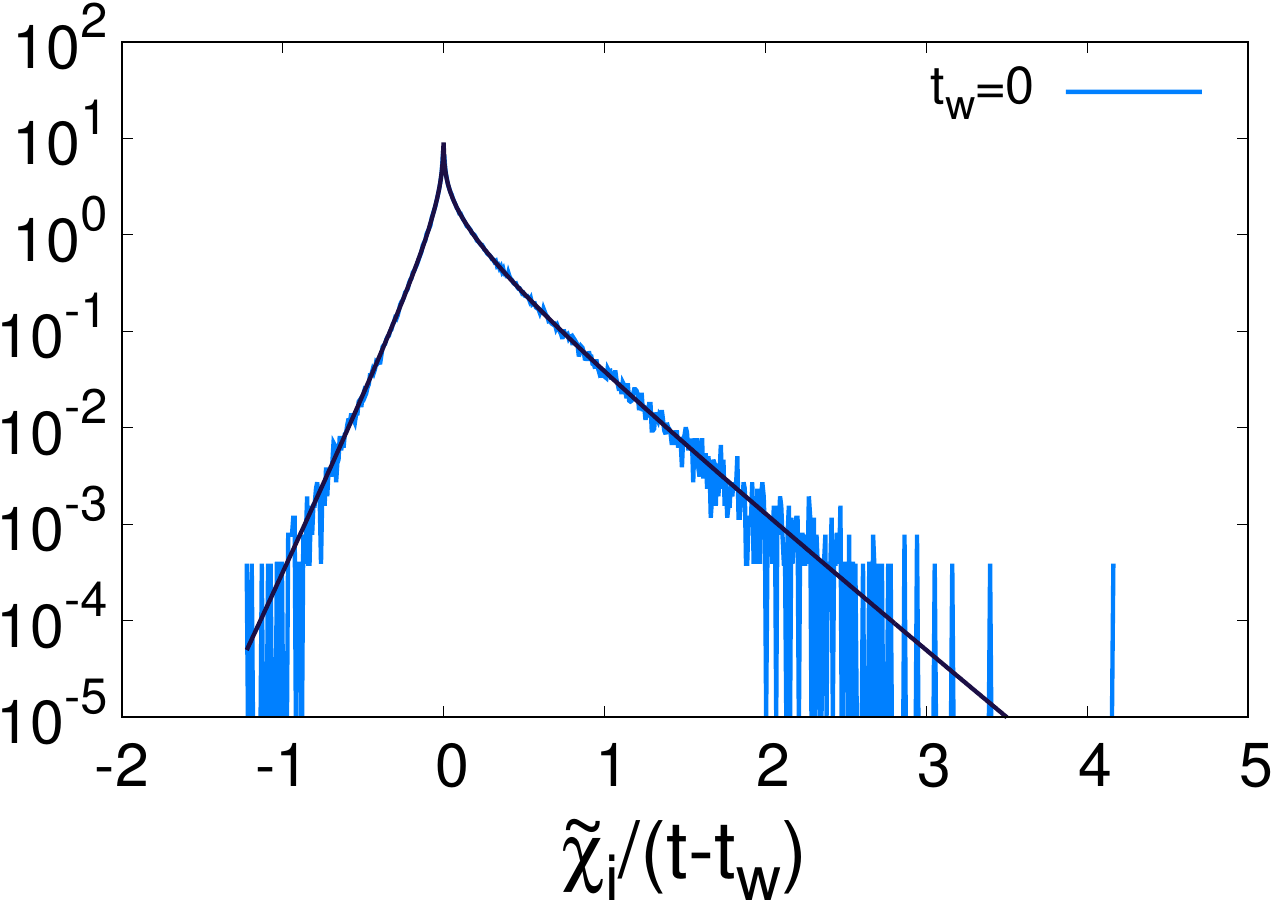}
\hspace{0.11cm}
\includegraphics[width=0.49\linewidth]{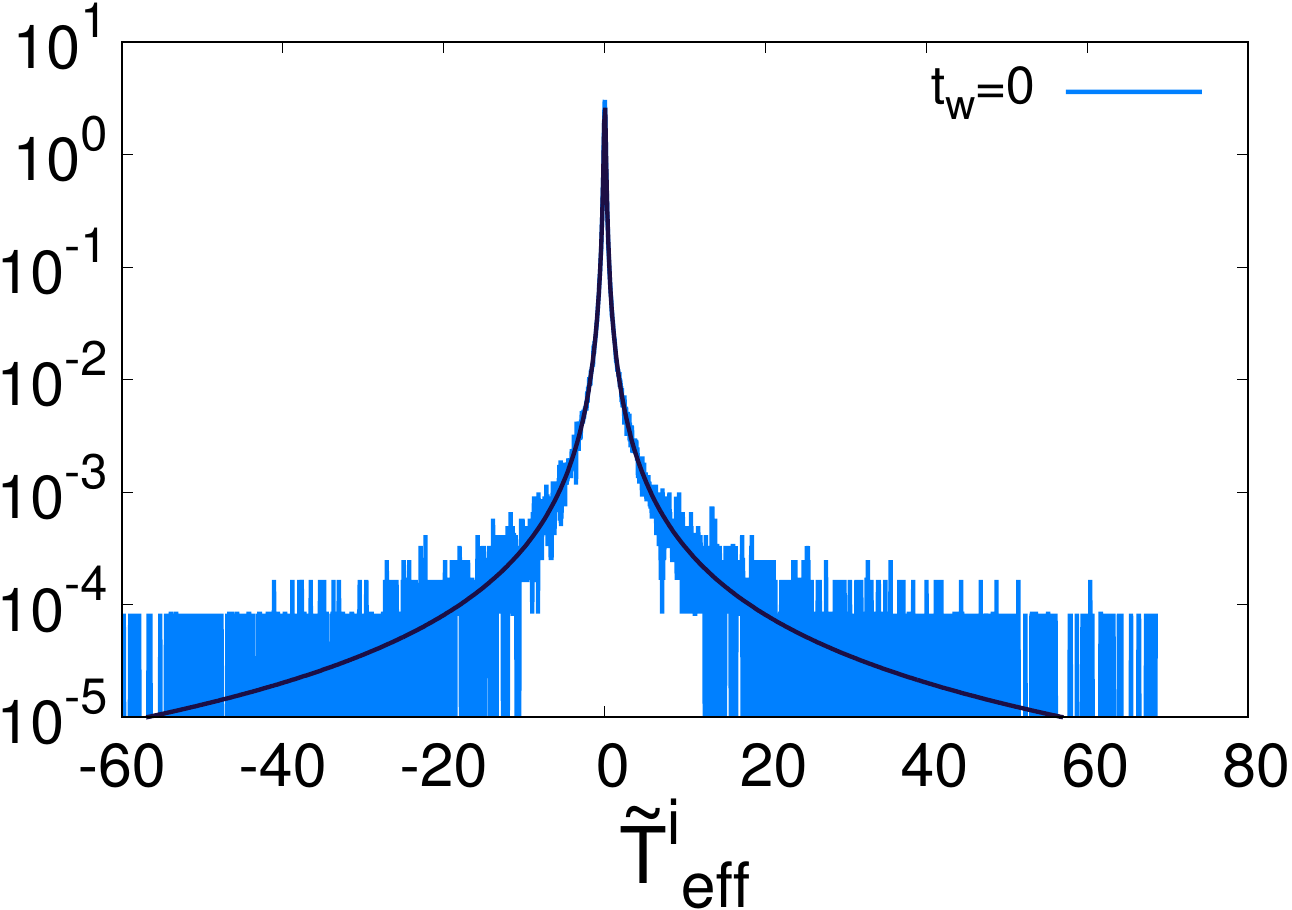}
}
\hspace{3cm}(a) \hspace{4cm} (b)
\caption{Probability distributions of (a) the 
time-integrated linear response and (b) the effective temperature, for $t_w=0 $,
in a very dilute active system approaching the single particle limit. Pe = 5 and $\phi=0.001$.
Solid lines: analytical predictions given by Eq.~(\ref{eqn:pdf_mu_active}) in (a) 
and Eq.~(\ref{eq:pdf_teff1_act}) in (b).
}
\label{fig:single_active}
\end{figure}

Notably, while in the passive case (if $t_w=0$) the effective
temperature is identically equal to the temperature of the heat bath, in the active 
case its distribution is not simply a delta function centered on its global value.
 In Fig.~\ref{fig:single_active}, we compare the numerical distribution of the response and the
 effective temperature for a single active particle with the
 analytical predictions in Eq.~(\ref{eqn:pdf_mu_active}) and Eq.~(\ref{eq:pdf_teff1_act})
 and once again the comparison is very favorable.

\subsection{Homogeneous phase}

Having clarified the behavior of the fluctuations in the single
particle case, we now turn to the problem of a dense homogeneous passive and  active system. 

In equilibrium, the averaged square displacement and
response function exhibit an explicit dependence on the density of the system.
In particular, these averages diminish when the density is increased, though in such a way that
their ratio remains independent of the packing fraction and always yields the
same effective temperature, equal to the temperature of the heat bath.

When considering the out of equilibrium active case, we found a dependence of the
effective temperature on the global density of the system. This effect is due to
two combined reasons: the
 mean value of the squared displacement decreases more than it does in the passive
  case with $\phi$, while the average of the response function also does but in a less pronounced way. 

We expect to retrieve similar features when studying the fluctuations
 of the displacement and the response in the homogeneous region of the phase diagram. Let us 
 see now if this is indeed so.

\begin{figure}[b!]
\vspace{0.25cm}
\centerline{
\includegraphics[scale=0.6]{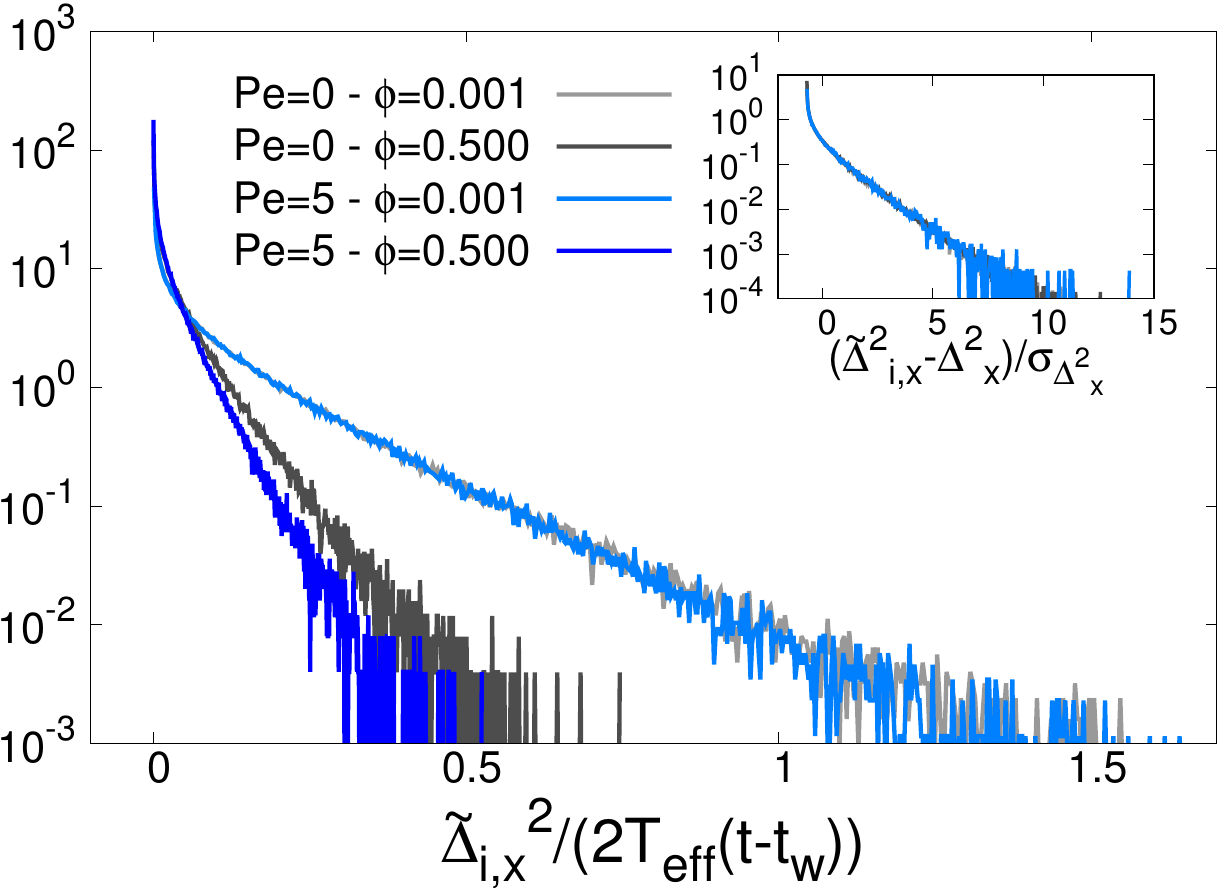}}
\caption{
Main plot: distribution of the individual square displacement divided by $2T(t-t_w)$
 in the passive case  (Pe = 0) and by $2T_{\rm eff}(t-t_w)$ in the active one  (Pe = 5). Comparison between the behavior in the very dilute ($\phi=0.001$) and in a dense
 case ($\phi=0.500$). Inset: probability distribution functions of the individual square
  displacement, $\tilde \Delta_{i,x}^2$, in normal form. All four curves collapse on top of each other.
 The waiting time for the measurement is $t_w=10^5 $ and the delay time $
\Delta t \equiv t-t_w=5\times 10^3 $.
}
\label{fig:pdfs_DxoverT_pe0_pe5}
\end{figure}

\subsubsection{The displacement}
\label{subsubsec:displacement}

The individual displacements are typically enhanced by the activity
and their distribution  develops a longer tail. On the contrary, increasing the 
packing fraction the dynamics get more sluggish and the individual displacements 
are reduced making their distribution narrower.

Still, under both kinds of changes, the displacement distributions preserve the 
functional form in Eq.~(\ref{eqn:pdf_mu_tw0}). 
We will make the working hypothesis that for weak activity and not too high density, 
$x(t)-x(t_w)$ remains Gaussian distributed 
(note that at sufficiently 
high density there could be deviations from this simple statistics~\cite{Phillies15}).
 Therefore, the distribution of the squared displacement is still given  by the
 one of the product of two perfectly correlated Gaussian variables. 

We confirm these claims in Fig.~\ref{fig:pdfs_DxoverT_pe0_pe5}, where we analyze the 
distribution of $\tilde \Delta_{i,x}^2/(2T(t-t_w))$ and $\tilde \Delta_{i,x}^2/(2T_{\rm eff}(t-t_w))$,
with $T_{\rm eff}$ the single particle effective temperature, in the 
passive and active cases, respectively. In the very dilute limit, the distributions 
collapse almost perfectly on top of each other, according to
 Eq.~(\ref{eqn:pdf_mu_tw0}). When the density is increased they keep the
 same functional form but their tails scale differently. The drop in the
  active dense case is more pronounced than in the passive dense one
  (compare the dark blue and gray curves in Fig.~\ref{fig:pdfs_DxoverT_pe0_pe5}), although notice that without dividing by $T$ and $T_{\rm eff}$ the displacements of the active particles are much larger than the passive ones, as $T_{\rm eff}$ 
is proportional to Pe$^2$.

In the inset of Fig.~\ref{fig:pdfs_DxoverT_pe0_pe5}, we study this same observable 
in its normal form. Although the displacements are enhanced by the activity, 
the distribution of $(\tilde \Delta_{i,x}^2 - \Delta_x^2)/\sigma_{\Delta_x^2}$, 
is independent of both the active force and the global packing fraction, as the various distributions of the individual displacements collapse on each other.

\subsubsection{The linear response}
\label{subsubsec:response}

First, we notice that turning on the activity and increasing the density produce similar
effects, in the sense that both of these changes reduce the 
correlation between the position of each particle at time $t$ and the integral of the
noise acting on it between $t_w$ and $t$. In fact, the correlation
 is maximal when the latter represents the only force exerted on the particle.

\begin{figure}[h]
(a) $\;$ \hspace{4cm}  $\;$
\vspace{0.15cm}
\centerline{
\includegraphics[scale=0.6]{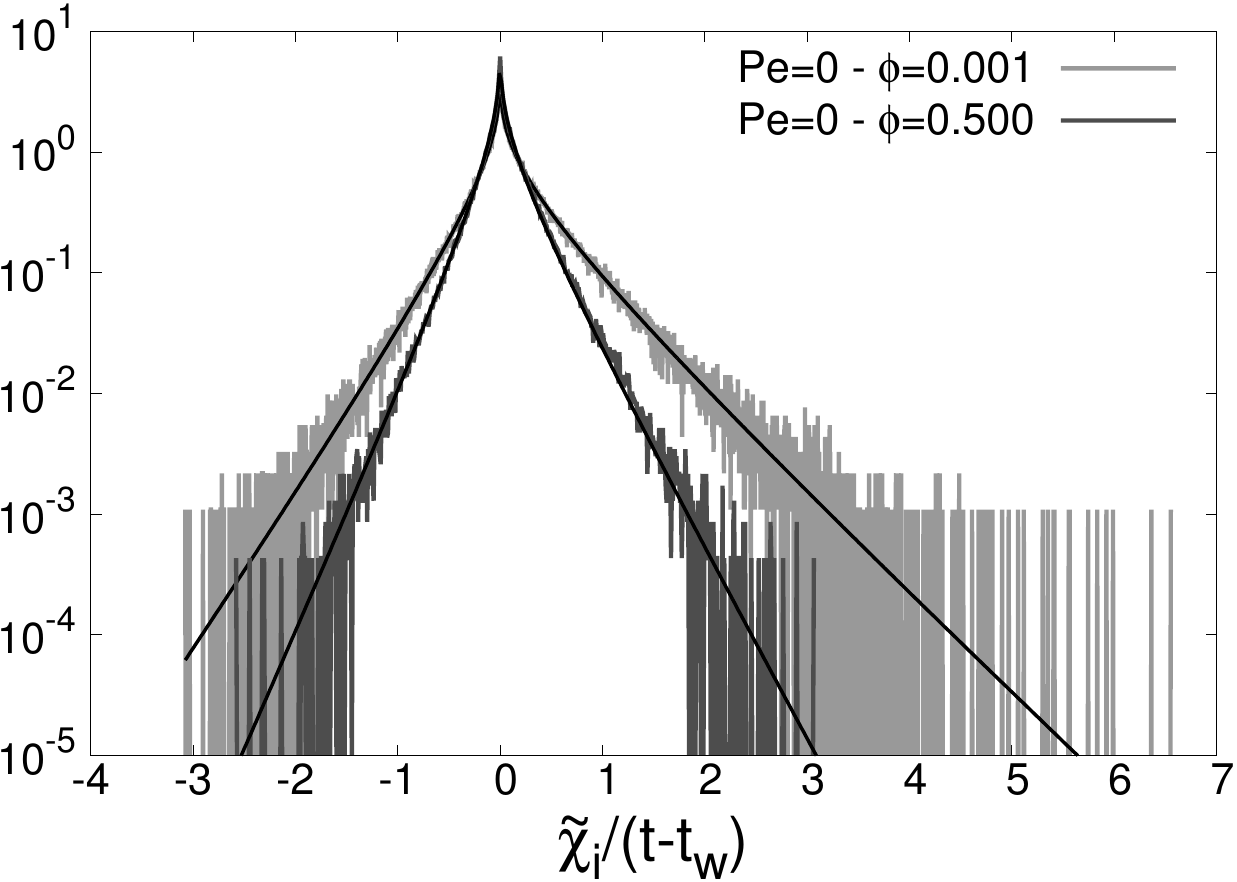}
}
(b) $\;$ \hspace{4cm} $\;$
\centerline{
\includegraphics[scale=0.6]{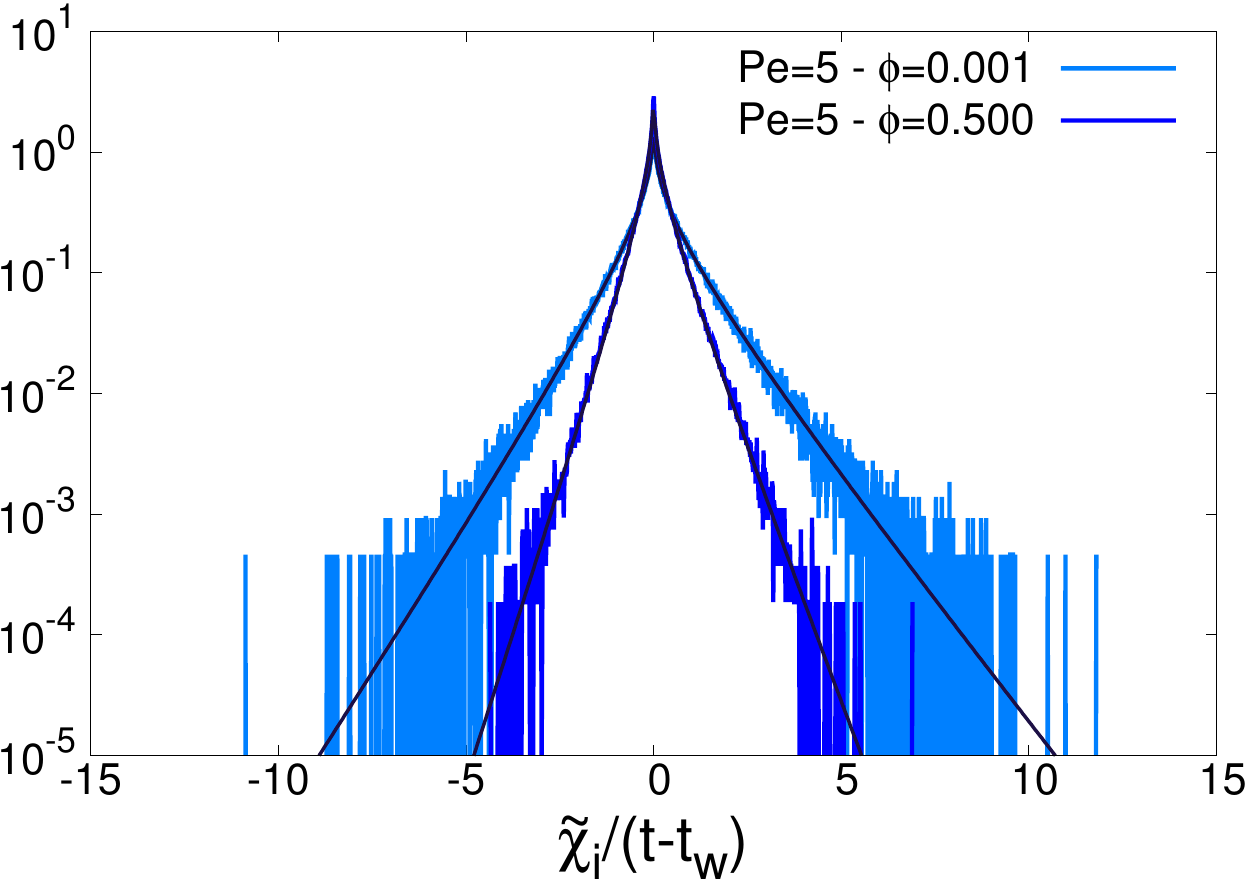}
}
\caption{Fluctuations of the time integral of the product of 
position and noise, $\tilde \chi_i$, divided by $t-t_w$, see Eq.~(\ref{eq:susc_i}). 
Passive (Pe = 0, (a)) and active (Pe = 5, (b)) systems: comparison between the
behavior in the dilute limit ($\phi=0.001$) and in the dense homogeneous systems 
($\phi=0.500$). The probability distribution functions are fitted using Eq.~(\ref{eqn:pdf_mu_fit}) with 
$t_w=10^{5}$ and $b=2T \, (t-t_w)/\gamma$ in all cases. In the dilute 
limit $c=2T\, (t-t_w)/\gamma$ while $a$ is obtained from the Gaussian
distribution of the displacement (not shown) in the four cases. 
Fitted parameters: $a=1050$ (Pe = 0, $\phi=0.001$), $a=359.4$ and $c=13.7$
(Pe = 0, $\phi=0.500$); $a=5425$ (Pe = 5, $\phi=0.001$), $a=1310$ and $c=19.6$
(Pe = 5, $\phi=0.500$).}
\label{fig:pdf_mu_fit}
\end{figure}

The shape of the probability distribution function of $\tilde \chi_i$ remains similar to the
reference case (Pe = 0 and $\phi=0.001$): it preserves a sharp peak at zero and two
 wings that depart from the peak. These features are shown 
in Fig.~\ref{fig:pdf_mu_fit}. The skewness of the susceptibility probability distribution function is 
also clear in all the plots that we present in this part of the paper. 

As in the single particle case, we found an explicit dependence
on the waiting time. However, the dependency is 
weaker due to, at fixed $t_w$, the presence of the active force (in the active dilute
case), or the presence of the inter-particle forces (in the dense passive case)
which reduce
the correlations between the position and the integral of the noise.
Therefore the value of the waiting time over which the
distributions (in normal form) collapse  is reached earlier.

In order to evaluate the probability distribution of the linear 
response, for each particle we rewrite it as
\begin{equation}
\frac{\tilde\chi_i(t,t_w)}{t-t_w}=\frac{X\cdot Y}{2T(t-t_w)}
\; ,
\end{equation}
with $X=x_i(t)$ and $Y=\sqrt{2T/\gamma}\int_{t_w}^t \xi_i(t')\,dt'$. We consider
$t_w\ne 0$ and we find
 \begin{eqnarray}
 &&
 p_{\frac{\tilde\chi_i(t,t_w)}{t-t_w}}(x)
 \nonumber\\
 && \qquad =
 \frac{2T(t-t_w)}{\pi\sqrt{|ab-c^2|}}\exp\left[2T(t-t_w)\frac{c}{ab-c^2}\, x\right]
 \nonumber\\
 && \qquad\qquad \times \; K_0\left(2T(t-t_w)\frac{\sqrt{ab}}{ab-c^2}\, |x|\right)
 \; ,
 \label{eqn:pdf_mu_fit}
 \end{eqnarray}
with 
$a=\braket{X^2}$, $b=\braket{Y^2}$ and $c=\braket{X\cdot Y}$.
 In the single particle case discussed in the previous subsection 
 $a= 2T_{\text{eff}} \, t/\gamma$, 
 $b=c=2T  \, (t-t_w)/\gamma$. 
 When the density is increased, we notice that the
 functional form of the distributions remains unchanged, while the value of the
 coefficients may depend on the density and the activity. In particular $b$,
  being simply the variance of the noise, is unaffected by these changes. We have 
  already noticed that $a$ (which is the mean squared displacement) is reduced 
  in the active case with the increase of $\phi$  more than it is in the equilibrium limit. 
  This effect would not explain the reason why the
  response function behaves in the opposite way. We expect that, in the active case, 
  $c$ 
  which is proportional to the average of $\tilde\chi$ and yields the linear response,
  be slightly less sensitive to the increase of the packing fraction. 
  Therefore, when $\phi$ is increased the correlation between the position of the
  particle and the integral of the noise 
  diminishes in the active case slightly less than it does in 
 the passive case. 

We support these claims with our numerical results.
We chose a long waiting time, $t_w=10^5$, and we measured the 
linear response in systems with Pe = 0, 5 and $\phi=0.001,\,0.500$. 
In Fig.~\ref{fig:pdf_mu_fit}, we compare the numerical data collected in
these four cases with the functional form in Eq.~(\ref{eqn:pdf_mu_fit}), using $c$ as the
only  fitting  parameter (we kept
 $b=2T \, (t-t_w)/\gamma$ and we used for 
  $a$ the value determined by the distribution of 
  the position). 

From the examples in Fig.~\ref{fig:pdf_mu_fit}, we can clearly
see that the shape of the linear response distribution is unchanged when the density
or the activity are increased with respect to the single Brownian 
particle case. The fitted parameters are written in the caption. 

\subsubsection{The effective temperature}
\label{subsubsec:temperature}

In the passive limit, the combination of the parameter effects on the displacements 
and  linear response fluctuations always lead to 
the same global effective temperature which is found to coincide 
with the temperature of the bath, in agreement with the results in Fig.~\ref{pe0_teff}.
 In the active case the global effective temperature
is found to be always higher than the bath temperature, but
lower than the one for the single active particle, see Fig.~\ref{fig:Teff-homo}.

In this subsection, we briefly explore how the distribution of the
single particle effective temperature, defined in Eq.~(\ref{eq:single-temperature0}),
changes when $\phi$ is increased. As we noticed when we evaluated the probability distribution function of the
effective temperature of a single active particle, the simplification we performed for a 
passive Brownian particle 
is no longer valid when other forces come into play. These considerations hold when we 
turn on the activity as well as when the particles interact among themselves. However,
if we choose $t_w=0$, we are allowed to simplify the expression of the single particle
effective temperature:
\begin{equation}
\tilde T_{{\text{eff}}_i} (t,0)=\sqrt{\frac{T\gamma}{2}}  \frac{x_i(t)}{\int_0^t dt' \, \xi_i(t')}\equiv T \, \frac{X}{Y}
\; .
\end{equation}
Under this particular condition, $\tilde T_{{\text{eff}}_i}$  is 
again the  ratio of two Gaussian variables. With the convention
$a=\braket{X^2}$, $b=\braket{Y^2}$ and $c=\braket{X\cdot Y}$, the probability distribution function of 
$\tilde T_{{\text{eff}}_i}$ is again a Cauchy distribution with coefficients:
\begin{equation}
A=T \, \frac{\sqrt{ab-c^2}}{b} \; ,  \qquad\quad x_0=T\, \frac{c}{b} \; .
\label{eqn:coeff_Cauchy}
\end{equation}
The coefficient $b$, being proportional to the variance of the thermal noise, remains
fixed whether the system is passive or active, no matter the density. Therefore we 
will keep $b=2Tt/\gamma$. Besides, $a$ can be estimated from the variance of the
displacement of the particle. Again, $c$ is the only fitting parameter. We expect
that the peak of the distribution, \emph{i.e.} $x_0$, be unaffected  by the presence of
the active force in the dilute limit (since the activity does not change
the correlation between the position of the particle and the integral of the noise 
acting on it), while it should be affected by the interparticle forces.
Since the covariance diminishes, the peak should be shifted toward zero. 
Notice that for a single Brownian particle, $a=b=c$ at $t_w=0$,  $A\to 0$  and the
distribution reduces to a delta function centered on $T$ as previously described.

\begin{figure}[t!]
\centerline{
\includegraphics[scale=0.6]{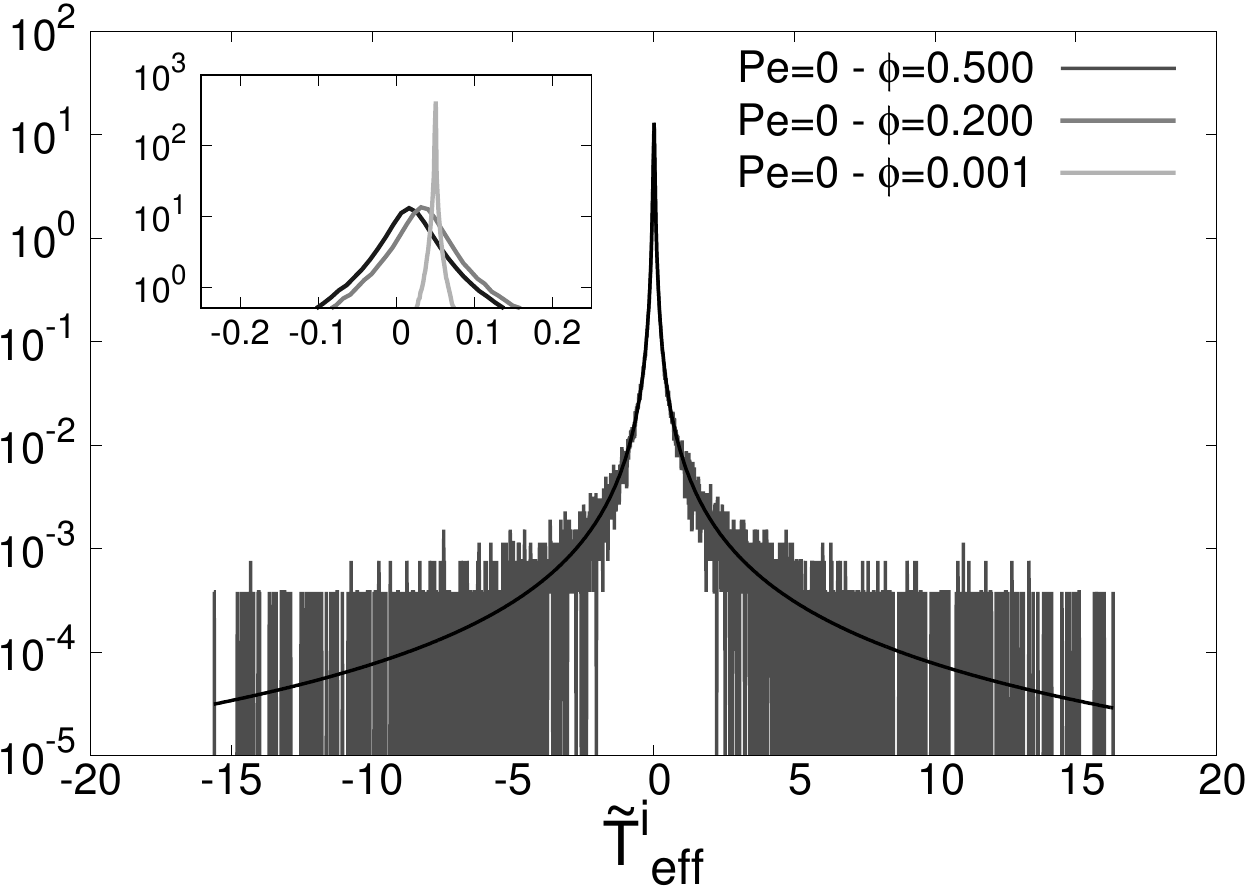}
(a)}
\centerline{
\includegraphics[scale=0.6]{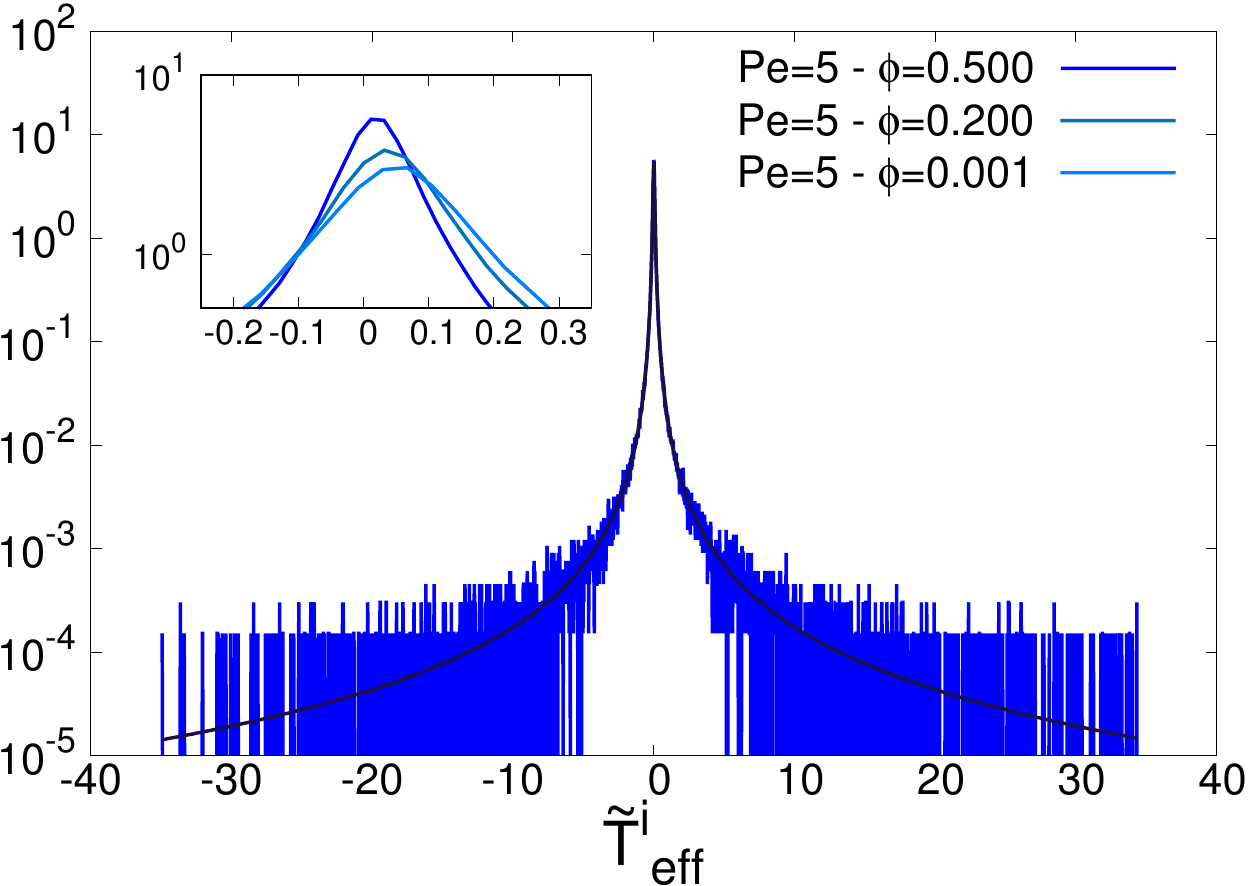}
(b)}
\caption{Fluctuations of the single particle effective temperature
$\tilde T_{{\rm eff}_i}$, see Eq.~(\ref{eq:single-temperature0}). The distributions in the 
main plots are fitted using a Cauchy distribution with coefficients given by Eq.~(\ref{eqn:coeff_Cauchy}), with  $b=2Tt/\gamma$, $a$ obtained from the Gaussian
distribution of the displacement (not shown) and $c$ a fitting parameter. 
All the distributions are evaluated at $t_w=0$. Fitting parameters: $c=17.1$ (Pe = 0, $\phi=0.5$) and $c=19.15$ 
(Pe = 5, $\phi=0.5$). In the insets, enlargements of the peaks of the distributions for
different values of the density are displayed. 
}
\label{fig:pdfs_teff}
\end{figure}

In Fig.~\ref{fig:pdfs_teff}, we show how the statistics  of the single
particle effective temperature, defined in Eq.~(\ref{eq:single-temperature0}),
changes when the density is increased from the very dilute limit both in the
 equilibrium (panel (a)) and in the active non equilibrium case (panel (b)). 
As anticipated above, the position of the peak is not affected by 
the active force and it remains around the temperature of the heat bath in the  dilute limit
(in simulation units $T=0.05$). When the density is increased the peak is shifted
towards zero in both cases.
On the other hand, in the equilibrium case the density increase does not change 
appreciably the width of the distribution, while in the active one increasing the 
density causes a shrinking of the probability distribution function. This feature is reflected in a decrease of the
coefficient $A$ in the latter case.
We note that the distributions for different $\phi$ and Pe collapse on each other 
when put in normal form (apart from the $\phi=0.001$, Pe = 0 one, which is close to a delta function.

As a concluding remark, we notice that if we consider $t_w\ne 0$,
the distribution of the effective temperature in the dense passive and active cases
cannot be fitted with a Cauchy distribution.

\subsection{Heterogeneous phases}
\label{subsec:heterogeneous}

In the heterogeneous phases, especially in MIPS, we need to consider long time
delays (of the order $\Delta t =5\times 10^3$) to ensure that 
a large number of
particles reach the diffusive regime.  The problem with this choice is that 
the bimodal structure of the pdf of the
time-delayed averaged hexatic parameter is smoothed and a lot of particles can not be
considered to belong to one and only one phase  during the full long time interval.
Although this does not affect sensibly the time average, it can interfere, as we will see, in the distribution form.
We therefore adopted a criterium with two different thresholds 
to separate the particles into those belonging to the dilute and dense phases:
the particles contributing to the dilute distributions are the ones with mean hexatic 
order parameter on the left side of the first peak, while the ones contributing to the dense 
distributions are the ones with mean hexatic 
order parameter on the right side of the second peak,
see panel (c) of Fig.~\ref{fig:classification} or panel (b) of 
Fig.~\ref{fig:separation-MIPS}. We then leave
out from the distributions the particles with mean hexatic order parameter in
between the two thresholds. 
For similar reasons, in this section we will always consider $t_w=0$.
 Since we can not access very long times keeping a satisfactory particles distinction,
 we prefer to maximize the measuring time $\Delta t=t-t_w$ to the detriment of
 the waiting time.
  
We now  study systems in their co-existence regions, be them close to Pe = 0 or 
at high Pe, with a packing fraction roughly in the middle of the MIPS region of the phase diagram. 
The aim is to compare the fluctuations of the individual
displacement and response in the two coexisting phases to better understand the origin of  the 
different effective temperatures at the global level (after averaging
over all the particles belonging to each phase).

\begin{figure}[ht]
\centerline{
\vspace{-0.25cm}
\includegraphics[scale=0.34]{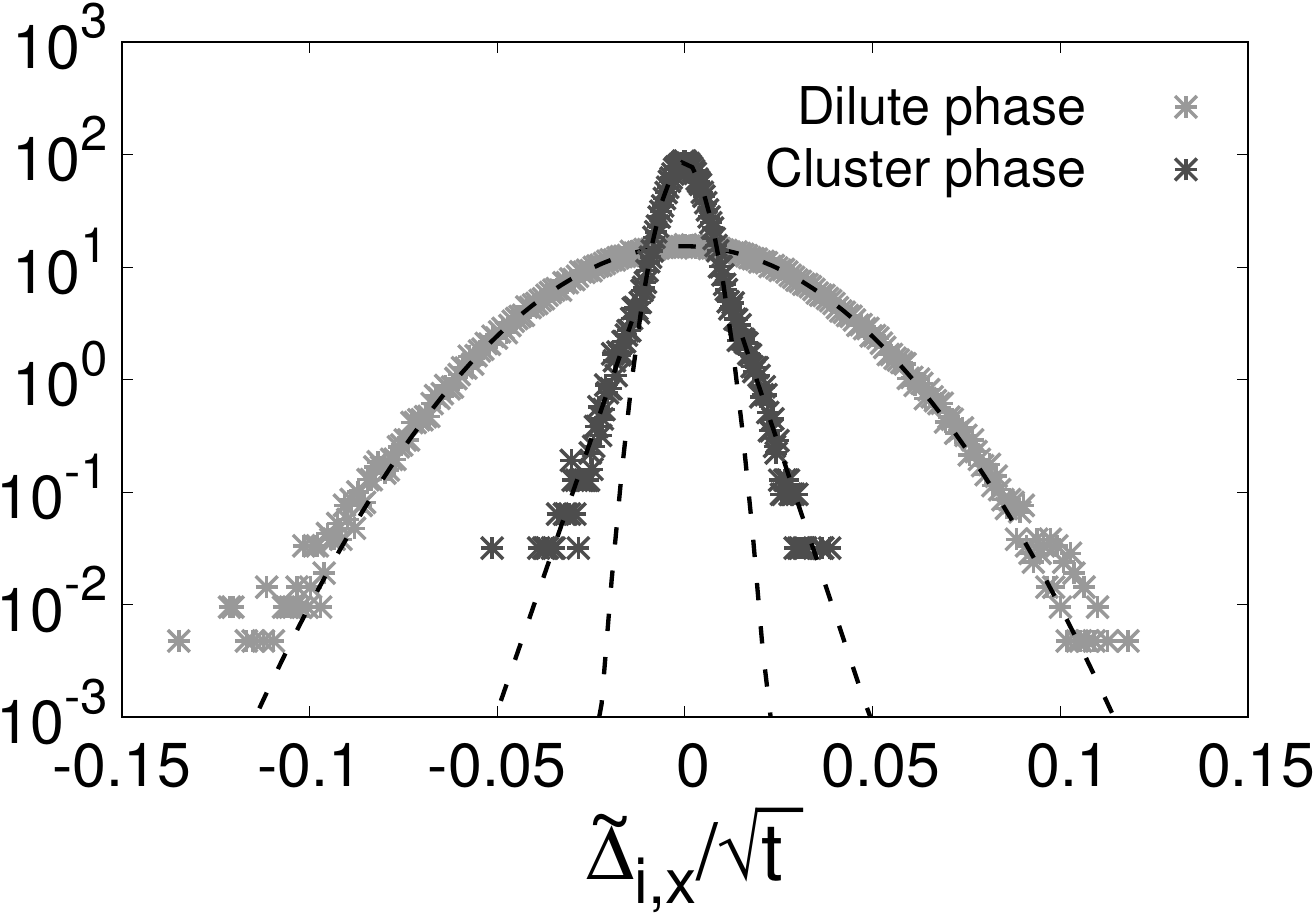}
\includegraphics[scale=0.34]{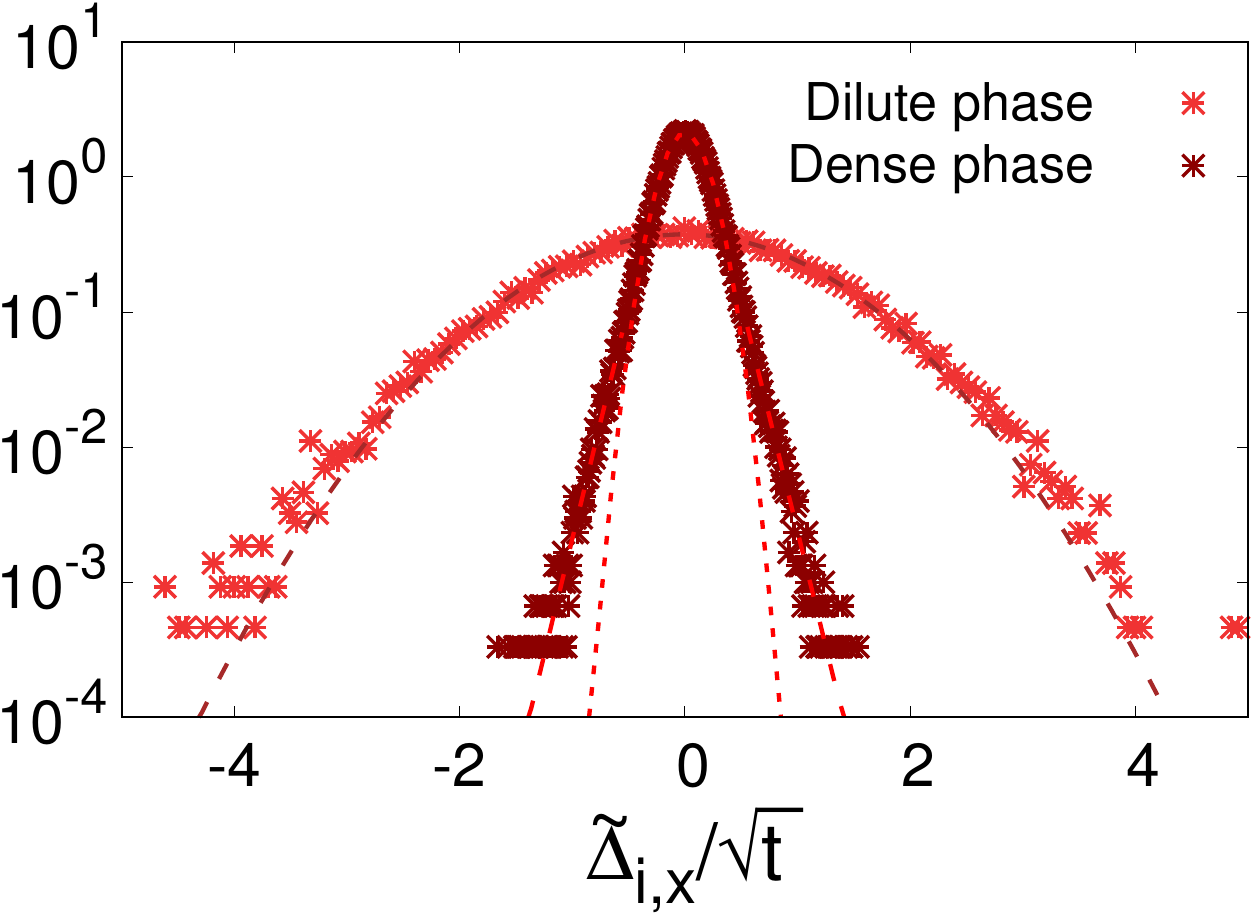}
}
\hspace{3cm}(a) \hspace{4cm} (b)
\caption{Fluctuations in the passive system in the co-existence region of the phase diagram (panel (a) Pe = 0 and $\phi=0.71$) and in the active system in the MIPS 
region of the phase diagram (panel (b) Pe = 50 and $\phi=0.6$).
Probability distribution functions of the individual displacements normalized with
 respect to the squared root of the elapsed time ($t=5\times 10^3 $) in the fluid and in the
 dense phase. The distribution in the fluid phase is fitted using a Gaussian function,
  while in the cluster phase the tails are fitted with exponential decays.
 }
\label{fig:pdfs_deltax_coex}
\end{figure}

\subsubsection{Passive case}
\label{subsubsec:passive}

In Fig.~\ref{fig:pdfs_deltax_coex} (a), we show the pdf of the single
particle displacement, $\tilde \Delta_{i,x}$, in the coexistence region. On the one hand, the data for the 
particles belonging to the fluid phase are almost perfectly well fitted by a Gaussian
function. Instead, the data for particles belonging to the denser
cluster phase are not only narrower but also different. While the central part  is well described by
a Gaussian distribution, the tails strongly deviate from this behavior and
decay exponentially (with slight 
 deviations possibly due to the presence of some residual particles with mixed 
 dynamics). 
 If we look at the distribution of the particle displacements in the 
 homogeneous hexatic phase (Pe = 0, $\phi=0.726$, not shown) this small effect
 disappears and the tails are perfectly described by an exponential distribution.
(Exponential tails are also observed, for example, in complex liquids, see~\cite{Phillies15}
 and references therein.)
\begin{figure}[h!]
\vspace{0.5cm}
\centerline{
\includegraphics[scale=0.6]{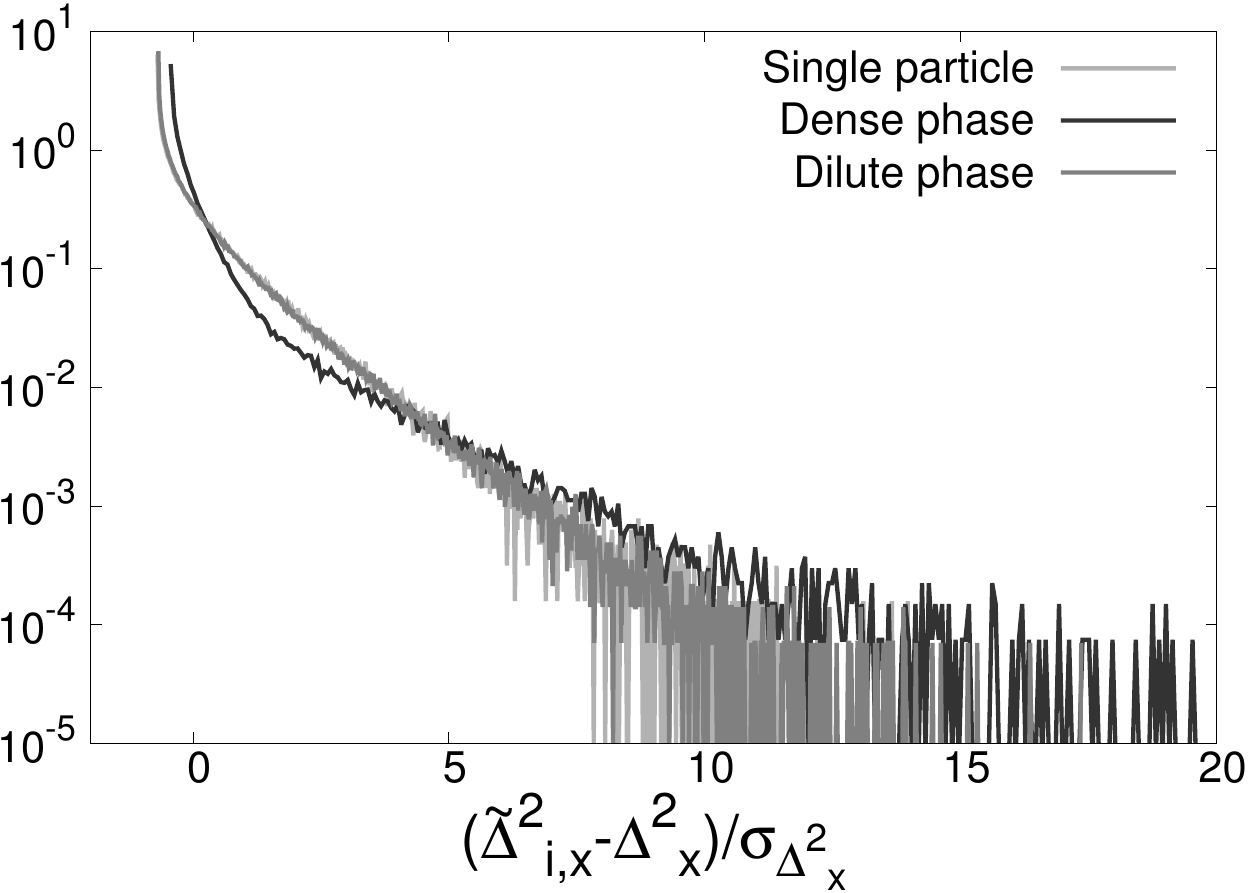}}
\caption{Fluctuations in the passive system,  Pe = 0 and $\phi=0.71$,
in the co-existence region of the phase diagram (same data as in Fig.~\ref{fig:pdfs_deltax_coex}, panel (a)). 
Probability distribution functions of the individual square displacement,
 $\tilde \Delta_{i,x}^2$, in normal form. 
}
\label{fig:pdf_msd_pe0_coex}
\end{figure}

The changes in the statistics of $\tilde \Delta_{i,x}$
are quite naturally reflected  in the distribution of the squared displacement that we 
show in Fig.~\ref{fig:pdf_msd_pe0_coex} for the same parameters.
In normal form, the distribution of the particles in the dilute component resembles strongly
the one in the homogeneous liquid state, with the pdf of the individual displacements falling 
on top of the single particle distribution. On the other hand, the 
exponential tails in the distribution of the displacement of the particles in the dense component 
gives rise to a wider exponential tail in the pdf of the square displacement which does not follow the trend 
of the reference case. 
\begin{figure}[ht]
\vspace{0.5cm}
\centerline{
\includegraphics[scale=0.34]{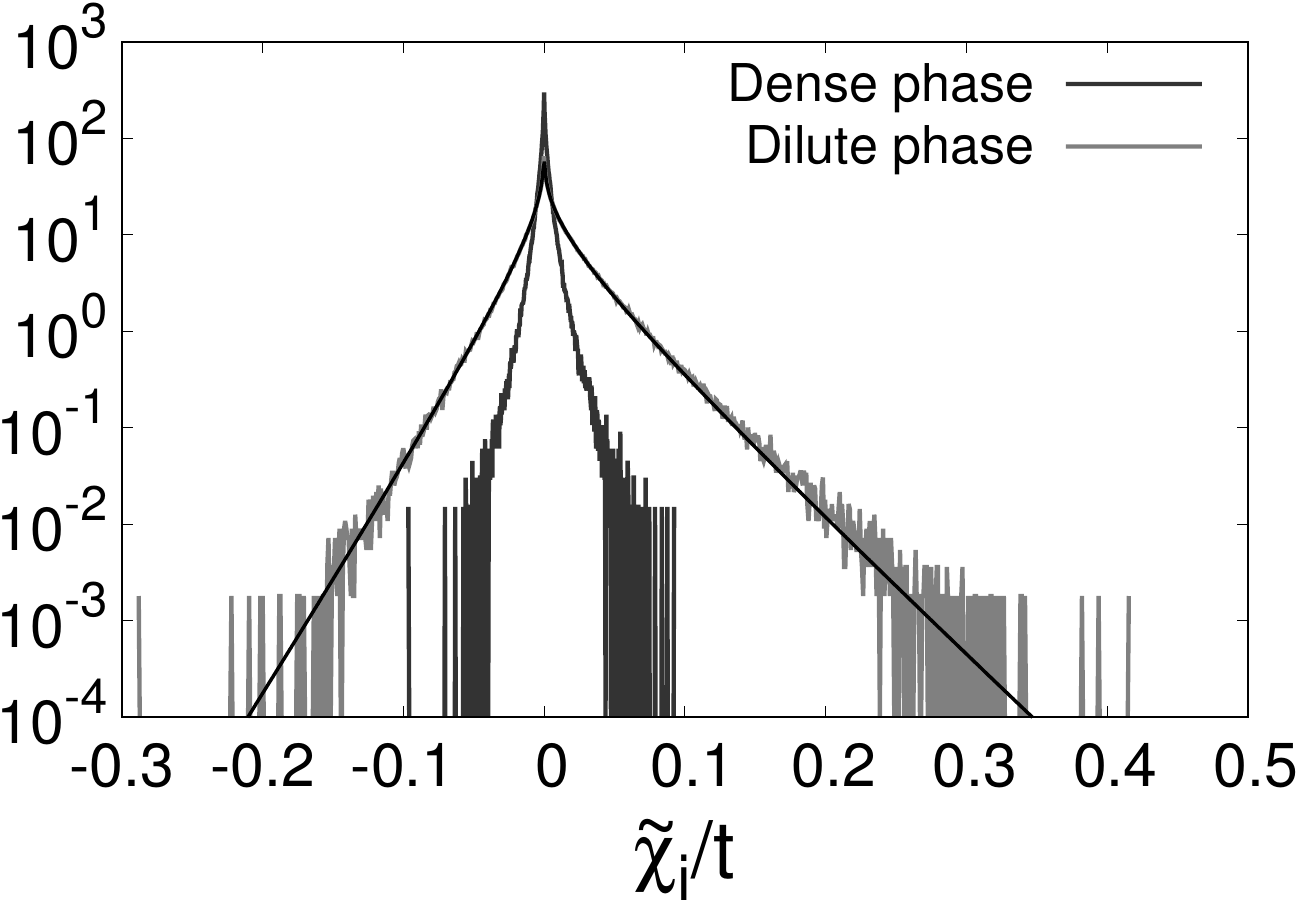}
\includegraphics[scale=0.34]{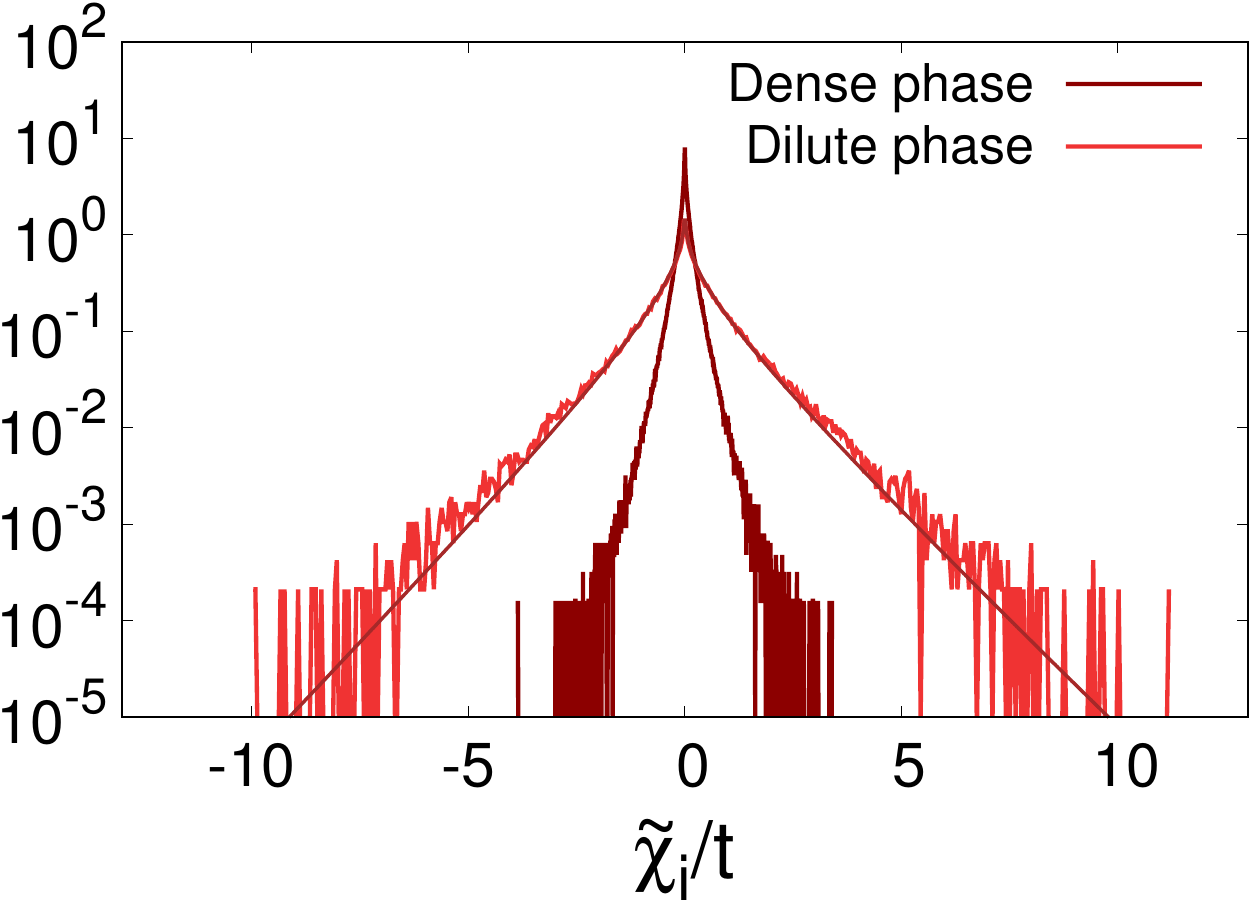}
}
\hspace{3cm}(a) \hspace{4cm} (b)
\caption{Fluctuations in the passive system,  Pe = 0 and $\phi=0.71$, in the co-existence region of the phase diagram (panel (a)) and in the active system, Pe = 50
and $\phi=0.6$, in the MIPS region (panel (b)).
Probability distribution functions of the response $\tilde \chi_i$ defined in Eq.~(\ref{eq:susc_i}) divided by the measuring time 
$t=5\times 10^3 $ in the dilute  and  dense phases. In the dilute case, the pdf is fitted using Eq.~(\ref{eqn:pdf_mu_fit}) with 
$t_w=0$ and $b=2T \, t/\gamma$, while $a$ is obtained from the Gaussian
distribution of the displacement. Fitted parameter: $c=3.3059$ (passive case)
and $c=17.97$ (active case). The shape of the distribution in the dense phase is 
different and it cannot be fitted using similar arguments.
}
\label{fig:pdf_mu_coex}
\end{figure}

The effects of the changes in the distribution of the particle
displacements are felt also by the fluctuations of the response functions. 
In Fig.~\ref{fig:pdf_mu_coex} (a), we show the fluctuations of the single particle
response $\tilde \chi_i$ defined in Eq.~(\ref{eq:susc_i}), divided by the measuring time,
in the co-existing dilute liquid and dense hexatic phases. 
The pdf for the particles in the dilute phase is similar to what we found in the
homogeneous fluid phase. Indeed, the distribution can be fitted using 
Eq.~(\ref{eqn:pdf_mu_fit}) with $t_w=0$ and the procedure explained above.
Importantly, even though the densities of the two phases
are very close, the shape of the distribution in the hexatic phase is different: the
pdf is tighter, its skewness is reduced, and it cannot be fitted using 
Eq.~(\ref{eqn:pdf_mu_fit}), mainly because of deviations in the tails.
 
\begin{figure}[h!]
\centerline{\includegraphics[scale=0.6]{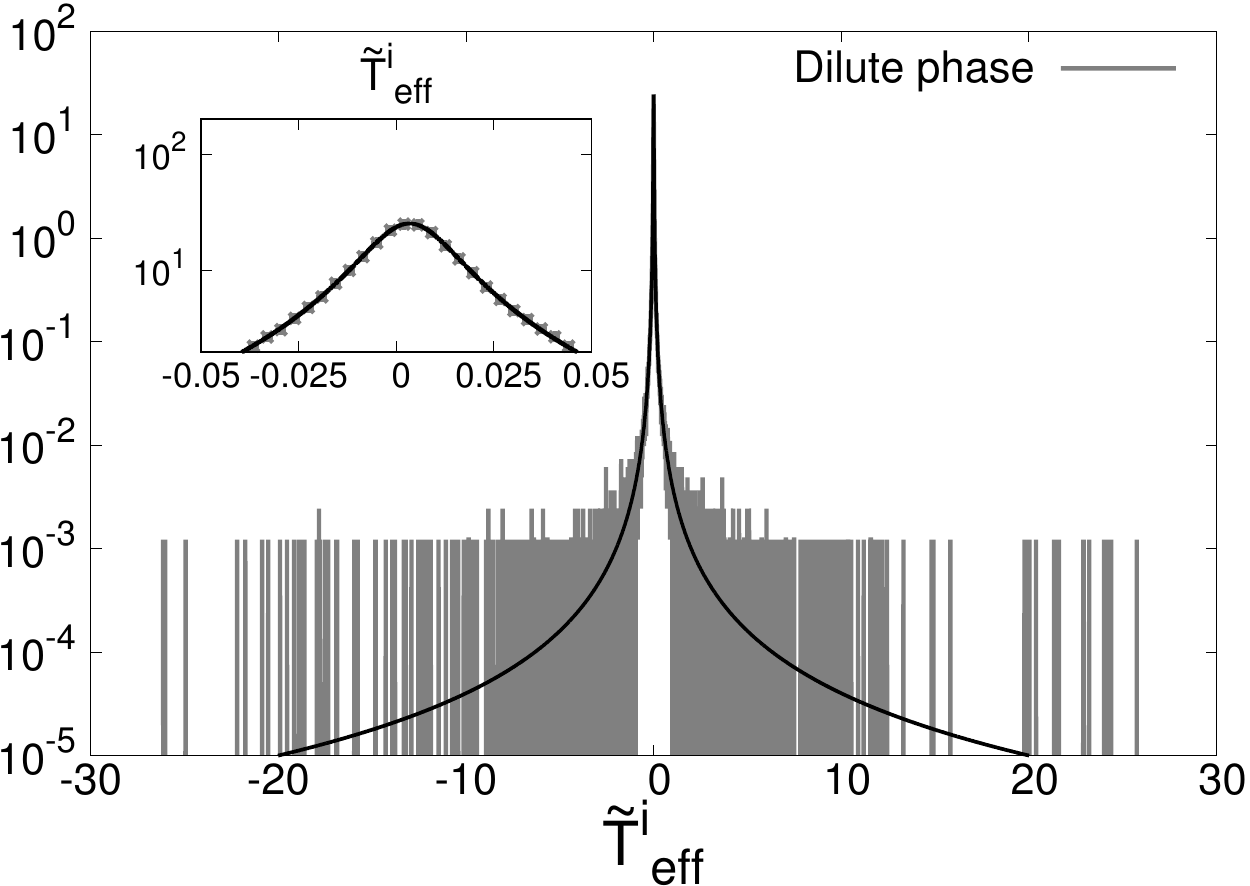}(a)}
\centerline{\includegraphics[scale=0.6]{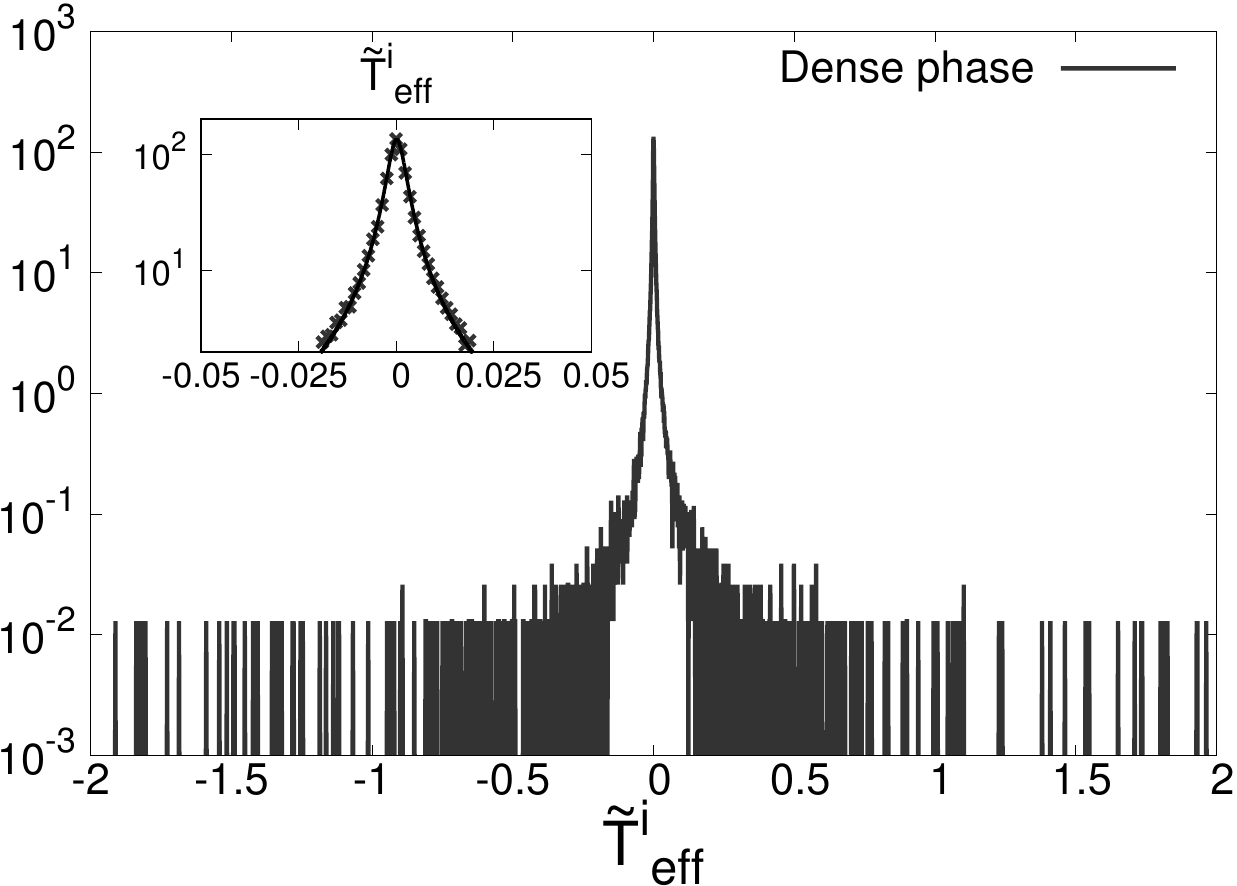}(b)}
\caption{Fluctuations of the single particle effective temperature
$\tilde T_{{\rm eff}_i}$, see Eq.~(\ref{eq:single-temperature0}), in the dilute (a)
and in the dense phase (b) coexisting at Pe = 0, $\phi=0.71$. 
The distributions are fitted using a Cauchy distribution with coefficients given by Eq.~(\ref{eqn:coeff_Cauchy}), with  $b=2Tt/\gamma$, $a$ obtained from the Gaussian
distribution of the displacement and $c$ a free parameter, taking values 
$c=3.38$ (dilute phase) and $c=0.11$ (dense phase).
All the distributions are evaluated at $t_w=0$. In the insets, enlargements of the peaks for
different values of the density. }
\label{fig:pdf_teff_pe0_coex}
\end{figure}

 In Fig.~\ref{fig:pdf_teff_pe0_coex}, we show the distribution of the
 single particle effective temperature defined in Eq.~(\ref{eq:single-temperature0})
 evaluated in the two phases separately. The distribution of the dilute phase appears
 broader, as expected. From the enlargement shown in the inset, 
 it is clear that  the peak of the distribution of the dense phase is 
 sharper than the one of the dilute phase and it is shifted around zero.
 
Following the same strategy that we used in the study of the 
 homogeneous system, we fitted the pdf of the dilute phase using a Cauchy distribution, Eq.~(\ref{eqn:pdf_teff}). The
fitted parameter is written in the caption. Notice that, naturally
enough, it is compatible with the fit performed for the response function.  We do not
 expect that the distribution in the denser hexatic phase preserves a similar shape. 
 In spite of that, we find that the central part of the pdf can still be described by a
 Cauchy distribution with violations in the tails. It is clear 
 the particles with Gaussian distruibuted displacement contribute to the peak,
 while those which contribute to the exponential tail of the displacement pdf
 also contribute to the tails of the pdf of the effective temperature.

\subsubsection{MIPS}
\label{subsubsec:MIPS}

We now turn to the study of the fluctuations under strong activity, in the MIPS region of the phase diagram.

In Fig.~\ref{fig:pdfs_deltax_coex} (b), we show the distribution of 
$\Delta_{i,x}$ in the dilute and dense phases.  The curves 
are similar to the ones in the  equilibrium heterogeneous phase (panel (a) of the
same figure): 
Gaussian for particles in the dilute component while, in the denser component, only the central
part of the distribution is normal and the tails decay exponentially. In spite of these
similarities, we also notice some differences. Firstly, there are some deviations from the
Gaussian behavior in the tails of the distribution of the dilute component, possibly due to a partial
mixture between the phases. Secondly, even considering the same measuring time
($t=5\times 10^3$), the  exponential tails are much more evident in the
active than in the passive case.

The Gaussian character of the displacements  in the
dilute phase leads to the collapse of the distribution of the squared displacement
$\tilde\Delta^2_{i,x}$ on top of the reference single particle one, when plotted in normal form (not shown), while  in the dense phase the distribution develops a longer exponential tail, similarly to the passive case.

 \begin{figure}[h!]
\centerline{
\includegraphics[scale=0.6]{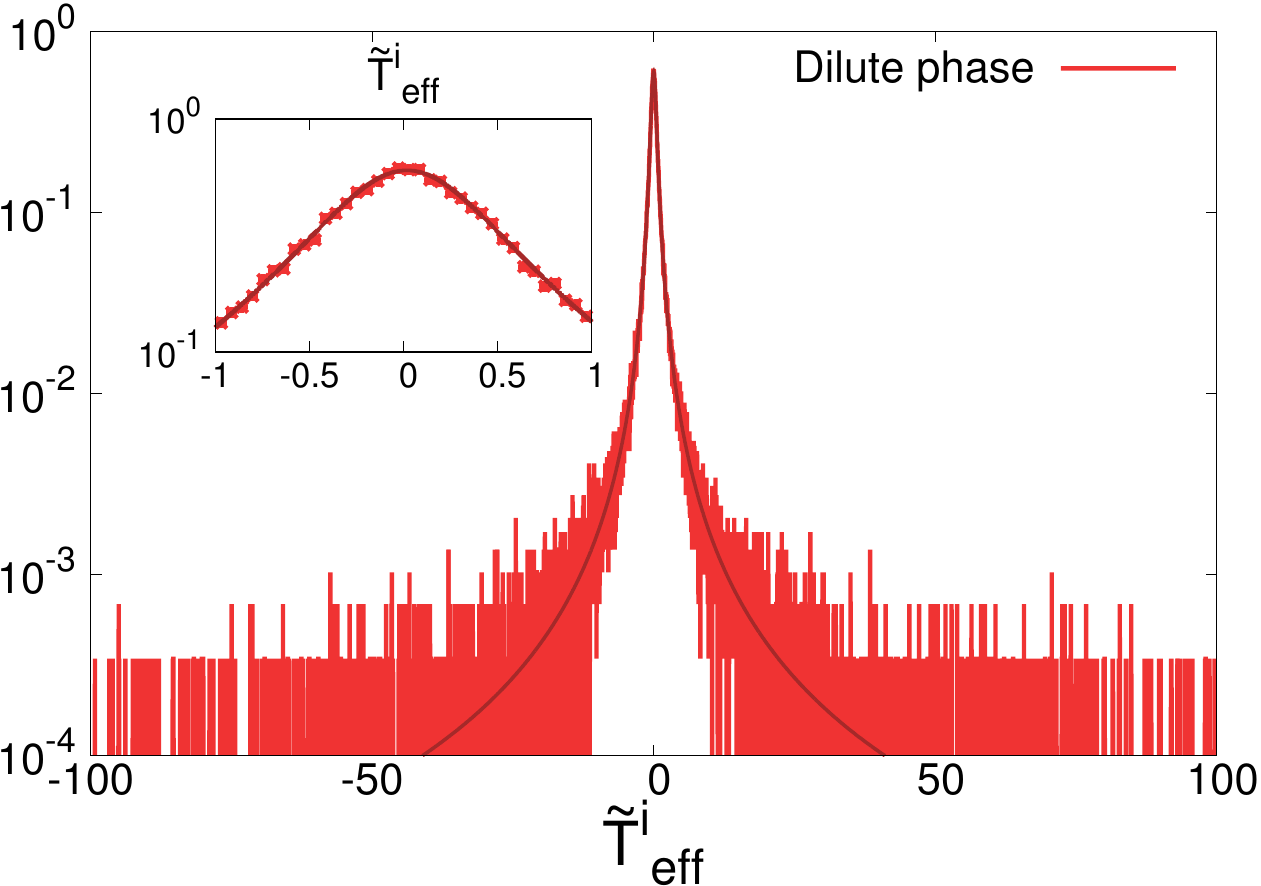}
(a)}
\centerline{
\includegraphics[scale=0.6]{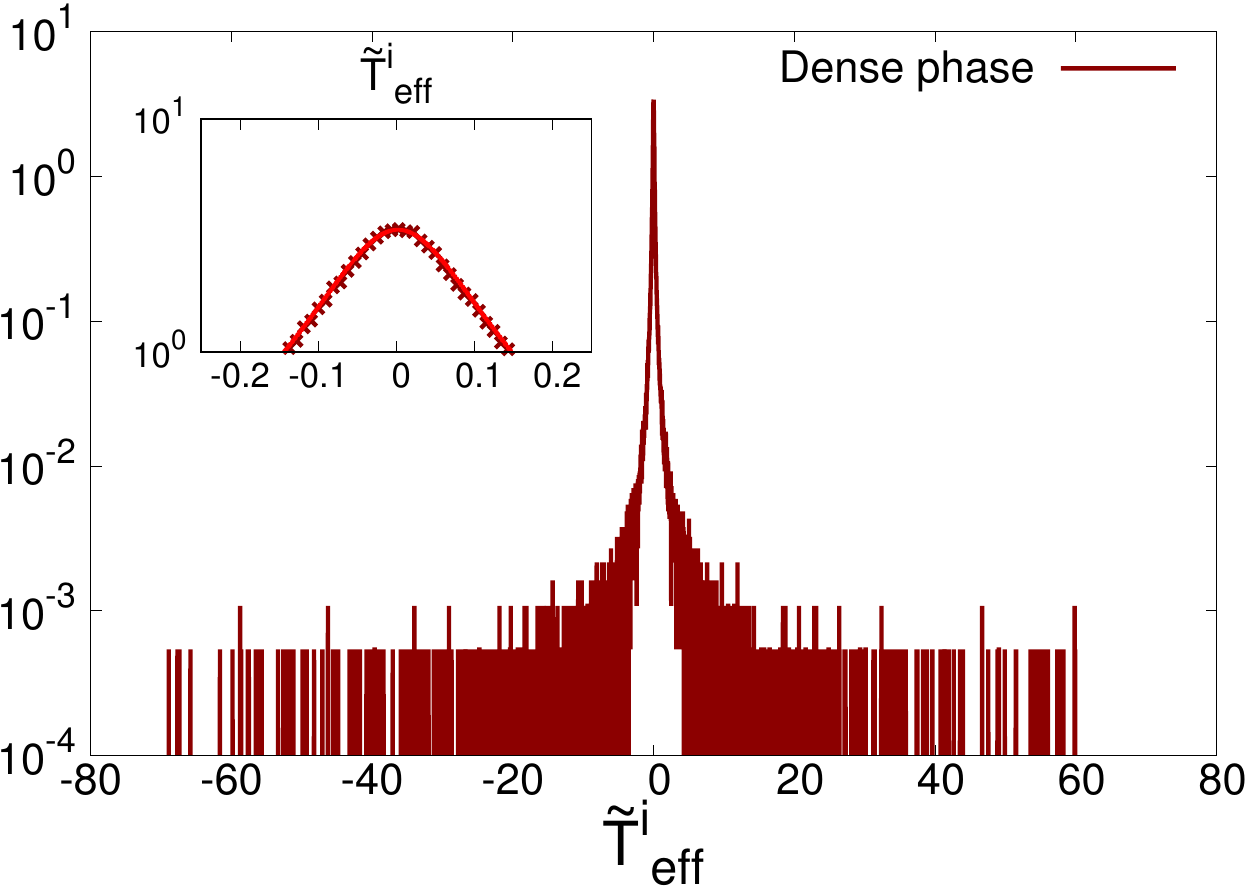}
(b)}
\caption{Fluctuations of the single particle effective temperature
$\tilde T_{{\rm eff}_i}$, Eq.~(\ref{eq:single-temperature0}), in the dilute (a)
and dense phase (b) coexisting at Pe = 50 and $\phi=0.6$. 
The distributions are fitted using a Cauchy distribution with coefficients given by Eq.~(\ref{eqn:coeff_Cauchy}), with  $b=2Tt/\gamma$, $a$ obtained from the Gaussian
distribution of the displacement and $c$ a fitting parameter. 
All the distribution are evaluated at $t_w=0$. Fitted parameters: $c=14.52$ (dilute
phase) and $c=2.09$ (dense phase). 
In the insets, enlargements of the peaks of the distributions for
different values of the density. }
\label{fig:pdf_teff_pe50_coex}
\end{figure}

The fluctuations of the individual linear response, see Fig.~\ref{fig:pdf_mu_coex} (b),  
also show differences when the particles belong 
to the dilute and dense phase. While the distribution of the 
former is well fitted using Eq.~(\ref{eqn:pdf_mu_fit}) and the
strategy adopted before, see the caption for the actual values of the parameters, 
for the latter a similar fit could be performed only in 
the central part of the distribution, with deviations in the tails.
Firstly, notice that, while
 in the passive case the two phases coexist with very similar densities, in the MIPS 
 region the significant difference in the densities of the phases causes the branches of
  the distributions to depart from the peak with very different slopes. Secondly, the 
  skewness of the  distributions of the individual response tends to disappear also
  in the dilute phase.

Before concluding the study of the fluctuations, we look at the
 behavior of the distribution of the single particle effective temperature.
 The shape of the distributions of $\tilde T_{{\rm eff}_i}$ in the MIPS region, 
 shown in Fig.~\ref{fig:pdf_teff_pe50_coex}, resembles the one found in the passive limit (see Fig.~\ref{fig:pdf_teff_pe0_coex}). The distribution
 evaluated in the dilute phase is broader than the distribution of the temperatures 
 in the dense phase and, while in the dilute phase the pdf follows a Cauchy 
 distribution, Eq.~(\ref{eqn:pdf_teff}), in the dense phase this holds only for the central part of the pdf, where
 the particles with Gaussian displacements are found.

\subsection{Summary}
\label{subsec:summary}

The detailed analysis of fluctuations that we presented in the Section can be summarized as follows.

In the homogenous active phase at low density, as in the dilute phase in MIPS, we found:
\begin{itemize}
\item[--]  Gaussian distributions
of the displacement $x_i(t)-x_i(t_w)$, that lead to the 
\vspace{-0.2cm}
\item[--] exponential (corrected by an algebraic factor) pdf of the square displacements, 
\vspace{-0.2cm}
\item[--] an exponential times a Bessel function for the linear response leading to different exponential 
decays for both positive and negative branches in the large argument limits,
\vspace{-0.2cm}
\item[--] and a Cauchy probability for the individual effective temperatures, in the 
way we defined them.
\end{itemize}
The origin of the statistics of $\tilde \Delta^2_i, \tilde \chi_i, \tilde T_{{\rm eff}_i}$  is in the fact that the 
first one is the product of the two same Gaussian variables, the second the product of two different 
Gaussian variables, and the third one is the ratio between two Gaussian variables. The parameters of 
these distributions depend on $\gamma$, $t$ and $t_w$. Once put in normal form, all these curves collapse
on three master curves that are, basically, the ones of the single active particle.

In the heterogeneous active phase close to the passive limit, it is hard to make conclusive
measurements as the two phases have very close density and are rather similar. Still, we have been
able to see that the fluctuations in the dense phase are different from the ones in the dilute one. This 
was more clearly seen in the study of fluctuations in the MIPS region, where the fluctuations of particles belonging
to the dilute phase behave just as in the homogeneous cases, while the ones in the dense component 
have different statistics. The latter are characterized by 
\begin{itemize}
\item[--] 
A Gaussian central behaviour and exponential tails for the distribution of 
$x(t)-x(t_w) $ that
\vspace{-0.25cm}
\item[--]  modify the tails of the distributions of the square displacement, 
linear response and effective temperature.
\vspace{-0.25cm}
\item[--] 
There is no collapse of the data for the dense and dilute phases on a single 
master curve even when put in normal form.

\end{itemize}

\section{Kinetic vs. effective temperatures}
\label{sec:kinetic}

In this Section we compare the (time-delayed) effective temperature
to the (instantaneous) kinetic temperature notion, that has been 
widely used in granular matter studies, but also recently in the 
context of active matter systems. We will first discuss in Subsec.~\ref{subsec:friction} the effect of the friction coefficient on the effective
temperature, and we will perform the proper comparison between the two quantities
 in Subsec.~\ref{subsec:kinetic-vs-effective}.

\subsection{The effect of the friction coefficient}
\label{subsec:friction}

The single particle effective temperature ${T_{\text{eff}}}$ depends on ${\gamma}$ only during a transient temporal regime, 
through the mean squared displacement, see Eq.~(\ref{delta_sp}), while its asymptotic value
does not. 
We wish to examine whether the effective temperature of the interacting system is also independent 
of $\gamma$ after a short irrelevant transient. 

\begin{figure}[h]
\centerline{
\includegraphics[width=8cm]{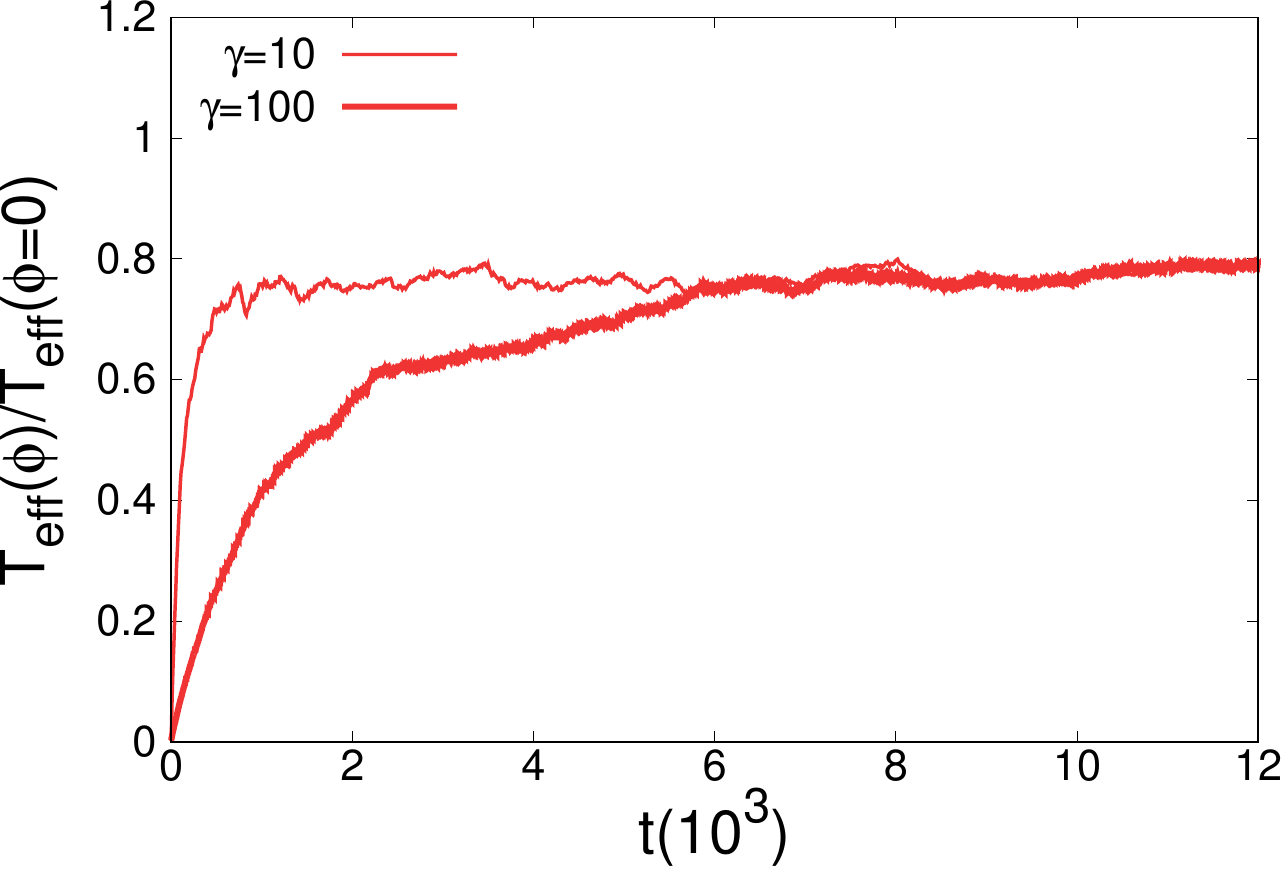}
}
\caption{(Color online.) 
Effective temperature \emph{vs} time in systems with different damping coefficients. 
There is a dependence on ${\gamma}$ only in the first regime, while the stationary value 
is $\gamma$-independent within numerical accuracy. ${\text{Pe}=50}$, ${\phi=0.200}$, ${\gamma=10,\, 100}$.}
\label{fig:Teff_200}
\end{figure}

As a first check, we evaluate the effective temperature in the homogeneous fluid and we change the value of the 
damping coefficient. In Fig.~\ref{fig:Teff_200}, we show 
the time-dependence of ${T_{\text{eff}}}$ in a system with ${\text{Pe}=50}$, ${\phi=0.200}$ and ${\gamma=10,\, 100}$. 
The single particle behavior is inherited in the sense that the effective temperature depends on $\gamma$  
only during a first regime, while the stationary value is independent of it.

\begin{figure}[h]
\centerline{
\includegraphics[height=3.05cm]{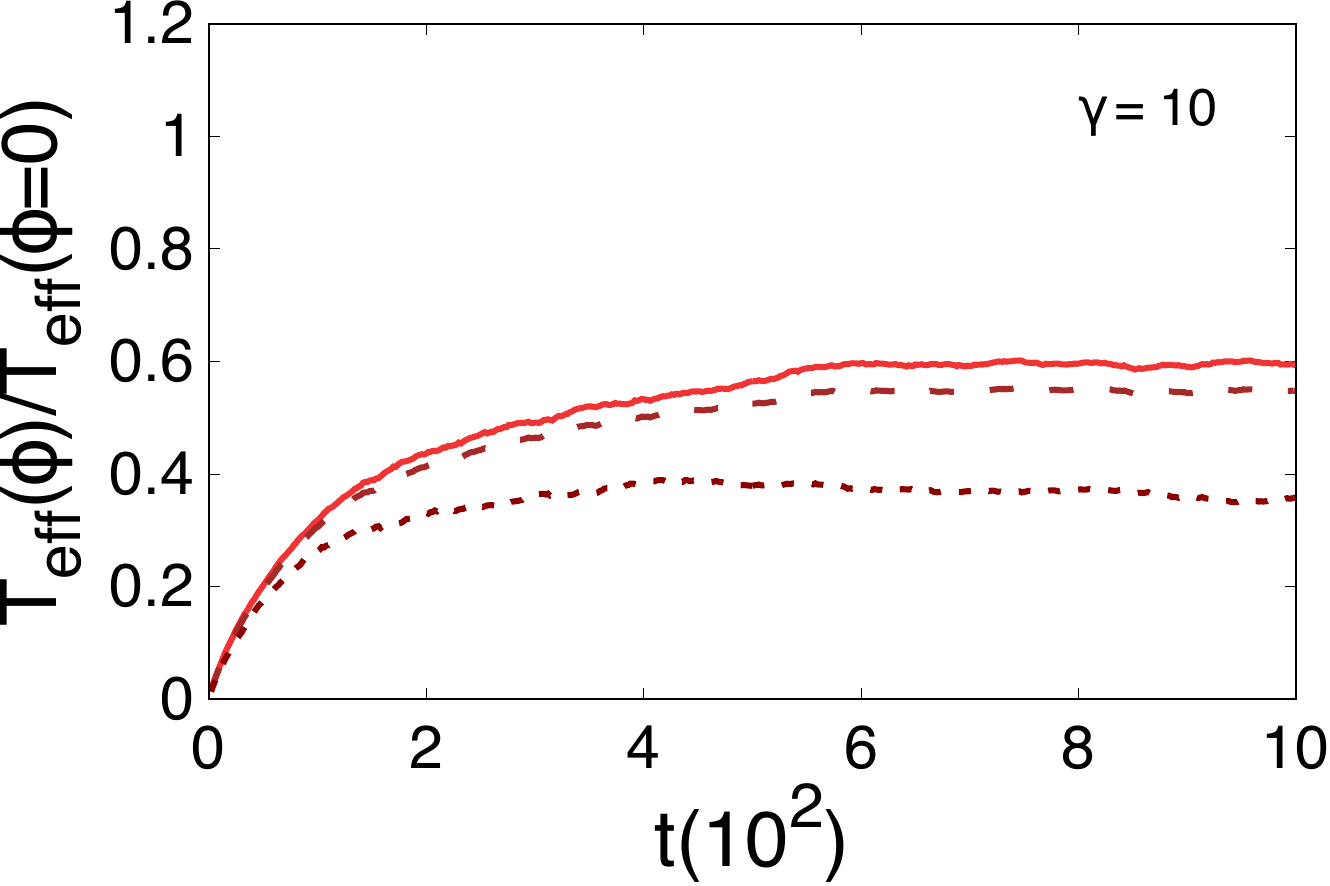}
\hspace{-0.1cm}
\includegraphics[height=3.05cm]{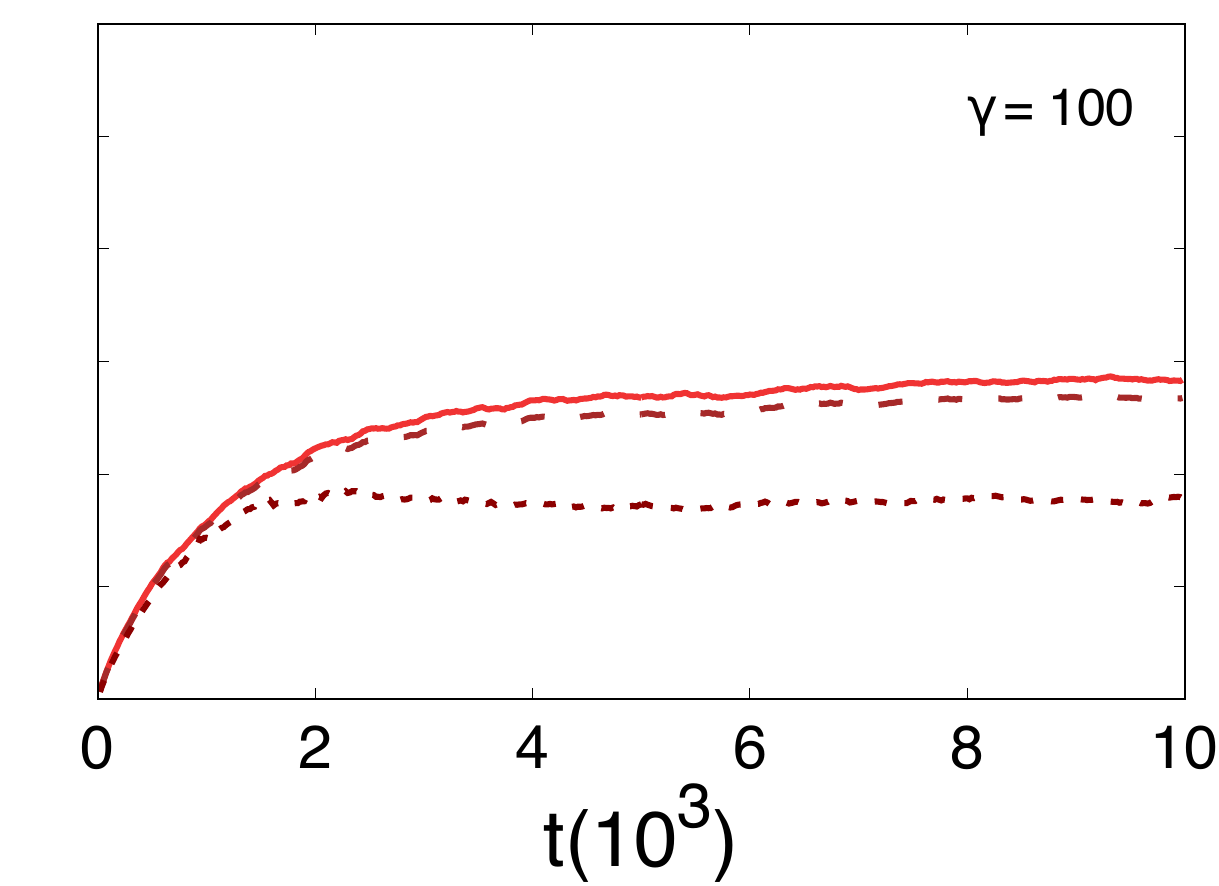}
}
\caption{Effective temperature in the MIPS region
(Pe=50, $\phi=0.5$) for different damping ratio, $\gamma=10$ (left), 
$\gamma=100$ (right), of the dilute and dense subsystems,  and the one of the full system (continous, dotted and dashed lines, respectively). The measurements with 
$\gamma=100$ have been performed using $\Delta t=10^4 $ instead of 
$\Delta t=10^3$. 
}
\label{fig:Teff_pe50_gamma}
\end{figure}

The measurements within MIPS, say for Pe = 50 and $\phi=0.500$, are harder since 
 since for stronger $\gamma$ the time needed to reach the steady state increases. 
This somehow counterintuitive statement can be understood by looking at the expression for 
the single particle $\Delta^2$, Eq.~(\ref{delta_sp}), where the time-scale for the decay of the exponential 
is $1/D_\theta$ with $D_\theta = 3T/(\gamma\sigma_{\rm d}^2)$.
We find again that the effective temperatures depend on the friction coefficient only
 during a transient, while they saturate to values that are independent of $\gamma$.  This is 
 shown in Fig.~\ref{fig:Teff_pe50_gamma}.
 
 \vspace{0.25cm}
 
\subsection{The kinetic temperature}
\label{subsec:kinetic-vs-effective}

The kinetic temperature ${T_i^{\text{kin}}}$ is the temperature extracted from the velocity fluctuations. As it is 
clear from its definition
\begin{equation}
T_i^{\text{kin}} \equiv \langle E_i^{\text{kin}}(t) \rangle =\frac{1}{2}m\langle v_i^2(t) \rangle
\; , 
\end{equation}
the kinetic temperature accesses the instantaneous properties of the system, and not the 
time-delayed ones, with $i$ 
the particle label. In the inhomogeneous 
cases we will distinguish the behavior of the particles in the dilute and dense phases. It is quite clear that
for Brownian particles one needs to study the stochastic dynamics in the under-damped limit, in which inertia is not 
neglected, to compute the kinetic energy and from it the kinetic temperature.

The kinetic temperature of a single active particle can be easily calculated by solving the full Langevin equation.
After a short transient after which the initial velocity is forgotten, one obtains
\begin{equation}
T_{\text{kin}}=
T
\left[1+\frac{T}{2\gamma\sigma^2}\frac{\text{Pe}^2}{\left(\frac{1}{t_I}+\frac{1}{t_A}\right)}
\right]
\; ,
\label{Tkin_sp}
\end{equation}
where ${t_I=m/\gamma}$ and ${t_A=1/D_{\theta}}$ are the relevant time scales, 
\emph{i.e.} the inertial and the active time. In the passive case Pe = 0 and $T_{\text{kin}}=T$ for all the possible values of $t_I$. In the 
active case with $t_I\to 0$ one also recovers $T_{\text{kin}}=T$, contrary to what happens with the effective 
temperature, Eq.~(\ref{eq:Teff-single}), that depends explicitly on Pe even in this limit.

\begin{figure}[h]
\centerline{
\includegraphics[width=8cm]{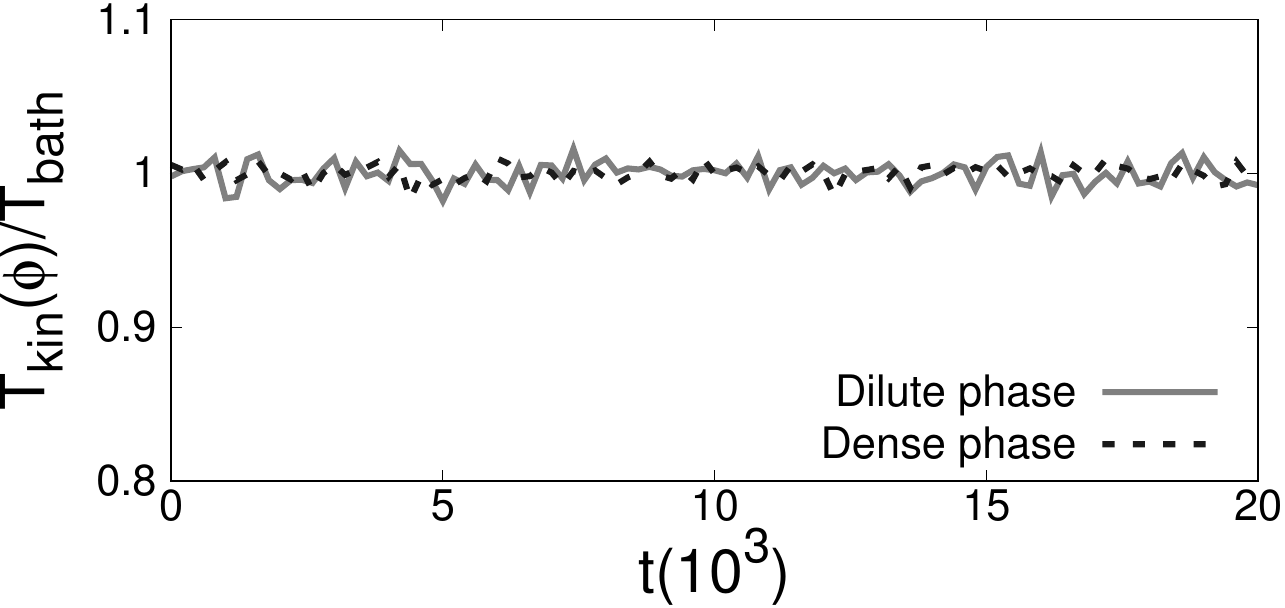}
}
\caption{(Color online.) 
Comparison of the kinetic temperature of the coexisting phases in the passive system. 
We normalize the kinetic temperature with respect to the bath temperature. 
${N=256^2}$, ${\text{Pe}=0}$,  ${\phi=0.710}$, and ${\gamma=10}$.
}
\label{Tkin_Pe0}
\end{figure}

In the passive case, we expect the coexisting phases to share the same kinetic temperature which, moreover, 
should equal the temperature of the thermal bath. In Fig.~\ref{Tkin_Pe0}, we show the kinetic temperature of the particles belonging to the hexatic and liquid phases separately: they are equal and, on average, they also coincide with the 
temperature of the bath. 

\begin{figure}[h!]
\centerline{
\includegraphics[scale=0.6]{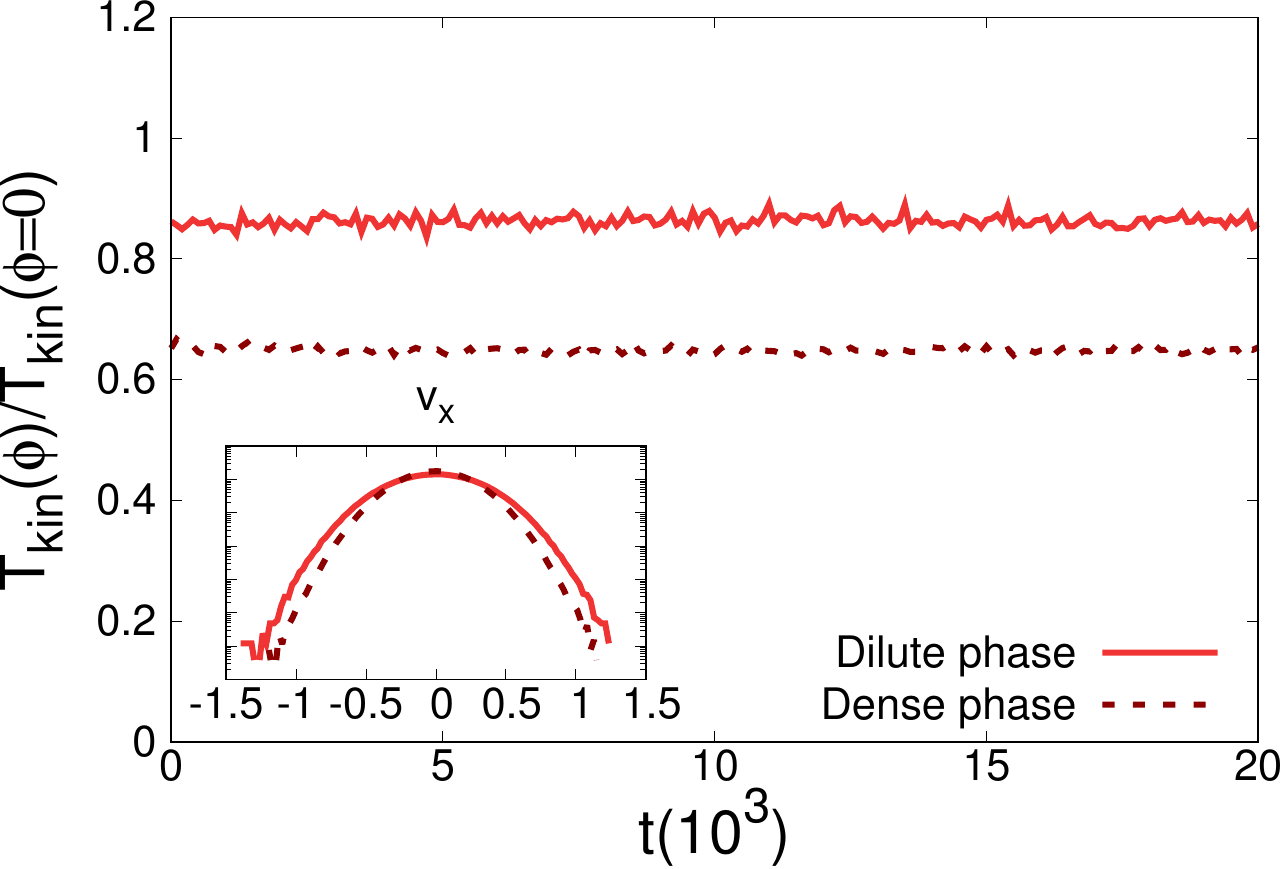}
(a)}
\centerline{
\includegraphics[scale=0.6]{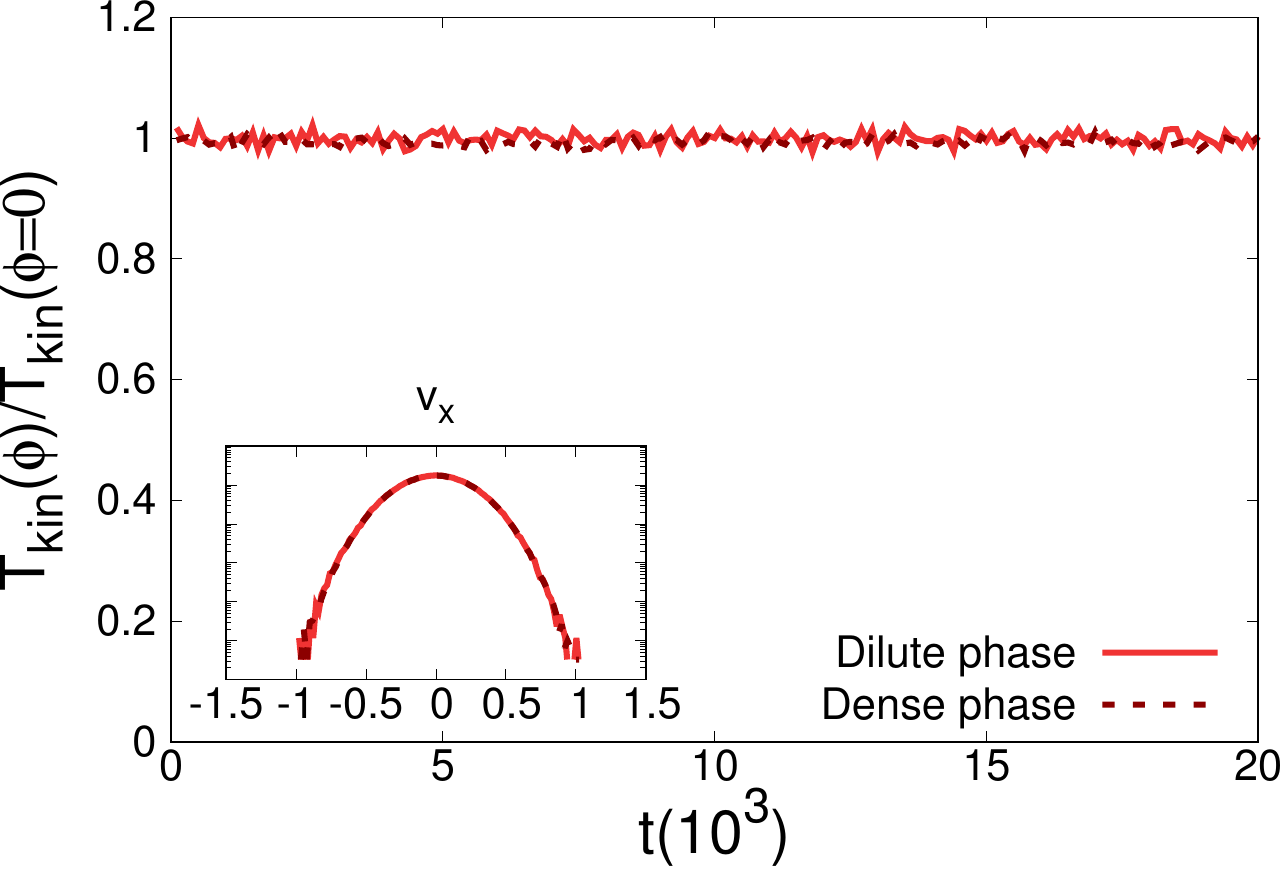}
(b)}
\caption{(Color online.) 
Comparison of the kinetic temperature of the two coexisting phases in the active system. 
We normalize the kinetic temperature with respect to the kinetic temperature of a single particle. 
Insets: probability distribution function of the particle velocity along one axis. 
${N=256^2}$,  ${\text{Pe}=50}$,  ${\phi=0.500}$, 
(a)  ${\gamma=10}$ and (b)  $\gamma=100$.
}
\label{Tkin_Pe50}
\end{figure}

We now reconsider the system with ${\text{Pe}=50}$ and ${\phi=0.500}$. We evaluate the kinetic temperature of the particles belonging to the two phases, and we present the results normalized by the single particle value given in  
Eq.~\eqref{Tkin_sp} for the same Pe. To investigate the effect of the inertia time-scale $t_I=m/\gamma$,  
we change the value of the friction coefficient from ${\gamma=10}$ to ${\gamma=100}$ and we repeat the analysis. 
The results are shown in Fig.~\ref{Tkin_Pe50}. With $\gamma=10$ (a) the kinetic temperatures of the two phases are different and both of them are lower than the single particle value (but higher than the bath temperature). Instead, 
with $\gamma=100$ (b) the kinetic temperatures fluctuate around the same value, the one of a single particle with the same activity (which is almost equal to the bath temperature). Differences are also seen at the 
level of the probability distribution functions (pdf) of the kinetic energies. In the insets we show the 
pdfs of, say, the horizontal velocity component ${v_x}$. While for small value of the friction coefficient the pdfs are different (a), for sufficiently strong friction forces, the curves collapse on the same curve (b). 

The results shown in this section are in qualitative agreement with those presented in~\cite{Mandal19}
where  it is reported that a system of
underdamped ABPs can separate into two coexisting phases at different kinetic
temperatures. This difference vanishes when the dynamics are overdamped and inertia
plays no role. Indeed, in this limit the kinetic temperature tends to the temperature of the heat bath 
and it becomes the same in both phases.

\section{Discussion}
\label{sec:discussion}

The main focus of this paper was the analysis of the effective temperature in a 
heterogeneous active  system. More precisely, we studied the stationary dynamics of the active Brownian particle bidimensional system in (i) homogeneous situations, (ii) the co-existence region between 
liquid and hexatic close to the passive limit and (iii) the Mobility Induced Phase Separation (MIPS) sector of the phase diagram.

The first type of measurement that we performed was the one of 
the effective temperature, using the fluctuation dissipation relation, 
as the ratio between the mean square displacement and (twice) 
the time integrated linear response,  either of the 
particles
in the selected phases or in the whole system. In normal diffusion problems, this is just the Einstein ratio, that is to say, the 
ratio between the diffusion coefficient and the 
mobility. In all cases we 
considered sufficiently long time delays so that the dynamics is diffusive and all 
transients have been surpassed.

In the homogeneous phases we found a $T_{\rm eff}(\phi, \mbox{Pe})$ that at fixed Pe decays linearly with increasing 
packing fraction.
Concerning the Pe dependence, at low density, $\phi<0.5$, $T_{\rm eff}(\phi, \mbox{Pe})$ increases as Pe$^2$, similarly to 
what was found in several other active systems, see {\it e.g.}~\cite{Loi1,Loi2,Loi3,SumaPRE1}.

In the heterogeneous phases, we showed that, while in equilibrium the effective temperatures of the two 
phases coincide and they are equal to the bath temperature with weak deviations 
from this behaviour for small values of Pe, the behaviour is very different, and way more interesting, 
in the MIPS region at high Pe.

In the MIPS sector of the phase diagram, 
at fixed Pe, the global effective temperature depends on the packing fraction: 
at the upper limit, $\phi_>$, it  approaches the one 
of the homogeneous dense phase while in the opposite limit, $\phi_<$, it reaches the one of the 
homogeneous gas. Looking more closely into the two co-existing phases, one 
sees that the global effective temperature can be decomposed into the 
contributions of the particles belonging to the dilute and dense phases, and that these are notably different.
Interestingly enough, in the whole interval $[\phi_<, \phi_>]$ at fixed Pe, 
the effective temperatures of the dilute and dense components are, within our numerical accuracy, constant and equal to the 
ones at the corresponding interval edges. Indeed, the curves $T_{\rm eff}(\phi)$
restricted to the particles belonging to each of the two phases,
are convincingly 
constant not too close to the borders of the interval,  while the measurements close to  $\phi_>$ and $\phi_<$ become 
harder and harder as one of the phases disappears (the 
dilute one close to $\phi_>$, 
and the dense one close to $\phi_<$). 
We have therefore  recovered the expectation that the effective temperature in each phase remains constant during the phase transition, but interestingly enough the temperatures of the two phases are different, being the system explicitly out of equilibrium.

Let us now compare these results to others in the literature. In particular, 
two studies  that addressed issues linked to the 
effects of heterogeneities on, and observable dependencies of, the effective temperatures.

Levis and Berthier~\cite{LevisBerthier}
performed a spectral analysis of the effective temperature
in the fluid, clustered, dense phase of a self-propelled particle 
model~\cite{Levis14} and found a non-trivial wave-vector dependence (over a factor of 2, approximately). Rightly so, 
these authors concluded that, contrary to what happens in glassy 
systems or weakly sheared liquids, a single effective temperature does not 
describe the dynamics of their active liquid. Instead, upon approach to a 
glassy arrest at very large densities, or to the dilute limit, the wave-vector dependence progressively 
disappears  and a single $T_{\rm eff}$ value is found with 
different measurements  in their system.  These findings are in line with our 
results, since they also reflect the strongly heterogenous nature of the system studied~\cite{Levis14}. 
A spatially resolved analysis was not conducted in this model.

Nardini {\it et al.}~\cite{Nardini17} also examined the wave vector dependence of the 
FDT violations in a scalar field theory with model B dynamics. We note that activity was added 
in this model with terms that act on the interfaces only. Interesting enough, exploiting an 
extension of the Harada-Sasa relation, they found that the entropy production is concentrated
on the interfaces between dense and dilute regions of the samples. In our model, there is 
violation of FDT in the whole system, that is to say, on the boundaries but also in the 
bulk of the dense and dilute regions, and  there should be entropy production in the latter as well. 

Sobolev {\it et al.}~\cite{Sobolev20} proposed an effective temperature of an active colloidal system written as 
the product of the single particle effective temperature extracted from the violations of the FDT and another factor 
that takes into account the collective motion and plays a similar role to the one of an order parameter. 
The difference in the effective temperatures of the coexisting phases that we found cannot be reduced to this scenario 
since, in our model, the drift velocity is zero in both phases. Anyway, it would be interesting to test this idea in, for example,
dumbbell~\cite{gonnella2014phase,SumaPRE1,SumaPRE2,Cugliandolo15,Digregorio17, Petrelli18}, rod~\cite{Peruani15,Bar20} or even Hamiltonian~\cite{Bore16,Casiulis20} systems in which there is collective motion of the dense phase.
 
 Having found that heterogenous active systems are characterized by two global effective temperatures, 
 linked to the dilute and dense components, it was quite natural to complement its study with the one of 
 local fluctuations or, more precisely, follow the statistical behavior of the individual particles.
 
In the homogeneous active liquid, the probability distribution functions of the 
displacement and integrated linear response put in normal form (that is to say, with the average 
subtracted and divided by the standard deviation) fall on two master curves that are independent of the 
global packing fraction and the strength of the activity. (The distribution of the 
integrated linear response for Pe = 0 and $\phi=0$  at $t_w=0$ has the peculiarity of not having a left wing.)
Therefore, $\phi$ and Pe only affect the mean values, $\Delta^2$ and $\chi$.

In the heterogeneous phases,  we saw that the particles separate in two populations also 
from the point of view of the fluctuations. The fluctuations of the square displacement and linear response
of the particles in the dilute phase, once presented in normal form, coincide with the ones of the 
single passive particle. As in the homogeneous phases, Pe and $\phi$ affect the means $\Delta^2$ and $\chi$ but not the 
fluctuations around them. 
Instead, the fluctuations of the displacement of the 
particles in the dense phase change form. They show a Gaussian core and 
exponential tails both in the dense passive and active cases.
The tails of the distributions of the square displacement, the linear response and
the effective temperature are then modified by the changes in the statistics of 
$\tilde\Delta_{i,x}$. Therefore, in the dense phase not only the mean values, 
$\Delta^2$ and $\chi$, but also the fluctuations of these observables differ from the
ones in the dilute phase.

Let us compare  the fluctuations in the ABP model with the ones found in other out of equilibrium particle systems.
Spatial fluctuations of the displacements, linear responses and effective temperature 
in glassy systems were
considered in~\cite{Castillo02,Castillo03,Avila13}, see [\onlinecite{Chamon07}] for a review.
In these articles an emerging time-reparametrisation symmetry, 
in the long time delay relaxation dynamics, 
was claimed to constrain the noise induced fluctuations to be such that the global $\chi(\Delta^2)$
parametric relation remains satisfied even locally (after coarse-graining). Solvable models with diffusive (and critical) 
but not glassy dynamics were considered in~\cite{Corberi12} and the fluctuations were shown to 
behave differently from the ones in glassy models. The simulations we show align with the results 
in~\cite{Corberi12}. 

In experiments~\cite{wu,Leptos09,Ortlieb19} the fluctuations at the scale of microorganisms 
are characterized by analyzing the trajectories of colloidal tracer particles dispersed in the active fluids.
In the diffusive regime, the pdf of the displacement exhibit similar features to the ones 
we measured: they are well fitted 
by the sum of a Gaussian and an exponential
and they can be rescaled by their standard deviation~\cite{Leptos09}.
We note that no dense-dilute phase separation was reported in these experiments
and that, therefore, they have most probably been performed in homogeneous phases.
Similar exponential tails were observed in the statistics of the displacement of the center of mass 
and orientation of active dumbbells in sufficiently dense systems at intermediate and long time delays~\cite{Cugliandolo15}.
    
Quenched disorder in spin-glass models were shown to induce
site dependent violations of the fluctuation dissipation theorem as applied to the 
noise averaged local correlations and linear response functions~\cite{Roma07}. More precisely,
in these models the spins acquire an ``identity'', correlated with the 
backbone of the ground states, itself determined by the random 
interactions. The spins then separate in two population, one behaving as in the paramagnetic phase and the 
other behaving as in domain growth problems. Our results bear some resemblance with the ones in~\cite{Roma07} although 
in our problem the particles do not have a static nature, but they behave in a way or another 
during the given observation time interval only (they can belong to the dilute or dense phase, or they can move from one to the 
other).

We ended the paper with a short analysis of the kinetic temperature. 
The difference between this concept and the effective temperature
has been extensively discussed in the context of granular and glassy systems. Here we wanted to stress the 
fact that the former depends on the friction coefficient while the latter does not (when the transient 
has been overcome and the measurements are done in the asymptotic diffusive regime). As long as the 
 friction coefficient is not too weak the kinetic temperature of the whole system, but also the ones of the 
 two components in heterogeneous cases, approach the single particle one with trivial 
 Gaussian fluctuations. Therefore, one can not use the instantaneous kinetic temperature as 
 means to study emerging thermodynamic properties of active systems.

\acknowledgements

Simulations ran at Bari ReCaS e-Infrastructure funded by MIUR through PON Research and Competitiveness 2007-2013 Call 254 Action I. GG acknowledges MIUR for funding (PRIN 2017/WZFTZP ``Stochastic forecasting in complex systems"). 
We warmly thank P.  Digregorio and D. Levis for very useful discussions.

\bibliography{sample}

\end{document}